\documentclass[a4paper,12pt]{article}
\usepackage{amsfonts,amsmath,amssymb,amsthm}
\usepackage{tabstackengine}
\usepackage{flafter}
\usepackage{array}
\usepackage{empheq}
\usepackage{mathrsfs}
\usepackage[most]{tcolorbox}

\usepackage{tikz}
\usepackage{tikz-cd}
\usepackage[framemethod=TikZ]{mdframed}
\usepackage{caption}

\usepackage[bottom]{footmisc}

\usepackage[colorlinks=true,urlcolor=posurl,anchorcolor=posurl,citecolor=posurl,filecolor=posurl,linkcolor=posurl,menucolor=posurl,linktocpage=true,pdfa=true]{hyperref}

\usepackage{graphicx}

\usepackage{float}

\usepackage{tensor}

\usepackage{accents}

\usepackage{pos}

\newtcbox{\mymath}[1][]{%
    nobeforeafter, math upper, tcbox raise base,
    enhanced, colframe=blue!30!black,
    colback=blue!30, boxrule=1pt,
    #1}

\newtheorem{exercise}{Exercise}[section]

\newtheorem{definition}{Definition}[section]
\newtheorem{theorem}{Theorem}[section]

\numberwithin{equation}{section}

\def\Dcal{{\cal D}}
\def\cF{{\cal F}}

\def\Mcal{{\mathcal{M}}}
\def\cN{{\cal N}}
\def\Ncal{\mathcal{N}}

\def\Vcal{{\cal V}}

\def\Lcal{{\cal L}}

\def\Pbb{\mathbb{P}}
\def\Rbb{\mathbb{R}}
\def\Zbb{\mathbb{Z}}

\def\Urm{\mathrm{U}}

\newcommand{\Es}{\mathrm{E}_{7(7)}}
\newcommand{\Ed}{\mathrm{E}_{d(d)}}
\newcommand{\SO}{\mathrm{SO}}
\newcommand{\SL}{\mathrm{SL}}
\newcommand{\SU}{\mathrm{SU}}
\newcommand{\GL}{\mathrm{GL}}
\newcommand{\Spin}{\mathrm{Spin}}

\newcommand{\es}{\mathfrak{e}_{7(7)}}

\def\ld{\left.}
\def\rd{\right.}
\def\lb{\left[}
\def\rb{\right]}

\newmdenv[
  linecolor=blue!30, 
  linewidth=2pt,    
  frametitle={Summary}, 
  frametitlefont=\bfseries\large, 
  frametitlerule=false, 
  frametitlebackgroundcolor=blue!10, 
  roundcorner=5pt 
]{summaryFramed}

\title{Modave Lecture notes: \\ Introduction to Exceptional Field Theory}
 \ShortTitle{Introduction to ExFT}

\author*[a]{Colin Sterckx}

\affiliation[a]{INFN, Sezione di Padova,\\
Via Marzolo 8, 35131 Padova, Italy.}

\emailAdd{colin.sterckx@pd.infn.it}

\abstract{These notes are based on lectures given at the XIX Modave School on Mathematical Physics and present an introduction to Exceptional Field Theory. We cover the standard Kaluza-Klein reductions on tori, with applications to supergravity. We review the supergravity action of type IIA/IIB and eleven-dimensional supergravities. We motivate the appearance of the hidden symmetry groups $\Ed$ through the lens of string dualities. We explore the construction and properties of $\Es$-ExFT and review other $\Ed$-ExFT for $3\leq d\leq 8$, as well as double field theories. These notes conclude with some applications of ExFT to consistent truncations.}

\FullConference{Modave Summer School in Mathematical Physics (Modave2023)\\
 4-8 September, 2023\\
Modave, Belgium}

\tableofcontents

\begin{document}

\maketitle

\newpage
\paragraph{Note to the reader}

These notes are an extension of the lectures given at the XIX Modave Summer School in Mathematical Physics, in September of 2023, and titled ``Introduction to exceptional field theory''. The goal of these notes is not to be a comprehensive review of the literature on Exceptional Field Theory (ExFT) (see for example \cite{Samtleben2023,Berman2020,Musaev2019}) or an in-depth formal approach to ExFT. These are introductory lecture notes aimed at master/first year PhD students. I tried to avoid heavy mathematical formalism, but a lot of interesting concepts are named and some references are given in case the reader wishes to deepen certain subjects according to his/her preferences.

The prerequisites are differential geometry, Riemannian geometry (i.e., the usual GR prerequisite), and Lie groups and their representations (a reminder is included in the appendices). It is not strictly necessary to know about supergravity and superstring theory, but this might make this subject more appealing and interesting. For the purpose of these notes, supergravity can be thought of theories of gravity coupled to specific matter content. The sections making use of string theory (\ref{subsecStringDualities} and \ref{subsectionSuperStringDualities}), while useful to understand the origin of the hidden symmetry groups, can be skipped on a first read.

\newpage

\section{Introduction}

The notion of symmetry is at the core of modern mathematical physics. For example, Poincaré symmetry is the organising principle behind Einstein's theory of special relativity. By recognising the central role of this symmetry, it was possible to spectacularly simplify the formulation of the laws of physics known at the beginning of the $20^{\,\text{th}}$ century. In particular, one can reorganise Maxwell's equations by unifying the electric and magnetic fields in a Poincaré-covariant tensor, $F$, known as the electromagnetic field strength. Expressed using this tensor, the four Maxwell's equations reduce to two simpler equations:
\begin{equation}
d\star F = \star J \hspace{5mm}\text{and}\hspace{5mm} dF = 0\,.
\label{eq:U1eomBianchi}
\end{equation}
This not only makes the equations more aesthetically pleasing but also simplifies the study of electromagnetism, e.g., its U(1) gauge symmetry, its conservation laws, or its generalisations (e.g., to curved space). Exceptional field theory (ExFT) aims at a similar unification of ``maximal supergravities'', revealing a hidden $\Ed$ symmetry, akin to how Poincaré symmetry appears in Maxwell's equations.

Supergravities (SUGRA) are theories invariant under a special type of \emph{gauge} symmetries: supersymmetries \cite{Freedman2012,DallAgata2021}. These are symmetries whose parameter is a spin-$\tfrac{1}{2}$ fermion. Lagrangians invariant under local supersymmetries are strongly constrained. They must contain a graviton (hence the name supergravity) and one or more spin-$\tfrac{3}{2}$ particles: the "gravitini". In eleven dimensions, this symmetry is so restrictive that it only admits one unique action called 11D SUGRA. In ten dimensions, there are only two inequivalent maximal supergravities called type IIA and type IIB. These two supergravities can be interpreted as the low-energy actions of two corresponding theories of superstrings, also called type IIA and type IIB \cite{Becker2007,Green_Schwarz_Witten_2012}. Solving the equations of motion of these supergravities can help us understand the behaviour of strings on complicated backgrounds, probe the AdS/CFT correspondence, or even test universal properties of low-energy effective field theories (EFT) of quantum gravity (i.e., test swampland program conjectures \cite{vanBeest:2021lhn}). 

To make the $\Ed$ symmetries of these supergravities manifest, we will use a trick due to Kaluza and Klein \cite{Kaluza2018, Klein1926}. In their seminal work, they consider a five-dimensional theory of pure gravity on a manifold $\mathbb{R}^{1,\,3}\times S^1$. They show that, by taking the limit of a small $S^1$ radius, the 5D theory is equivalent to an effective four-dimensional theory of gravity coupled to a specific matter content whose action is
\begin{equation}
S = \int \mathrm{d}^4x \,e \left(R(e) - \frac{1}{4} e^{3\phi} F_{\mu\nu}F^{\mu\nu} - \frac{3}{2} \partial_\mu \phi \partial^\mu \phi\,\right)\,.
\end{equation}
This procedure is known as ``Kaluza-Klein compactification''. We will show in this notes how to perform these KK-compactification of pure gravity on $T^d$, showing that the lower dimensional theory admits an $\SL(d)$ global symmetry group, originating form diffeomorphisms on $T^d$. Furthermore, we will see that this symmetry group gets enhanced $\SL(d)\ltimes \Lambda^p \mathbb{R}$ when we add $p$-forms to the higher dimensional theory (with $p\leq d$). For very specific choices of higher-dimensional theories, i.e. specific choices of field content and coupling constants, these symmetries can get enhanced, without having a geometric interpretation in the higher-dimensional theory. We call such symmetries ``\emph{hidden symmetries}''. In the context of compactification of type II supergravities on $T^d$ we observe a hidden $\Ed$ global symmetry. 

The goal of Exceptional Field Theory (ExFT) is to reformulate type II and 11D SUGRA in a manifestly $\Ed$ covariant way. In the context of the ten- and eleven-dimensional supergravities, the ExFT reformulation is done in two steps. For simplicity of notation, we focus on the ten-dimensional case (the 11D case follows the same logic). The first step consists in organising the field content in $\Ed$-representations. To do so, we break the $\SO(1,\,9)$ Lorentz symmetry of the original theory into an $\SO(1,\,10-d) \times \SO(d-1)$ subgroup. In other words, we consider that the underlying manifold $\mathcal{M}_{\text{tot}}$ can be split as a product of two smaller manifolds, $\mathcal{M}_{\text{tot}}=\mathcal{M}_{\text{ext}} \times \mathcal{M}_{\text{int}}$. This splits coordinates into external, $x^\mu$, and internal, $y^m$, coordinates. Due to the specific field content and couplings of the type II supergravities, we can reorganise their fields into $\Ed$-representations. This is done by grouping together certain $\SO(d-1)$-representations to build $\Ed$-representations. Still, the equations of motion of the resulting, reorganised, theory are not $\Ed$-covariant. The issue is that the $y^m$ coordinates on $\mathcal{M}_{\text{int}}$ transform only under $\SO(d-1)$ and not under $\Ed$. To deal with this issue, we embed the internal coordinates $y^m$ into a larger set of coordinates $Y^M$, admitting an $\Ed$-representation. This permits the construction of the manifestly $\Ed$-covariant equations of motion. 

To summarise, in the first step we have broken the $\SO(1,\,9)$ Lorentz symmetry to $\SO(1,\,10-d) \times \SO(d-1)$. In the second step, we have enhanced it to $\SO(1,\,10-d) \times \Ed$. On the surface, the extra $Y^M$ coordinates play the same role as the extra circle in the KK compactification. The main difference, and feature of ExFT, is that the extra $Y^M$ coordinates are simply a trick to geometrise the $\Ed$ symmetry of ExFT. As we will see in the main text, the fields only depend on the original internal coordinates $y^m$. This is due to a series of constraints called the ``section constraints''. These constraints and their geometrical meaning are one of the specificities of generalised field theories, of which ExFT is just one example.

\paragraph{Plan of the Notes} We will start with three sections of introductory material before starting the study of ExFT properly speaking. In section \ref{sec:DimRed}, we review the basics of KK-compactifications on tori. This includes the notions of consistent truncation, effective field theory and homogeneous manifolds. We will present formulas for compactifications of gravity and $p$-forms on tori and we will provide a first example of hidden symmetry: Ehlers symmetry. In section \ref{sec:SUGRA}, we will give a brief introduction to the main concepts of supergravity. Focussing on bosonic sectors, we will review the action and equations of motion of the maximal supergravities in ten dimensions: type IIA and type IIB. We will also present the unique supergravity in eleven dimensions (11D SUGRA). We will show how 11D SUGRA compactified on a circle reduces to type IIA. In section \ref{sec:HiddSym}, we will present a brief review of dualities in the context of string theory. We will then show how these dualities can be used to predict the symmetry groups of maximal supergravity on tori. We will explicitly compute the reduction of the gravity limit of the bosonic string on $T^d$ and show that it enjoys an $\SO(d,\, d)$ symmetry group. We will motivate how the exceptional symmetry group appear in the context of superstrings. The goal of this section is to demystify the appearance and role of the exceptional Lie groups in string compactifications. These first sections contain a few lengthy ``exercises'' with solutions.

The lectures on ExFT start properly in section \ref{sec:E7ExFT}. We will explain how to build the $\Es$-ExFT. We begin by defining the notions of generalised Lie derivative, as well as that of internal and external generalised diffeomorphisms. We will then build manifestly $\Es$-covariant equations of motion. In section \ref{sec:genExFT}, we will present other types of generalised field theory such as the $\Ed$-ExFT for $3\leq d \leq 8$. But we will also touch on double field theory (DFT), the $\SO(d,\,d)$ cousin of ExFT, as well as an $\SO(d,\,d+N)$ generalised field theory. We will summarily present the challenges that appears for low number of external dimensions. Finally, section \ref{sec:Applications} is devoted to applications of ExFT. We will focus on  consistent truncations methods making use of the hidden symmetry groups. Several ``Summary'' bubbles are sprinkled throughout these notes and should help you get the main points as efficiently as possible. References to either textbooks or the original papers are available throughout the main text.

\newpage

\section{Dimensional reduction, truncations and consistency}
\label{sec:DimRed}

As explained in the introduction, certain information on string theory phenomenology can be obtained from solving a specific set of partial differential equations in ten dimensions (the SUGRA equations). These equations are often hard to solve as they originate from the Einstein-Hilbert action coupled to $p$-forms and scalar fields. A way to simplify these equations can be obtained by \emph{dimensional reduction}. The seminal idea is due to Kaluza and Klein \cite{Kaluza2018, Klein1926} and is to assume that the theory we are studying is defined on a $D+d$ dimensional manifold $\mathcal{M}_{\text{Tot}}$ which is the product of an ``\emph{external}'' $D$-dimensional manifold and an ``\emph{internal}'' $d$-dimensional manifold:
\begin{equation}
	\mathcal{M}_{\text{Tot}} = \Mcal_{\text{ext}} \times \Mcal_{\text{int}}\,.
\end{equation}
Although all manifolds are locally of this form, this is a strong assumption that cannot be made for any geometry (e.g. a sphere is not the Cartesian product of two lower-dimensional manifolds). However, it is a useful one. It will allow us to reinterpret the ($D+d$)-dimensional theory in terms of a $D$-dimensional one which might be simpler.

This can be done in two ways. The first one is to consider the internal manifold to be compact and small compared to some energy scale $\Lambda$. Then, the physic observed at scales below $\Lambda$ will be effectively independent on any internal coordinate and the low energy effective theory will only be sensitive to the external directions. The other way is to choose an ansatz fixing the internal coordinate dependence of the fields in a (very) particular way. It may happen that for an appropriate choice of ansatz, the internal coordinate dependence factors out of the equations of motion, resulting in a theory which is effectively $D$-dimensional. 

In this section we will sketch how to compactify theories. First, we will study the simple model of a scalar field reduced on $S^1$. This will allow us to introduce notions of effective field theory and consistent truncations. We will then show how to dimensionally reduce Einstein theory on a torus by studying the interplay between symmetries and dimensional reduction. As a first example, we will reduce pure Einstein gravity to three dimensions and have a first encounter with ``hidden symmetries''. Finally, we will study the reduction of a theory containing $p$-forms. We followed conventions of \cite{Samtleben2023}. A classic review on KK reduction in the context of supergravity is \cite{Duff1986}.

\subsection{\texorpdfstring{Lessons from scalars on $S^1$}{Lessons from scalars on the circle}}

Let us start with an example: the free complex scalar compactified on a circle. In this setup, we consider the manifold 
\begin{equation}
\Mcal_{\text{Tot}} = \Rbb^{1,\,D-1}\times S^1\,.
\end{equation} 
The internal manifold is the circle $S^1$ of radius $R$, and the external space is the flat Minkowski space. The action for the free massless scalar is
\begin{equation}
	S_0[\phi(x,\,y)] = \int_{\Mcal_{\text{Tot}}} d^{D}x\, dy \, \left(\partial_\mu \phi \partial^\mu \bar{\phi} + \partial_y \phi \partial^y \bar{\phi}\right)\,, \label{eq:freeScalarS1}
\end{equation}
where we have introduced coordinates $x^\mu$ on $\mathbb{R}^{1,D-1}$ and $y \in \left[0,\,2\pi R\right[$ on $S^1$. Having singled out a specific compact direction, one can perform a Fourier mode expansion for $\phi$ along the $S^1$:
\begin{equation}
	\phi(x,\,y) = \sum\limits_{n\,\in\, \mathbb{Z}} e^{i\, \frac{n}{2\pi R}y} \, \phi_n(x)\,.\label{eq:FdecScalarS1}
\end{equation}
This replaces the unique, $y$-dependent, scalar field $\phi(x,\,y)$ by a ``tower'' of fields $\phi_n(x)$. These only depend on the external coordinates $x^\mu$ and are thus $D$-dimensional scalar fields. Notice that the ``momenta'': $n/(2\pi R)$ take discrete values because of the periodicity of the $y$ variable. It is the discreteness of the parameter $n$ (which reflects the compactness of $S^1$) which will allow us to reduce to a proper $D$-dimensional field theory in a moment.

Plugging \eqref{eq:FdecScalarS1} in the equations of motion of the action \eqref{eq:freeScalarS1}, one gets
\begin{equation}
	\frac{\delta S_0}{\delta \phi} = 0 \hspace{5mm}\Leftrightarrow\hspace{5mm} \partial_\mu \partial^\mu \phi_n  - \frac{n^2}{(2\pi R)^2} \phi_n = 0\,,\,\,\forall \, n \in \mathbb{Z}\,.
	\label{eq:freeScalarS1eom}
\end{equation}
We recognise the equations of motion of free scalars, of masses $m^2_n = \frac{n^2}{(2 \pi R)^2}$. Interestingly, one can also obtain this result by plugging the Fourier decomposition directly in the action \eqref{eq:freeScalarS1} and then by computing its variation with respect to each of the $\phi_n(x)$ fields independently:
\begin{align}
	S_0[\phi(x,\,y)] &= \int d^Dx\,dy \sum\limits_{(n,\,m)\in\mathbb{Z}^2}\,\exp\left(i\frac{(n-m) y}{(2\pi R)}\right) \left[\partial_\mu \phi_n \partial^\mu \bar{\phi}_m - \frac{n\,m}{(2\pi R)^2} \phi_n \bar{\phi}_m\right]\\
	&= (2\pi R) \int d^Dx \sum\limits_{n\in\mathbb{Z}}\left(\partial_\mu\phi_n\partial^\mu\bar{\phi}_n - \frac{n^2}{2\pi R}|\phi_n|^2 \right)\,.
	\label{eq:ActionFreeScalarS1}
\end{align}
In the first line, we just plugged the Fourier decomposition of $\phi$ in the action. In the second line we integrated over $S^1$. We find the action for an infinite set of free scalars in $D$ dimension each of mass $m^2_n = \frac{n^2}{(2\pi R)^2}$. This shows that, in the case of a compact internal space, and without putting any care to mathematical rigour, a theory in $D+d$ dimensions is equivalent to a theory in $d$ dimensions. The price to pay is that the $d$-dimensional theory will contain an infinite number of degrees of freedom: the full tower of excitations, $\phi_n$.

\subsubsection*{Integrating out vs. consistent truncations}

To obtain a proper $D$-dimensional field theory, i.e., one with a finite number of degrees of freedom, we must specify a procedure to keep only a finite number of fields $\phi_n$. There are two very different methods to do this. The first one is to make an approximation. Understanding \eqref{eq:ActionFreeScalarS1} as an effective theory, with a cut-off scale $\Lambda_{UV}$, the excitations $\phi_n$ become very massive for $|n| > \Lambda_{UV} (2\pi R)$. One could then \emph{integrate out} these d.o.f. à la Wilson. This approximation would only retain a finite number of fields. Intuitively, a low energy experiment would not be able to see the extra dimension $S^1$ without accessing to length smaller than $R$. The effective theory would be a $D$-dimensional field theory. This approximation relies on the fact that $\mathcal{M}_{int}$ is compact and thus there is a discrete number of excitations, otherwise there would always be an infinite number of modes below any cut-off scale.

There is a more subtle way to remove the infinite tower of particles which consists in performing a ``consistent truncation''. The truncation is simply the operation setting almost all fields to a certain fixed value (in our case zero). This effectively removes, or ``\emph{truncates}'', degrees of freedom. To understand what a ``consistent'' truncation is, we will illustrate it with an example. Let us add interactions to our free scalar theory by introducing a $\lambda \,|\phi|^4$ term in the action. The relevant action is now
\begin{equation}
	S_\lambda[\phi(x,\,y)] = S_0[\phi(x,\,y)]+ \int d^Dx \,dy \, \lambda |\phi|^4\,,
\end{equation}
whose equations of motion are
\begin{equation}
	\frac{\delta S_\lambda}{\delta \phi} = 0 \hspace{2mm}\Leftrightarrow \hspace{2mm} \partial_\mu\partial^\mu \phi + \partial_y\partial^y \phi - 2 \lambda |\phi|^2\phi = 0\,.
\end{equation}
Plugging the Fourier decomposition of $\phi$ in these equations of motion, one gets
\begin{equation}
	\partial_\mu\partial^\mu \phi_n - \frac{n^2}{(2\pi R)^2}\phi_n - \sum\limits_{k +l -m =n} 2 \lambda \phi_k \phi_l \bar{\phi}_m =0\,\label{eq:eomScalarS1}.
\end{equation}
On the other hand, plugging the Fourier decomposition in the action one gets
\begin{equation}
	S_\lambda[\phi_n(x)] = S_0[\phi_n(x)] - (2\pi R)\,\lambda\, \int d^Dx\, \sum\limits_{(n,\,m,\,k,\,l) \in \mathbb{Z}^4} \delta(n + m - k - l) \, \phi_n\phi_m \bar{\phi}_k\bar{\phi}_l\,.
\end{equation}
You can observe that the different modes now source each other. If we perform a random truncation, e.g. by selecting exactly two non-zero modes, $\phi_{n_1}$ and $\phi_{n_2}$, the equations of motion in the $D+1$ dimensional theory for the mode $\phi_{2n_1 - n_2}$ is
\begin{equation}
	\partial_\mu \partial^\mu \phi_{2 n_1 -n_2} - \frac{(2 n_1 -n_2)^2}{(2\pi R)^2}\phi_{2 n_1 -n_2}  + 2 \lambda \phi_{n_1}\phi_{n_1} \bar{\phi}_{n_2} =0\, ,
\end{equation}
where we have used that only $\phi_{n_1}$ and $\phi_{n_2}$ are dynamical (i.e., can be non-zero). This equation shows that  $\phi_{2n_1 - n_2} = 0$ is not implied by the e.o.m. for $\phi_{n_1}$ and $\phi_{n_2}$. In other words, $\phi_{n_1}$ and $\phi_{n_2}$ \emph{source} $\phi_{2n_1 - n_2}$. This example illustrates the fact that one cannot simply truncate any modes at will and expect the e.o.m. of the truncated theory to solve those of the full theory. The reason being that a mode truncated away can be sourced by a mode that was kept in the lower-dimensional theory. 

Truncations with the property that solutions of the truncated theory also solve the e.o.m. of the full theory are called ``\emph{consistent}''. In general, it is hard to build a consistent truncation or to prove that a certain truncation is consistent. However, the following result can help us in many cases:
\begin{mdframed}
        For an action $S$, invariant under a group $G$ (not necessarily Lie), the truncation to the $G$-invariant sector is consistent \cite{Coleman1985App}.
\end{mdframed}
Intuitively, it is true because the variation of the full action w.r.t. a truncated mode will transform non-trivially under the action of $G$. Thus, any source term in this equation must also transform in a non-trivial $G$-representation. Since it is impossible to make a non-invariant term with only the invariant modes kept in the theory, setting all the non $G$-invariant terms to zero makes the source term vanish\footnote{Very schematically, the idea of a proof would go as follow. To any element $g\in G$ corresponds an automorphism $\phi_g$ of the phase space. Since $\phi_g$ is a symmetry of the action we have that $S\circ \phi_g = S$. Thus, at a fix point $p=\phi_g(p)$, we have $\delta S_{|p} = \phi_g^* \delta S_{|\phi_g(p)=p}$. For any vector not in the kernel of $\text{Id} - \phi_{g\,*\,|p}$, i.e., a variation with respect to a non-invariant mode, we must have that $\delta_v S = 0$. For any vector $s$ such that $\text{Id} = \phi_{g\,*\,|p}$ we obtain the equations of motion of the truncated theory at $p$: $\delta_s S$. Applying the argument for any $g \in G$, we obtain the result.}. We emphasised that, for this argument to hold, one needs to keep \emph{all} the $G$-singlets of the initial theory.

As an example, let us build $G$-invariant truncations for the interacting scalar on $S^1$. First, observe that our theory is invariant under translations along the $y$ coordinate. This corresponds to the action of a $\Urm(1)$ group under which the field $\phi_n$ has a charge $n$. Consider a truncation keeping only fields invariant under this $\Urm(1)$ symmetry. This keeps only the zero mode $\phi_0$ which yields a consistent truncation, indeed, $\phi_0$ alone does not source any other field. This procedure can be applied to any subgroup of $\Urm(1)$. For example, considering the modes invariant under $\mathbb{Z}_k \in \Urm(1)$, i.e. the modes $\phi_{n \,k}$ for $n\in\mathbb{Z}$, we can build other consistent truncations of the full interacting theory (even if they retain an infinite number of fields). Notice that in the $\mathbb{Z}_k$ truncation, we do not keep only the modes of lowest masses in the tower of excitations. Finally, this method does not rely on the compactness of $\mathcal{M}_{int}$ as was the case in the EFT approach (e.g., it works for reduction on $\mathbb{R}^d$ in almost the same way it works for reductions of $\mathbb{R}^d$, replacing the role of the $\Urm(1)$ symmetry by a $\mathbb{R}$ symmetry).

\subsubsection*{ }
\begin{exercise}
Consider the following action for $\phi$, a complex scalar, and $\lambda$, a real scalar:
\begin{equation}
	S[\phi,\,\lambda] = \int d^dx\, \partial_\mu \phi \partial^\mu \bar{\phi} + \partial_\mu \lambda \partial^\mu \lambda + g \,\lambda \phi \bar{\phi}\,.
\end{equation}
\begin{itemize}
\item Compute the equations of motion of $S[\phi,\,\lambda]$.
\item Check that $\lambda = 0$ is not a consistent truncation.
\item Check that $\phi = 0$ is a consistent truncation. Can you explain this by a symmetry argument?
\end{itemize}
\end{exercise}
\paragraph{Solution}
The e.o.m. for this action are
\begin{align}
	\frac{\delta S}{\delta \lambda}=0 \Leftrightarrow \square \lambda - g |\phi|^2 = 0 \, ,\label{eq:dSdlambda}\\
	\frac{\delta S}{\delta \phi} = 0 \Leftrightarrow \square \phi - g \lambda \phi = 0 \, .
\end{align}
We consider the two truncations:
\begin{equation}
\begin{array}{cc}
	\hspace{1cm}\lambda = 0 \hspace{1cm}&\hspace{1cm} \phi =0\hspace{1cm}\\[2mm]
	S[\phi,\,0] = \int \mathrm{d}x \partial_\mu \phi \partial^\mu \bar{\phi}\hspace{1cm} & \hspace{1cm} S[0, \lambda] = \int \mathrm{d}x \partial_\mu \lambda \partial^\mu \lambda\\[2mm]
	\frac{\delta S[\phi,\,0]}{\delta \phi} = 0 \Rightarrow \square \phi =0 \hspace{1cm} & \hspace{1cm} \frac{\delta S[0, \lambda]}{\delta \lambda} = 0 \Rightarrow \square \lambda = 0
\end{array}
\end{equation}
\begin{itemize}
\item Setting $\lambda = 0$, the equations of the truncated theory impose $\square \phi =0$ whereas those of the full theory \eqref{eq:dSdlambda} impose that $\phi = 0$. This truncation is thus \emph{not} consistent.
\item Setting $\phi = 0$ imposes $\square \lambda =0$ which implies the full e.o.m. and makes the truncation consistent. 
\end{itemize}
The action admits a U(1) global symmetry acting on $\phi$. This second truncation corresponds to keeping all the U(1)-invariant modes, i.e. fixing $\phi=0$.

\begin{summaryFramed}
\begin{itemize}
	\item Defining a $D+d$ theory on the product of a compact ``internal'' $d$-dimensional manifold and an ``external'' $D$-dimensional manifold allows one to re-express it as a $D$-dimensional theory with a countable, infinite, number of degrees of freedom.
	\item One can get rid of the infinite degrees of freedom by taking a limit where characteristic length of the internal manifold goes to zero and then integrate out massive modes. This requires introducing a cut-off scale above which this approximation breaks down.
	\item One can truncate modes, independently of their masses, by keeping the modes invariants under a certain symmetry of the full theory. This is an exact procedure, which means that any solution of the truncated theory uplifts to a solution of the full theory. This method is called a ``\emph{consistent truncation}''.
\end{itemize}
\end{summaryFramed}

\subsection{\texorpdfstring{Einstein theory on $T^d$}{Einstein theory on tori}}

Now that we have understood basic properties of dimensional reductions, we will perform the reduction of the Einstein-Hilbert (E-H) action on a torus. This is the first step in studying more complicated reductions of theories of gravity coupled to matter.

The E-H action depends on a vielbein\footnote{The $\SO(1,\,D+d-1)$ quotient identifies two vielbein related by a local Lorentz transformation}
\begin{equation}
{E_M}^A(Y) \in \GL(D+d)/\SO(1,\,D+d-1)
\end{equation}
where $A = 1,\,\cdots,\,D+d$ is the $\SO(1,\,D+d-1)$ index of Local Lorentz Transformations (LLT) and $M$ is the $D+d$ coordinate index. The action is
\begin{equation}
	S[E] = \int\,\mathrm{d}^Dx\,\mathrm{d}^dy\,|E|\, R_{(D+d)}(E)\,,
\end{equation}
where we have introduced the coordinates $x^\mu$ on $\mathcal{M}_{\text{ext}}$ and $y^m$ on $T^d$, splitting the coordinates $Y^M$ into external and internal coordinates. The function $R_{(D+d)}(E)$ is the Ricci scalar of ${E_M}^A$ and $|E|$ is the absolute value of the determinant of the vielbein $E$.
 
At this stage it would be possible to decompose the vielbein in Fourier modes and compute the full action in terms of those modes. However, this would be a tedious task. It is more interesting to directly study a consistent truncation of this theory. Since translations along the torus directions are symmetries of the Einstein action, keeping only the zero Fourier modes, i.e. removing the $y^m$ dependence in our fields, is consistent\footnote{These are the modes invariant under the group $\textrm{U}(1)^d$, corresponding to $y^m$ translations.}. It is now manageable to directly compute the action of the truncated theory. Studying first the symmetries that the truncated theory must inherit from the full theory simplifies this computation. Since studying symmetries is more fun than brute-force computations, this is what we will do in this section.

\subsubsection{The KK ansatz and symmetries}

We first bring the vielbein to an upper diagonal form (using LLT) and write it as
\begin{equation}
	{E_M}^A = \begin{pmatrix}
		e^{-\phi/(D-2)} {e_\mu}^a & e^{\phi/d}  {A_\mu}^m{V_m}^{\underline{a}}\\ 0 & e^{\phi/d} {V_m}^{\underline{a}}
	\end{pmatrix}\,.
	\label{eq:AnsatzVielbein}
\end{equation}
We have defined ${e_\mu}^a \in \GL(D)/\SO(1,\,D-1)$ and ${V_m}^{\underline{a}} \in \GL(d)/\SO(d)$. This choice of vielbein explicitly breaks the LLT down to $\SO(1,\,D-1)\times \SO(d)$ local transformations acting, respectively, on ${e_\mu}^m$ and ${V_m}^{\underline{a}}$. The goal of this parametrisation is to simplify the study of symmetries in the context of dimensional reduction.

The $D+d$ E-H action is invariant under the diffeomorphisms, which are generated by the Lie derivative
\begin{equation}
	\mathcal{L}_\xi {E_M}^A = \xi^N\partial_N {E_M}^A + (\partial_M \xi^N) {E_N}^A\,,
\end{equation}
as well as the LLT, acting on the index $A$. The truncated theory inherits a subset of these symmetries, the ones generated by a vector $\xi$ such that 
\begin{equation}
\partial_m (\mathcal{L}_\xi {E_M}^A) =0\,.
\end{equation}
If this condition is not satisfied the transformation will turn on modes which lie outside of the truncation, i.e., which have $y^m$ dependencies. There are three families of $\xi$ whose associated symmetry descend to the lower dimensional theory:
\begin{itemize}
	\item $\xi_{(D)} = \xi^\mu(x) \partial_\mu$. This corresponds to a coordinate change in the $D$-dimensional theory. We can indeed verify that the transformations $\delta_{\xi_{(D)}} = \mathcal{L}_{\xi_{(D)}}$ reduce to the Lie derivative in $D$-dimensions. As such, ${e_\mu}^a$ and ${A_\mu}^m$ are $D$-dimensional vectors, whereas ${V_m}^a$ and $\phi$ are scalars, as they should.
	\item $\xi_{\text{gauge}} = \xi^m(x) \partial_m$. This corresponds to $\Urm(1)^d$ gauge transformations of the vector ${A_\mu}^m$:
	\begin{equation}
		\delta_{\xi_{\text{gauge}}} {A_\mu}^m = \partial_\mu \xi^m\,,
	\end{equation}
	with all other fields uncharged under the $\Urm(1)^d$ gauge group.
	\item $\xi_{\text{SL}} =  {\Lambda^{m}}_n y^n \partial_m$ where $\Lambda \in \mathfrak{sl}(d,\,\mathbb{R})$ and which corresponds to a global $\SL(d)$ symmetry of the truncated theory\footnote{A subtlety: note that the $\xi_{(SL)}$ vectors are not well defined on $T^d$, due to their non-periodicity in $y^m$. The group of symmetries of the truncated sector is actually the subgroup of $\text{Diff}(\mathcal{M}_{\text{ext}} \times T^d)$ preserving the truncation. It contains a $\SL(d,\,\mathbb{Z})\subset \text{Diff}(T^d)$ group whose action on the fields is that of \eqref{eq:SLdAction}. This group is enhanced in the lower dimensional theory to $\SL(d,\,\mathbb{R})$. This is an artefact of the truncation. Indeed, since we only keep $\Urm(1)$-invariant modes, the same $D$-dim theory could have been obtained from a reduction on $\mathbb{R}^d$ instead of $T^d$. On this space, $\xi_{(\SL)}$ is well defined and this predict the enhancement of symmetry from $\SL(d,\,\mathbb{Z}) \rightarrow \SL(d,\,\mathbb{R})$. The same behaviour appears in the context of exceptional duality group later in these notes.}.
	\begin{equation}
		\delta_{\xi_{(\SL)}} {A_\mu}^m = - {\Lambda^m}_n {A_\mu}^n \hspace{5mm} \delta_{\xi_{(\SL)}} {V_m}^{\underline{a}} = {\Lambda^n}_m {V_n}^{\underline{a}}\hspace{5mm} \delta_{\xi_{(\SL)}} {e_\mu}^a = 0 = \delta_{\xi_{(\SL)}}\phi
	\label{eq:SLdAction}
	\end{equation}
\end{itemize}
The full LLT breaks down to a lower dimensional LLT generated by the group $SO(1,\,D-1)$, acting on the indices $a$, and a gauge invariance with the group $\SO(d)$ acting on the fundamental $\underline{a}$ indices. 

This study shows that the $D$-dimensional theory's field content will be the $\tfrac{d(d+1)}{2}$ scalars from the internal vielbein (${V_m}^{\underline{a}}$ and the ``\emph{dilaton}'' $\phi$), the $d$ $U(1)$-vectors from the off-diagonal block of the metric (${A_\mu}^m$), and a $D$-dimensional vielbein (${e_\mu}^a$). This implies that the reduced action should be written in terms of a field strength 
\begin{equation}
{F_{\mu\nu}}^m = \partial_{[\mu}{A_{\nu]}}^m
\label{eq:defF}
\end{equation}
and a metric on the scalar manifold
\begin{equation}
	M_{mn} = {V_m}^{\underline{a}} {V_n}^{\underline{b}} \delta_{\underline{a}\underline{b}}\,,
	\label{eq:defM}
\end{equation}
(whose inverse is denoted $M^{mn}$) which are both invariant under all gauge transformations of the reduced theory.

\subsubsection{The reduced action}
For the reader, we have collected useful formulas concerning the vielbein and the metric of the KK ansatz.
\begin{equation}
	{E_M}^A = \begin{pmatrix}
		e^{-\phi/(D-2)} {e_\mu}^a & e^{\phi/d}  {A_\mu}^m{V_m}^{\underline{a}}\\ 0 & e^{\phi/d} {V_m}^{\underline{a}}
	\end{pmatrix}\,.
	\label{eq:AnsatzVielbein2}
\end{equation}
\begin{equation}
	(E^{-1})\indices{_A^M} = \begin{pmatrix}
		e^{\phi/(D-2)} {(e^{-1})_a}^\mu & - e^{\phi/(D-2)}  {(e^{-1})_a}^\mu{A_\mu}^m\\ 0 & e^{-\phi/d} {(V^{-1})_{\underline{a}}}^{m}
	\end{pmatrix}\,.
	\label{eq:AnsatzVielbeinInverse}
\end{equation}
\begin{equation}
\text{det}({E_M}^A) = e^{-\frac{2}{D-2}\phi} \text{det}({e_\mu}^a)
\end{equation}
\begin{equation}
	G_{MN} = \begin{pmatrix} e^{-2\frac{\phi}{D-2}} g_{\mu\nu} + e^{\frac{2\phi}{d}} {A_\mu}^m M_{mn} {A_\nu}^n & e^{\frac{2\phi}{d}} {A_\mu}^m M_{mn} \\e^{\frac{2\phi}{d}}{A_\mu}^m M_{mn}  & e^{\frac{2\phi}{d}} M_{mn}\end{pmatrix}
\end{equation}
\begin{equation}
	(G^{-1})^{MN} = \begin{pmatrix} e^{2\frac{\phi}{D-2}} g^{\mu\nu}  & -e^{2\frac{\phi}{D-2}} g^{\mu\nu} {A_\mu}^m  \\-e^{2\frac{\phi}{D-2}}g^{\mu\nu}{A_\mu}^m & e^{-\frac{2\phi}{d}} M^{mn} + e^{2\frac{\phi}{D-2}} {A_\mu}^m g^{\mu\nu} {A_\nu}^n\end{pmatrix}
\end{equation}
The symmetries of the full theory fix completely the two-derivatives action of the reduced theory, up to coefficients. A tedious computation yields the Lagrangian
\tcbset{highlight math style={boxsep=5mm,colback=blue!30!red!30!white}}
\begin{empheq}[box=\mymath]{align}
    \label{eq:ReductionMetricS1}
	\mathcal{L}_{(D)} e^{-1} = &\, R_{(D)} + \frac{1}{4} \partial_\mu M_{mn} \partial^\mu M^{mn} - \beta \partial_\mu \phi \partial^\mu \phi\\
	\nonumber &- \frac{1}{4} e^{2 \beta \phi} M_{mn} {F_{\mu\nu}}^m {F^{\mu\nu}}^n + \text{total derivatives}\,,
\end{empheq}
where $R_{(D)}$ is the Ricci scalar of ${e_\mu}^a$ in $D$ dimensions and $M_{mn}$ and $F_{\mu\nu}{}^m$ are defined in \eqref{eq:defM} and \eqref{eq:defF}. The coefficient $\beta$ is
\begin{equation}
\beta = \frac{D+d-2}{d(D-2)}.
\end{equation}
Perhaps surprisingly, the reduced KK action is invariant under an extra $\mathbb{R}$-symmetry, enhancing SL($d$) to GL($d$). This symmetry acts as a rescaling of the fields of the theory
\begin{equation}
	\delta \phi = \lambda \hspace{5mm}\text{and}\hspace{5mm}\delta {A_\mu}^m = -\lambda\, \beta \, {A_\mu}^m\,.
	\label{eq:ReducedTrombone}
\end{equation}
This enhancement of the symmetries is due to the ``\emph{trombone}'' symmetry of the Einstein action. Under a constant rescaling of the vielbein, the equations of motion are still satisfied; and the action is just rescaled. This is this symmetry (of the e.o.m. and not of the action) that reduces to the $\mathbb{R}$-symmetry in $D$ dimensions.

\subsubsection{A word about coset spaces}

The scalar fields of a reduced theory are going to be the coordinates of a manifold of the form
\begin{equation}
	\Mcal_{\text{scal}}=G/H
\end{equation}
where $G$ and $H$ are two Lie groups (not necessarily semi-simple). Manifolds of this type are called \emph{homogeneous} (see Chapter 1 of \cite{Cap2009}). For these types of scalar manifolds, we want to build a $G$-invariant kinetic term. This is completely equivalent to building a $G$-invariant Riemannian metric on $\Mcal_{\text{scal}}$. Since this issue will appear more than once in these notes, we add this small section to present the main formulas concerning certain homogeneous manifolds, with a very pedestrian approach. That being said, we name several concepts and add references to allow the interested reader to explore the subject in more depth.

\paragraph*{SL(2)/SO(2)} Let us start by an example to fix ideas before generalising to other coset spaces. Our first goal is to build a kinetic term for the scalars parametrising the space $\SL(2)/\SO(2)$ which is invariant under the global $\SL(2)$ group. A good way to parametrise the SL(2)/SO(2) coset space is to consider this space as upper-triangular matrices of determinant one. This fixes uniquely the SO(2) gauge, and a standard parametrisation is
\begin{equation}
	\mathcal{V} = \exp\left(\chi\, e_+\right)\cdot \exp\left(\frac{\phi}{2} h\right)
\end{equation}
where we used the basis of $\mathfrak{sl}_2$:
\begin{equation}
	h = \begin{pmatrix}
		1 & 0 \\ 0 & -1
	\end{pmatrix}\hspace{5mm},\hspace{5mm}e_+ = \begin{pmatrix}
		0 & 1 \\ 0 & 0
	\end{pmatrix} \hspace{5mm}\text{and}\hspace{5mm}e_- = \begin{pmatrix}
		0 & 0 \\ 1 & 0
	\end{pmatrix} \,.
\end{equation}
From the matrix $\mathcal{V}$, we can build the $\SO(2)$-gauge invariant symmetric matrix $M = \mathcal{V}\mathcal{V}^T$ (this should remind you of how to build a metric out of a vielbein). This matrix $M$ is invariant under the right-action of $\SO(2)$ on $\mathcal{V}$ \footnote{In this context, a mathematician would just say ``well-defined'' instead of gauge ``invariant''.}, and it is positive definite. Now, we can build the kinetic term
\begin{equation}
	\mathcal{L}_{kin}\,e^{-1} = \frac{1}{4} \text{Tr}\left[\partial_\mu M \partial^{\mu}(M^{-1})\right]\, = -\frac{1}{2} \left(\partial_\mu \phi \partial^\mu \phi + e^{-2 \phi} \partial_\mu \chi \partial^\mu \chi\right)\,,
	\label{eq:SL2kineticTerm}
\end{equation}
which is SL(2)-invariant and well defined, as required. The metric on $\SL(2)/\SO(2)$ reads
\begin{equation}
ds^2 = d\phi^2 + e^{-2\phi} d\chi^2\,.
\end{equation}

\paragraph*{SL(n)/SO(n)} We can generalise this method to the coset-space $\SL(n)/\SO(n)$. Starting with a good choice of Cartan generators $\vec{h}$ of $\mathfrak{sl}_n$\footnote{For example we can take $h_i = \text{diag}(\underbrace{0,\cdots,\,0}_{i-1 \text{ times}},\, 1,\,-1, 0\cdots)$ for $i=1,\,\cdots,\,n-1$.}, we can build the set of positive roots $\left\{k^\alpha_+\right\}$ (and the negative roots $\left\{k^\alpha_-\right\}$). The vector space generated by the positive roots and the Cartan subalgebra is the solvable algebra of traceless upper triangular matrices. An element of $\SL(n)/\SO(n)$ can be parametrised as
\begin{equation}
	\mathcal{V} = \exp\left(\chi_\alpha \,k^\alpha_+\right)\cdot\exp\left(\vec{\phi}\cdot\vec{h}\right)\,.
	\label{eq:SolvablePSLn}
\end{equation} 
Once again, we build the SO($n$) invariant matrix $M = \mathcal{V}\mathcal{V}^T$ and the kinetic term
\begin{equation}
	\mathcal{L}_{\text{kin}}\, e^{-1} = \frac{1}{4} \text{Tr}\left[\partial_\mu M\partial^\mu(M^{-1})\right]
	\label{eq:kinScalM}
\end{equation}
is well defined, since we have fixed uniquely a representative in each $\SL(n)/\SO(n)$ coset by \eqref{eq:SolvablePSLn}, and it is invariant under a global $\SL(n)$ action.

\paragraph{G/H} For a generic homogeneous manifold $G/H$, one would be tempted to generalise the method we used for the $\SL(n)/\SO(n)$ coset-space. However, this method relied on the (semi-)simplicity of $G$ and the fact that $H$ was its maximal compact subgroup. If those conditions are not satisfied, the gauge fixing procedure (which relied on a choice of Cartan generators and roots) might not be well defined or the obtained metric might not be Riemannian. In other words, the kinetic term for the scalars might have zero-modes or ghosts. 

There exists a sufficient condition\footnote{Corollary 1.4.4 of \cite{Cap2009}} for generic Lie groups $G$ and $H$, however, to simplify the discussion, we will impose certain constraints on $G$ and $H$. We will first work in the case where $G$ is (semi-)simple and $H$ is its maximal compact subgroup (this is an example of \emph{symmetric reductive manifold} see \cite{Helgason1978}). In this case, we have a non-degenerate metric on $\mathfrak{g}$: the Cartan-Killing metric. This metric allows us to split $\mathfrak{g}$ into two orthogonal pieces $\mathfrak{g} = \mathfrak{h} \oplus \mathfrak{K}$ where $\mathfrak{h}$ is the Lie algebra generating $H$ and $\mathfrak{K}$ is its orthogonal complement. For symmetric reducible spaces we have the property that $\mathfrak{h}$ and $\mathfrak{K}$ satisfy the conditions
\begin{equation}
	\underbrace{[\mathfrak{h},\,\mathfrak{h}]\subset \mathfrak{h}}_{\mathfrak{h}\text{ is a subalgebra of } \mathfrak{g}},\hspace{8mm}\underbrace{[\mathfrak{h},\,\mathfrak{K}]\subset\mathfrak{K}}_{\text{the pair }(\mathfrak{h},\,\mathfrak{K}) \text{ is reductive}}\hspace{8mm} \text{ and }\hspace{8mm} \underbrace{[\mathfrak{K},\,\mathfrak{K}] \subset \mathfrak{h}}_{G/H \text{ is symmetric}}\,.
\end{equation}
This implies that $\mathfrak{K}$ admits a $H$-representation. One can canonically identify the tangent space on $G/H$ with $\mathfrak{K}$ and the Cartan-Killing metric on $G$ induces a Riemannian metric on $G/H$. This suggests that we can build a unique representative of each class of the quotient space by exponentiating $\mathfrak{K}$. This is essentially correct. 

We review two useful parametrisations of $G/H$. The first one consists in defining an element of $G/H$ as an element of $\exp(\mathfrak{K})$. In this parametrisation:
\begin{equation}
	\mathcal{V}(\phi^r) = \exp(\phi^r K_r)
\end{equation}
for $\left\{K_r\right\}$ a basis of $\mathfrak{K}$. The set of the $\mathcal{V}(\phi^r)$ is not a group because $\mathfrak{K}$ does not close as an algebra. The advantage of this parametrisation is that the action of $H$ on the scalar fields is linear. 

The second useful parametrisation is the \emph{solvable} parametrisation. It is built by first selecting a maximal abelian subspace of $\mathfrak{K}$. These will generate a non-compact maximal abelian subgroup of $G$. Let us denote $\mathcal{S}^+$ the vector space generated by the positive roots of $\mathfrak{g}$ with respect to this choice of non-compact Cartan generators. The union of $\mathcal{S}^+$ and the Cartan generators yields a solvable algebra $\mathcal{S}$. In this parametrisation 
\begin{equation}
\mathcal{V} = \exp(\phi^r T_r)
\label{eq:SolvableParam}
\end{equation}
for $\left\{T_r\right\}$ a basis of $\mathcal{S}$. Since the algebra $\mathcal{S}$ closes and is solvable, this parametrisation has a natural group structure. However, $\mathcal{S}$ is no longer orthogonal to $\mathfrak{h}$ and thus does not admit a natural representation of $H$.

For any of these parametrisation we define $M = \mathcal{V}\mathcal{V}^T$, which is invariant under $H$ since it is compact. Then as before, $\text{Tr}\left[\partial_\mu M \partial^\mu M^{-1}\right]$ gives us the appropriate kinetic term. As a last remark, notice that taking the trace implies that we are working in a representation of $G$. The prefactor in front of the kinetic term will depend on this choice of representation.

\subsubsection{First contact with hidden symmetries}
\label{subsubsec:HiddenSym}
We apply the results of the last sections to the specific case of the reduction of pure gravity from four to three dimensions. According to \eqref{eq:ReductionMetricS1}, the three-dimensional action is 
	\begin{align}
	\mathcal{L}_{(3)} e^{-1} = &\, \mathcal{R}_{(3)} - 2\partial_\mu \phi \partial^\mu \phi - \frac{1}{4} e^{4 \phi} {F_{\mu\nu}} {F^{\mu\nu}} \,. 
	\label{eq:3DactionEinsteinS1}
\end{align}
At first glance, this action admits only the $\mathbb{R}$-global symmetry \eqref{eq:ReducedTrombone}. We will show that the e.o.m. of \eqref{eq:3DactionEinsteinS1} admit a $\SL(2,\,\mathbb{R})$ symmetry which is not manifest in the action. To show this, we will ``dualise'' the vector fields using the Bianchi identities. This subsection serves as an introduction to both \emph{hidden symmetries} and \emph{dualisation}. We will revisit both of these concepts in section \ref{sec:HiddSym}.

The trick to dualising the vector field is to consider its field strength $F$, and not the potential $A$, as the fundamental dynamical field. To retain the same physics, we must supplement the standard equation of motion, derived from \eqref{eq:3DactionEinsteinS1}, with a \emph{Bianchi identity}. Together, they yield two equations that $F$ must satisfy. They are
\begin{equation}
	d( e^{4 \phi} \star F) = 0 \hspace{1cm} dF= 0\,.
\end{equation}
The $\star$ operator is the \emph{Hodge dual}. It sends $p$-forms to $(d-p)$-forms and is defined as
\begin{equation}
	(\star A)_{\mu_1\cdots\mu_{n-p}} = \frac{1}{p!} \epsilon^{\nu_1\cdots \nu_p}{}_{\mu_1\cdots \mu_{n-p}} A_{\nu_1\cdots\nu_p}\,. 
\end{equation}
It has the property that $\star \star A_{(p)} = (-1)^{s+p(d-p)} A_{(p)}$, where $s$ is the number of minus signs in the signature of the metric. The first equation implies that $e^{4 \phi} \star F$ is closed, and thus locally exact\footnote{This is known as the \emph{Poincaré lemma} and states that any $p$-form $F$ which is closed (i.e. $dF = 0$) is also locally exact, i.e. there exists a neighbourhood $U$ of any point $p$ such that $F_{|U} = dA$ for $A$ a $(p-1)$-form on $U$.}. With this, we can introduce a scalar field $\chi$ such that
\begin{equation}
d\chi=e^{4 \phi} \star F \,.
\label{eq:dualRel4to3}
\end{equation} 
This scalar field is defined up to an additive constant. Plugging this back into \eqref{eq:3DactionEinsteinS1} yields
\begin{equation}
\tilde{\mathcal{L}}_{(3)} e^{-1} = R_{(3)}  - 2 \partial_\mu \phi \partial^\mu \phi - \frac{1}{4} e^{-4\phi} \partial_\mu \chi \partial^\mu \chi
\label{eq:3DactionEinsteinS1dual}
\end{equation}
We have, in a way, exchanged the role of the Bianchi identity and the equations of motion. This action is invariant under the trivial $\chi\rightarrow \chi + \text{cste}$ as well as the simultaneous rescaling of $\chi$ and translation of $\phi$. But now, the scalar kinetic term is that of the $\SL(2)/\SO(2)$ sigma-model of \eqref{eq:SL2kineticTerm}\footnote{a possible rescaling $\phi\rightarrow \phi/2$ and $\chi \rightarrow \sqrt{2} \chi$ might make this more obvious}. In particular, the global symmetry group of \eqref{eq:3DactionEinsteinS1dual} is $\SL(2,\,\mathbb{R})$. This means that, although the full four-dimensional Einstein theory has no apparent $\SL(2,\,\mathbb{R})$ symmetry, the space of solutions to the four-dimensional Einstein equations with a U(1) Killing vector admits a non-trivial $\SL(2,\,\mathbb{R})$ action. Actually, out of the three generators of $\mathfrak{sl}_2$, only one has a non-trivial and non-geometric action. Indeed, the Cartan generator rescales $\phi$, thus is geometric, and the positive root acts as a translation on $\chi$, which does modify the 4d uplift. Only, the negative root of $\mathfrak{sl}_2$ has no geometric interpretation in 4d and is truly the only ``hidden'' symmetry of the compactified action.

The dualisation procedure, encoding the d.o.f. of $F$ in the scalar field $\chi$, can also be done more conveniently by imposing the Bianchi identities via a Lagrange multiplier:
	\begin{align}
	\mathcal{L}_{(3)} e^{-1} = &\, \mathcal{R}_{(3)} - \frac{1}{2} \partial_\mu \phi \partial^\mu \phi - \frac{1}{4} e^{2 \phi} {F_{\mu\nu}} {F^{\mu\nu}} -\frac{1}{2} \epsilon^{\mu\nu\rho} \partial_\mu F_{\nu \rho} \chi\,, 
\end{align}
where we have now rescaled the dilaton. The dynamical fields of this action are the scalars $\phi,\,\chi$ and the 2-form $F$. Solving for $F$ yields back the action \eqref{eq:3DactionEinsteinS1dual}. We also want to emphasise that the duality relation \eqref{eq:dualRel4to3} is non-local when expressed in terms of $A$ and $\chi$ (because one must ``integrate'' the exterior derivative in front of $\chi$ in \eqref{eq:dualRel4to3}).

\begin{exercise}
\label{exe:sln}
Compute the symmetry group of the $3+d$-dimensional E-H action reduced on $T^d$ (after dualisation). This can be done in the following steps:
\begin{itemize}
\item To the 3D truncated action, add the Lagrange multipliers $c_m$ to impose the Bianchi identities $dF^m =0$. Solve for $F^m$.
\item Show that the obtained kinetic term is that of the $\SL(d+1)/\SO(d+1)$ coset space. You might want to use the parametrisation
\begin{equation}
	\tilde{M}_{mn} = \begin{pmatrix}
	e^{-2 d \gamma\, \phi} & - e^{-2 d \gamma\, \phi} c_m\\
	-e^{-2 d \gamma\, \phi} c_m & e^{-2 d \gamma\, \phi} c_m c_n + e^{2\gamma \phi} M_{mn}
	\label{eq:MetricSLdSOd}
	\end{pmatrix}\,,
\end{equation}
where $\gamma$ must be fixed and $M_{mn}$ is a symmetric matrix with unit determinant.
\end{itemize}
\end{exercise}
\paragraph{Solution} The three-dimensional truncated action with the Lagrange multiplier is
\begin{align}
	\mathcal{L}_{(3)} e^{-1} = &\, \mathcal{R}_{(D)} + \frac{1}{4} \partial_\mu M_{mn} \partial^\mu M^{mn} - \beta \partial_\mu \phi \partial^\mu \phi\\
	&- \frac{1}{4} e^{2 \beta \phi} M_{mn} {F_{\mu\nu}}^m  {F^{\mu\nu}}^n - \alpha\, \epsilon^{\mu\nu\rho}  c_m \partial_\mu{F_{\nu\rho}}^m
	\label{eq:Einstein3DDual}
\end{align}
with $\alpha$ a real number to be fixed and $\beta = (d+1)/d$. We can solve for $F^m$ imposing
\begin{equation}
	{F_{\mu\nu}}^m = 2\alpha e^{-2\beta \phi}M^{mn} {\epsilon_{\mu\nu}}^\rho \partial_\rho c_n\,.
\end{equation}
Plugging the solution back in the action one gets
\begin{align}
	\label{eq:Einstein3DDual2}
		\mathcal{L}_{(3)} e^{-1} = &\, \mathcal{R}_{(D)} + \frac{1}{4} \partial_\mu M_{mn} \partial^\mu M^{mn} - \beta \partial_\mu \phi \partial^\mu \phi\\
	\nonumber&-2\alpha^2 e^{-2 \beta \phi} M^{mn}  \partial_\mu c_m\,\partial^\mu c_n\,.
\end{align}
Using the metric on $\SL(d+1)/\SO(d+1)$ presented in \eqref{eq:MetricSLdSOd} we compute its inverse: 
\begin{equation}
\tilde{M}^{mn} = \begin{pmatrix} e^{2 \gamma d\, \phi} + e^{-2\gamma \phi} c_m M^{mn} c_n & e^{-2\gamma \phi} c_m M^{mn}\\
e^{-2\gamma \phi} c_n M^{nm} & e^{-2\gamma \phi} M^{mn}
\end{pmatrix}\,.
\end{equation}
This allows us to compute that
\begin{align}
4 e^{-1}\mathcal{L}_{kin} = \partial_\mu \tilde{M}_{mn} \partial^\mu \tilde{M}^{mn} =& \partial_\mu (e^{2d \gamma \,\phi} + e^{-2\gamma \phi} c_m M^{mn} c_n)\partial^\mu(e^{-2d \gamma \, \phi})\\
\nonumber&- 2 \partial_\mu(e^{-2\gamma\phi} M^{mn} c_n)\partial^\mu(e^{-2d \gamma\, \phi}c_m)\\
\nonumber&+ \partial_\mu(e^{-2\gamma\phi} M^{mn}) \partial^\mu(e^{-2d \gamma\,\phi} c_m c_n + e^{2 \gamma \phi} M_{mn})
\end{align}
organising terms by number of $\partial_\mu \phi$ derivatives we get
\begin{align}
	4 e^{-1}\mathcal{L}_{kin} =\,& \partial_\mu \phi \partial^\mu \phi \left(-4 (d^2 \gamma^2 + d \gamma^2) + e^{-2(\gamma+d \gamma)\phi} c_m M^{mn} c_n d \gamma \gamma (4 - 8 + 4)\right)\\
	\nonumber&- \partial_\mu\phi\big(2d\gamma \,e^{-2\gamma(1+d)\phi}  \left[\partial^\mu (c_m M^{mn} c_n) - 2 \partial^\mu(M^{mn}c_n) c_m -  c_m c_n\partial^\mu M^{mn}\right]\\
	\nonumber&\phantom{\partial_\mu\phi\big(}+e^{-2\gamma(1+d)\phi} \gamma M^{mn} \left[-4 c_n\partial_\mu c_m + 2 \partial_\mu (c_m c_n)\right]\\
	\nonumber &\phantom{\partial_\mu\phi\big(}+ 2 \gamma \left[M^{mn}\partial^\mu M_{mn} - M_{mn}\partial^\mu M^{mn}\right]\big)\\
	\nonumber&+ \partial_\mu M^{mn} \partial^\mu M_{mn} - 2 e^{2(\gamma+ d\gamma)} M^{mn}\partial_\mu c_m \partial^\mu c_n\,.
\end{align}
Using the fact that $M^{mn}\partial_\mu M_{mn} = \partial_\mu \log(\det M) = 0$, this simplifies to
\begin{align}
	4 e^{-1} \Lcal_{kin} =& - (d^2\gamma^2 + d \gamma^2) \partial_\mu\phi\partial^{\mu}\phi + \frac{1}{4} \partial_\mu M^{mn}\partial^\mu M_{mn}\\ 
	\nonumber&-\frac{1}{2} e^{-2\gamma (1+d) \phi} M^{mn} \partial_\mu c_m \partial^\mu c_n\,,
\end{align}
which reduces to \eqref{eq:Einstein3DDual2} for $\gamma=1/d$ and $\alpha=1/2$, and we do observe again a hidden symmetry enhancing $\SL(d)$ to $\SL(d+1)$. 

Notice that this symmetry enhancement is sensitive to the relative coefficient in the action and would not appear if the vectors did not rescale with the weight they do under the trombone symmetry.

\begin{figure}[ht!]
\begin{summaryFramed}
\begin{itemize}
	\item Einstein theory, compactified on $T^d$, reduces to a theory of $D$-dimensional gravity coupled to a scalar (the ``dilaton''), $d$ vector fields, and scalars parametrising the coset space $\SL(d)/\SO(d)$.
	\item The higher-dimensional diffeomorphisms reduce to lower-dimensional diffeomorphisms, a $\Urm(1)^d$ gauge symmetry (under which no scalar field is charged), and a $\SL(d)$ global symmetry.
	\item The higher-dimensional LLT reduce to the lower-dimensional LLT and an $\SO(d)$ ``gauge'' invariance.
	\item In specific cases, upon dualising degrees of freedom, one can extend these geometrical symmetries to new ``hidden'' symmetries. The prototypical example being the reduction of pure gravity down to three external dimensions.
\end{itemize}
\end{summaryFramed}
\end{figure}

\newpage
\subsection{\texorpdfstring{A $p$-form on $T^d$}{A p-form on a torus}}
\label{subsec:pFromReduction}

In the next sections, we will be mainly interested in the dimensional reduction of the bosonic sector of supergravities. The action of these supergravities contains not only the Ricci scalar, whose dimensional reduction we have already studied, but also $p$-forms. Following the same strategy as before, we start by analysing how the symmetries of the full theory reduce on $T^d$. 

In $D+d$ dimensions, a $p$-form transforms under diffeomorphisms in the $p$-antisymmetric representation of $\GL(D+d)$. It also admits a ($p-1$)-form gauge symmetry acting as
\begin{equation}
	C_{(p)} \rightarrow C_{(p)} + d \Lambda_{{(p-1)}}\,,
	\label{eq:gaugepform}
\end{equation}
where $\Lambda_{{(p-1)}}$ is the gauge transformation parameter (this generalises the usual $U(1)$ gauge transformation of vector fields). The presence of a $p$-form does not change the fact that the diffeomorphism invariance of the $(D+d)$-dimensional theory reduces to $D$-dimensional diffeomorphisms, a $U(1)^d$ gauge symmetry, and an $\SL(d)$ global symmetry, as shown in the previous sections. However, we must also understand the action of these symmetries on the reduced $p$-forms as well as the lower-dimensional interpretation of the gauge symmetry \eqref{eq:gaugepform}. To do this, we will proceed in two steps. First, we will describe the simpler case of the reduction of a 1-form. Secondly, we will generalise what we will have found to the reduction of $p$-forms.

\subsubsection{Reduction of a 1-form} 

\subsubsection*{Reduction Ansatz}

\noindent The KK procedure reduces the 1-form $C_M$ to a $D$-dimensional 1-form, $B_\mu$, and a series of scalars, $B_m$. The reduction ansatz is
\begin{equation}
	C_M(x^\mu,\,y^m) \rightarrow (B_\mu := C_\mu - {A_\mu}^m C_m,\, B_m = C_m)\,,
	\label{eq:split1form}
\end{equation} 
where ${A_\mu}^m$ is the vector field of the KK-ansatz \eqref{eq:AnsatzVielbein}. This ansatz is made to obtain reduced fields which transform nicely under the lower-dimensional symmetries. We can derive the transformation rules of $B$ under the three families of reduced diffeomorphisms (i.e., $\delta_{\xi_{(D)}}$, $\delta_{\xi_{\text{gauge}}}$ and $\delta_{\xi_{\text{(SL)}}}$). Under $\delta_{\xi_{(D)}}$ the $B_\mu$ transforms as a $D$-dimensional 1-form whereas the $B_m$ transform as scalars. The other transformations are
\begin{equation}
\begin{array}{ll}
\delta_{\xi_{\text{gauge}}} B_\mu = \partial_\mu \xi^m C_m - \delta_{\xi_{\text{gauge}}}({A_\mu}^m C_m) = 0 \,,\phantom{aaa} &\delta_{\xi_{\text{gauge}}} B_m = 0\,,\\[2mm]
\delta_{\xi_{(SL)}} B_\mu = 0 \,,\hspace{5mm} &\delta_{\xi_{(SL)}} B_m = {\Lambda^n}_m B_n\,.
\end{array}
\end{equation}
We thus obtain a single 1-form $B_\mu$ in $D$ dimensions which is U$(1)^d$-invariant and SL($d$)-invariant. We also obtain $d$ scalars $B_m$ transforming in the fundamental representation of $\SL(d)$. 

How did we\footnote{Not me specifically, but the ones who did.} guess the proper ansatz \eqref{eq:split1form}? Since $C_{(1)}$ is a $D+d$ tensor, we can act with the inverse vielbein ${(E^{-1})_{A}}^{M}$ to obtain 
\begin{equation}
C_A ={(E^{-1})_{A}}^{M} C_M 
\end{equation}
which is a scalar under $D+d$ coordinates change. Splitting the index $A$ in $\left\{a,\,\underline{a}\right\}$ and using the appropriate ${e_\mu}^a$ vielbein, we now obtain a $D$-dimensional one-form: 
\begin{equation}
B_\mu = e^{-\phi/(D-2)}{e_\mu}^a C_a\,.
\end{equation}
In the same spirit, acting with ${V_m}^{\underline{a}}$ we obtain $d$ $D$-dimensional scalars: 
\begin{equation}
B_m = e^{\phi/d}{V_m}^{\underline{a}} C_{\underline{a}}\,.
\end{equation}
This trick of going to flat indices to guess the ansatz will be useful when computing the reduction of Maxwell-like actions for $p$-forms.

\subsubsection*{Reduction of the gauge symmetry}

\noindent The reduction of diffeomorphism being understood, let us turn to the reduction of the gauge symmetry. In this case, from the 0-form symmetry in $D+d$ dimension, we can read the transformations of the $D$-dimensional fields. The fact that the $D$-dimensional fields must be independent of $y^m$ constraints the allowed gauge parameters. They must belong to one of the two families
\begin{equation}
\Lambda_{(0)}(x^\mu,\,y^m) = \Lambda(x^\mu)\hspace{5mm}\text{ or }\hspace{5mm}\Lambda_{(-1)} = a_m y^m\,,
\end{equation} 
with $a_m\in\mathbb{R}$ constants. The first transformation $\delta_{\Lambda_{(0)}}$ acts as the usual $D$-dimensional gauge symmetries on $B_\mu$ and does not act on the scalars $B_m$. The second family $\delta_{\Lambda_{(-1)}}$ corresponds to a \emph{global} shift symmetry acting as
\begin{align}
	&\delta_{\Lambda_{(-1)}} B_m \rightarrow B_m + a_m\,,\\
	&\delta_{\Lambda_{(-1)}} B_\mu \rightarrow B_\mu - {A_\mu}^m a_m\,.
\end{align}
The global symmetry group of the $D$-dimensional theory is enhanced to ${\SL(d) \ltimes \mathbb{R}^d}$ (not considering the trombone symmetry). Note that this $\mathbb{R}^d$ symmetry does not commute with $\SL(d)$ but does not enhance it to new a semi-simple group either. Under this shift symmetry, the would-be field strength ``$\partial_{[\mu}B_{\nu]}$'' is not an invariant quantity. The proper invariant field strength (also invariant under the global symmetries) is the ``\emph{improved field strength}'':
\begin{equation}
	\tilde{F}_{\mu\nu} = 2 \partial_{[\mu}B_{\nu]} - F_{\mu\nu}^m B_m\,.
\end{equation}
It can be obtained from the 'flattening of indices' trick:
\begin{equation}
	e^{-\frac{2}{D-2}\phi}{e_\mu}^a {e_\nu}^b {(E^{-1})_a}^M {(E^{-1})_b}^N F_{MN} = 2\partial_{[\mu}C_{\nu]} - {A_{[\mu}}^i \partial_{\nu]} C_i = \tilde{F}_{\mu\nu}\,.
\end{equation}

\subsubsection*{Reduction of Maxwell action}

\noindent We have the proper definitions to reduce the usual Maxwell action:
\begin{align}
	S_{(D+d)\text{, 1}} =&- \int d^Dx d^dy \, \sqrt{-G}\,  \frac{1}{4}\, \partial_{[M}C_{N]}\partial^{[M}C^{N]}	
	\end{align}
	which yields
	\begin{align}
S_{(D)\text{, 1}} 	=&- (2\pi R)^d \int d^Dx \sqrt{-g}\,\left(\frac{1}{4}e^{\frac{2\phi}{(D-2)}}\tilde{F}_{\mu\nu}\tilde{F}^{\mu \nu} + \frac{1}{2} e^{-\frac{2\,\phi}{d}} M^{mn} \partial_\mu B_m \partial^\mu B_n\right)\,.
\end{align}
We recognise the improved field strength $\tilde{F}$ and the inverse internal metric $M^{mn}$. Finally, we can make the $\GL(d)\ltimes \mathbb{R}^d$ (taking trombone symmetry into account) invariance of the action more explicit by defining
\begin{equation}
	\hat{M}_{\hat{m}\hat{n}} = \begin{pmatrix}
		e^{2 \beta \phi}M_{mn} + e^{\phi\frac{2}{D-2}} B_m B_n & -e^{\phi\frac{2}{D-2}} B_m\\-e^{\phi\frac{2}{D-2}} B_n & e^{\phi\frac{2}{D-2}}
	\end{pmatrix}= e^{-2\left(\frac{1}{d+1} + \frac{1}{D-2}\right) \phi} \tilde{M}
\end{equation}
where $det(\tilde{M})=1$. The matrix $M$ can be written as $\mathcal{V}\mathcal{V}^T$ for $\mathcal{V} \in (\GL(d)\ltimes \mathbb{R}^d)/ \SO(d)$. We define $F_{\tilde{m}} = (F_m,\,d B)$ and we compute that
\begin{align}
\nonumber	(\mathcal{L}_{(D)} + \mathcal{L}_{(D),\,\text{1}})\sqrt{-g}^{-1} =&\,  R_{(D)} + \frac{1}{4} \partial_\mu M_{mn} \partial^\mu M^{mn} - \beta \partial_\mu \phi \partial^\mu \phi\\
\nonumber	&- \frac{1}{2} e^{-\phi\,2/d} M^{mn} \partial_\mu B_m \partial^\mu B_n\\
\nonumber	&- \frac{1}{4} e^{2 \beta \phi} M_{mn} {F_{\mu\nu}}^m {F^{\mu\nu}}^n -\frac{1}{4}e^{\phi\,2/(D-2)}\tilde{F}_{\mu\nu}\tilde{F}^{\mu \nu} \\
	=& \mathcal{R}_{(D)} + \frac{1}{4} \partial_\mu\left( e^{2\alpha \phi} \tilde{M}_{\hat{m}\hat{n}}\right) \partial^\mu \left(e^{-2\alpha \phi}\tilde{M}^{\hat{m}\hat{n}}\right) \label{eq:reduced1formAction}\\ 
\nonumber	&\phantom{aaaa}- \frac{1}{4} \hat{M}_{\hat{m}\hat{n}} {F_{\mu\nu}}^{\hat{m}} F^{\mu\nu\,\hat{n}}
\end{align}
where $\alpha^2 = \frac{1}{d+1}\left(\frac{1}{d+1} + \frac{1}{D-2}\right)$ and we recall from exercise \ref{exe:sln} that
\begin{align}
\frac{1}{4} \partial_\mu\hat{M}_{\hat{m}\hat{n}} \partial^\mu\hat{M}^{\hat{m}\hat{n}} = - \frac{1}{d(d+1)} \partial_\mu\phi\partial^\mu\phi &- \frac{1}{2} e^{2\phi/d} M^{mn}\partial_\mu B_m \partial^\mu B_n\\ 
\nonumber&+ \frac{1}{4} \partial_\mu M_{mn} \partial^\mu M^{mn}\,.
\end{align}
It almost looks like the action \eqref{eq:reduced1formAction} which admits a $\mathfrak{sl}_{d+1}$ symmetry. Unfortunately, the specific weights of the vector $B_\nu$ (equivalently, of the scalars $B_m$) are not compatible with such an enhancement. This is made clear by the extra $(\partial_\mu \phi)^2$ kinetic term as well as the fact that $\hat{M}$ is not of determinant one, i.e. $\hat{M} \notin \SL(d+1)$. This is to show that a symmetry enhancement to a simple group is a very peculiar behaviour which does not happen for any dimensional reduction.

\subsubsection{Reduction of a $p$-form} 
 We are now ready to generalise the previous results to the case of a $p$-forms $C_{(p)}$ reduced on $T^d$. The $p$-form reduces to a series of $\left(\begin{smallmatrix}d\\p-k\end{smallmatrix}\right)$ $k$-forms, for $k$ such that $\text{Max}(0,\,p-d)\leq k \leq p$. To obtain $D$-dimensional $k$-forms, transforming appropriately under diffeomorphisms, we first go to flat indices using the inverse vielbein $(E^{-1})\indices{_A^M}$ before using the $D$-dimensional metrics $(e_\mu)^a$ and ${V_m}^{\underline{a}}$ to obtain the ``curved-indices'' back. This procedure is equivalent to defining
\begin{equation}
	B_{\mu_1\cdots \mu_k m_{1} \cdots m_{p-k}} = {P_{\mu_1}}^{M_1}\cdots {P_{\mu_p}}^{M_p}C_{M_1\cdots M_pm_{1} \cdots m_{p-k}}\,,
	\label{eq:AnsatzPform}
\end{equation}
where $P = ({\delta_\mu}^\nu,\,-{A_\mu}^m)$. Because of these definitions, we are sure that the fields $B_{\mu_1\cdots \mu_k m_{1}\cdots m_{k}}$ transform as $k$-forms under $D$-dimensional diffeomorphisms and as $(p-k)$-antisymmetric tensors under $\SL(d)$. 

We must identify how the higher-dimensional gauge symmetries reduce. As in the case of a 1-form, for gauge parameters $\Lambda_{\mu_1\cdots\mu_{k-1} \,m_{1} \cdots m_{p-k}}(x^\mu)$, which only depend on the external coordinates, we get $(k-1)$-form symmetries. Because of the ansatz \eqref{eq:AnsatzPform}, these will act on the $k$-forms, but also on the reduced $q$-forms with $q\geq k$ forms.

As for the reduction of a 1-form, there can be an enhancement of the $\mathfrak{gl}_d$ symmetry if scalar degrees of freedom are produced by the reduction, i.e., if $d\geq p$. In that case, the gauge transformation with parameter
\begin{equation}
	\Lambda_{(-1)} = a_{\left[m_1\cdots m_p\right]}y^{m_p}
\end{equation}
reduces to a global symmetry. This symmetry acts as a shift on the scalars, but it also acts on the other reduced $k$-forms. Without entering the details, this requires to improve the fields strength $dB_{(k)}$ using the ``going to flat indices'' trick. 

Finally, one can check that the reduced $k$-forms transform with weight \begin{equation}
	\left(-\frac{p}{D-2} + (p-k) \beta\right)
	\label{eq:pformWeight}
\end{equation}
 under the trombone symmetry. We have thus an enhancement of the global symmetry algebra to an algebra of the form
\begin{equation}
	\left(\mathfrak{gl}(d)\ltimes \Lambda^p\mathbb{R}^d\right).
\end{equation}

\paragraph{Reduction of the action} We compute the reduction of the standard action for a $p$-form. Starting with
\begin{equation}
	S_{(D+d)\text{, }p} = - \frac{1}{2} \int d^Dx \,d^dy \sqrt{-G}\,\, |F_{(p)}|^2.
	\end{equation}
We can compute the KK-reduction. There is a trick to simplify this computation which is to go to the flat indices:
\begin{align}
	|F_{(p)}|^2 &:=  \frac{1}{p!}\,F_{A_1\cdots A_{p}} F_{B_1\cdots B_{p}} \eta^{A_1 B_1} \cdots \eta^{A_{p} B_{p}}\\
\nonumber	&=\frac{1}{p!}\, \sum_{k=0}^{p} \frac{p!}{k!(p-k)!}F_{a_1\cdots a_k\,\underline{a}_1 \cdots \underline{a}_{p-k}} F^{a_1\cdots a_k\,\underline{a}_1 \cdots \underline{a}_{p-k}}\\
\nonumber	&= \sum_{k=0}^{p} \frac{e^{2 k\phi/(D-2)}e^{-2 (p-k)\phi/d}}{k!(p-k)!}  \tilde{F}_{\mu_1\cdots\mu_k m_1 \cdots m_{p-k}}\tilde{F}^{\mu_1\cdots\mu_k m_1 \cdots m_{p-k}}\,\\
\nonumber	&= e^{-2 \phi p/d}\sum_{k=0}^{p} \frac{e^{2 k\beta\phi}}{k!(p-k)!}  \tilde{F}_{\mu_1\cdots\mu_k m_1 \cdots m_{p-k}}\tilde{F}^{\mu_1\cdots\mu_k m_1 \cdots m_{p-k}}\,.
\end{align}
In the second line we have separated underlined and normal flat indices corresponding to internal and external flat indices. The combinatorial coefficient $\tfrac{(p+1)!}{k!(p-k+1)!}$ comes from sorting the indices. In the last line, reintroducing the curved indices corresponds to defining the improved field strengths $\tilde{F}$. The Greek indices are raised and lowered using $g_{\mu\nu}$ while the Latin indices are raised and lowered using $M_{mn}$. Rearranging all the terms we obtain:
\begin{empheq}[box=\mymath]{equation}
\label{eq:ReductionpFormS1}
	S_{(D)\,p} = -\frac{1}{2}\int d^Dx \,\sqrt{-g}\,\,e^{\frac{2(p-1)}{(D-2)}\phi}\sum_{k=0}^{p} \frac{e^{-2 (p-k)\beta\phi}}{k!(p-k)!}  \tilde{F}_{\mu_1\cdots\mu_k m_1 \cdots m_{p-k}}\tilde{F}^{\mu_1\cdots\mu_k m_1 \cdots m_{p-k}}\,.
\end{empheq}
Note that the prefactors $e^{\cdots \phi}$ match the weights of the reduced $p$-form after dimensional reduction \eqref{eq:pformWeight}.

\paragraph{Extending the scalar manifold} 
Reducing $p$-forms on $T^d$ with $d\geq p$ will add scalars to the $D$-dimensional theory. These scalars transform in the $p$-antisymmetric representation of $\SL(d)$ and admit a transitive action of the abelian groups $\Lambda^p \mathbb{R}^d$ (with the addition as a group product). If we have several $p_i$-forms, $i\in I$, with $d\geq p_i$, the scalar manifold will admit the action of the group
\begin{equation}
G_{p_i}=\GL(d)\ltimes \left(\sum\limits_{i\in I} \Lambda^{p_i}\mathbb{R}^d\right)\,.
\end{equation}
Each $\Lambda^{p_i}\mathbb{R}^d$ factor has a positive weight $\tfrac{p_i}{d}$ under the $\mathbb{R}^+ \subset \GL(d)$. Finally, the scalar manifold itself is just
\begin{equation}
	\mathcal{M}_{\text{scal}} = \left[\GL(d)\ltimes \left(\sum\limits_{i\in I} \Lambda^{p_i}\mathbb{R}^d\right)\right]/\SO(d)
\end{equation}
Let us look at the Lie algebra of $G_{p_i}$. It can be written, as a vector space, as
\begin{equation}
	\mathfrak{g}_{p_i} = \mathfrak{sl}_d \oplus \mathbb{R} \underset{i\in I}{\oplus} \Lambda^{p_i}\mathbb{R}^d\,.
\end{equation}
This looks a lot like the decomposition of a simple Lie group in Cartan subalgebra $\mathfrak{h}$, positive $\mathfrak{K}_+$ and negative roots $\mathfrak{K}_-$
\begin{equation}
	\mathfrak{g} = \mathfrak{K}_- \oplus \mathfrak{h} \oplus \mathfrak{K}_+\,,
\end{equation}
if we removed the negative roots. We recall that in the solvable parametrisation of $G/H$, with $H$ the maximal compact subgroup of $G$, one does not need $\mathfrak{K}_-$. However, one can still act in a non-trivial way on the scalar with the full group $G$. 

The point is that, if the $p_i$ are chosen such that they correspond to a subalgebra $\mathfrak{n}_+$, of a semi-simple algebra of the form
\begin{equation}
	\mathfrak{g} = \mathfrak{n}_- \oplus \mathfrak{sl}_d \oplus \mathbb{R} \oplus \mathfrak{n}_+\,,
\end{equation} 
we can obtain a symmetry enhancement from $\mathfrak{g}_{p_i}$ to $\mathfrak{g}$. When we reduced pure gravity to three dimensions (with the extra step of dualising vector fields), this is the pattern of symmetry enhancement that we observed. The algebra $\mathfrak{gl}_d$ is enhanced, as a vector space, to $\mathfrak{gl}_d \oplus \mathbb{R}^d$. This vector space can be identified with the Cartan generators and the positive roots of $\mathfrak{sl}_{d+1}$. Because the space $\mathbb{R}^d$ had the proper weight under the $\mathfrak{gl}_d$ rescaling, it can further be given the appropriate Lie bracket and we have symmetry enhancement. We will see more complicated examples of this phenomenon in the next sections.

\begin{figure}[h]
\begin{summaryFramed}
\begin{itemize}
	\item The reduction of a $p$-forms on $T^d$ yields a set of ${\left(\begin{smallmatrix}d\\ p-k\end{smallmatrix}\right)}$ $k$-forms in $D$ dimensions ($\text{Max}(0,\,p-d)\leq k \leq p$). These $k$-forms transform under $(k-1)$-form gauge symmetries as well as under the $p-k$ antisymmetric representation of the global $\SL(d)$ group.
	\item The reduced action is written in terms of improved field strengths which we can easily define by the ``going to flat indices trick''.
	\item When $d\geq p$, the reduction produces new scalar degrees of freedom. These scalars admit global shift symmetries which enhance the global symmetry algebra from $\mathfrak{gl}_d$ to
	\begin{equation}
		\mathfrak{gl}_d \ltimes \mathfrak{n}_+
	\end{equation}
	where $\mathfrak{n}_+$ is a nilpotent algebra, positively graded under the rescaling $\mathbb{R}\subset \mathfrak{gl}_d$.
\end{itemize}
\end{summaryFramed}
\end{figure}
\newpage

\section{Supergravities}
\label{sec:SUGRA}

Supergravity (SUGRA) is at the junction of two fundamental ideas of mathematical physics: gauge symmetry and supersymmetry (susy). Broadly speaking, supersymmetries are symmetries whose parameter, $\epsilon^i$ transform as a spin-$\tfrac{1}{2}$ particle. Supergravities are defined by their invariance under \emph{local} supersymmetry. I.e., they have ``gauged supersymmetry'' transformations parametrised by $\epsilon(x)$. To gauge supersymmetries, one must introduce spin-$\tfrac{3}{2}$ fields called gravitini, $\psi_\mu$. It is a fermionic field carrying an additional vector index. Gravitini play the same role as that of a gauge connection for usual gauge theories. 
The gravitini transformation rules are of the form
\begin{equation}
	\delta_\epsilon\psi_\mu = D_\mu \epsilon\,.
\end{equation}
The super-partner of these gravitini (i.e., the field completing the irrrep of the susy algebra) is a spin-2 particle, parametrised by a vielbein $e_\mu{}^a$. The vielbein transforms schematically as
\begin{equation}
	\delta_\epsilon e_{\mu}{}^a= \frac{1}{2} \bar{\epsilon} \gamma^a \psi_\mu\,.
\end{equation}
The fact that gauged supersymmetry must contain gravity explains the name ``supergravity''. An exposition on the basics of supergravity in four dimensions is contained in \cite{Freedman2012}.

There is a whole zoo of supergravities in different dimensions. We can even have more than a single SUSY transformation (and thus work with an ``\emph{extended}'' supersymmetry algebra). The parameters of such transformations are denoted $\epsilon^I$ for $I= 1,\,\dots,\,\mathcal{N}$. The matter content of SUGRAs must fit into irreps of these extended SUSY-algebras. When these theories have global symmetries, they can be gauged (modulo some constraints arising from possible non-compatibility of gauge and SUSY transformation). As such, a specific theory of supergravity is often labelled by its dimension, the number of supersymmetries $\Ncal$ and its gauge group $G_g$. Whenever different matter contents are possible, we also use them to label the supergravity. 

A supergravity can be \emph{maximal} in the sense that it has the maximal number of independent supersymmetry transformations such that the associated superalgebra admits non-trivial irreps in which all excitations have spin less than 2 (this is required to have an interacting theory with finite degrees of freedom). This imposes an upper limit on the number of ``supercharges'' $\mathcal{N}$ which is $\mathcal{N} \times \text{Dim}(\mathbf{spin}(1,\,D-1)) \leq 32$, where $\mathbf{spin}(1,\,D-1)$ is an irreducible spin representation in $D$ dimensions with Minkowski signature. Alternatively, we can think of it as an upper bound for the dimensionality of space-time when fixing $\Ncal=1$. This upper limit is reached at $D=11$ which is what makes supergravity in 11D a particularly interesting theory: it is the highest dimensional non-trivial theory of supergravity with Minkowski signature. Moreover, it happens that this theory is uniquely fixed by supersymmetry. 

In ten dimensions, there are two different \emph{maximal} supersymmetry algebras which give rise to two theories of supergravity called \emph{type IIA} and \emph{type IIB} supergravities. They are no less special than their higher-dimensional cousin because they each describe the background corresponding to the low energy, weakly coupled limit of type IIA and type IIB superstrings, respectively. 

In this section, we will review the bosonic sector of these three theories of supergravity. The end goal is to study which hidden symmetries arise when compactifying these theories on tori and how these three theories are related to one another. For a recent survey of supergravities see \cite{Sezgin2023}. For ten- and eleven-dimensional supergravity equations of motion see \cite{Hamilton2016}. Concerning more detailed constructions of gauged supergravity in lower dimensions, see the review \cite{Trigiante:2016mnt} (we also included a crash-course on some aspects of gauged supergravity in \ref{subsec:GaugeSugra}).

\subsection{11D supergravity}

The matter content of 11D supergravity \cite{CREMMER1978409} consists of a metric $G$, a spin-$3/2$ gravitino $\Psi_M$ and a three-form $A_{(3)}$. The action of this theory is uniquely specified by supersymmetry and its bosonic sector is
\begin{equation}
	S = \frac{1}{2\kappa_{11}^2}\int \mathrm{d}^{11}Y \left[ \sqrt{-G}\left(R - \frac{1}{2} |F_4|^2 \right) - \frac{1}{6}\,A_3 \wedge F_4\wedge F_4\,\right].
\end{equation}
The equations of motions are
\begin{align}
	&R_{MN} = \frac{1}{2\cdot 3!} F_{MPQR}{F_N}^{PQR} - \frac{1}{6\cdot 4!} g_{MN}\,  F_{PQRS} F^{PQRS}\,,\\
	&d\star F_4 + \frac{1}{2}\, F_4\wedge F_4 = 0\,,
\end{align}
where we have defined $F_4=dA_3$. This can also be imposed by the Bianchi identity:
\begin{equation}
	d F_4 = 0\,.
\end{equation}

\subsection{\texorpdfstring{10D $\mathcal{N}=2$ supergravity}{10D N=2 supergravity}}

There are two supergravities in ten dimensions because there are two different chiralities possible for spinorial representations in ten dimensions. This gives the freedom to choose a $\mathcal{N}=2$ superalgebra with both fermionic parameters of the same chirality or of the opposite chirality. Choosing them of opposite chirality, the resulting theory is non-chiral and is type IIA supergravity. Choosing them of the same chirality, the resulting theory is chiral and is type IIB supergravity.

Both are similar in their structure. Their bosonic field content and action can be organised into two parts called the Neveu-Schwarz-Neveu-Schwarz (NS-NS) sector, the Ramond-Ramond (R-R) sector. Only the R-R sector differs between type IIA and type IIB.

\subsubsection{Universal NS-NS action}

The NS-NS sector contains a metric $G$, a two-form $B_2$ and a dilaton $\phi$. Its action can be written as
\begin{equation}
\label{action_NS-NS}
S_{\textrm{NS-NS}}  =  \dfrac{1}{2 \kappa^{2}_{10}}  \int d^{10}x \, \sqrt{-G} \, \left(   R - \tfrac{1}{2} \, \partial^{M} \Phi \partial_{M} \Phi  - \tfrac{1}{2} e^{-\Phi}  |H_{3}|^{2}  \right) \ .
\end{equation}
We have used the standard definitions
\begin{equation}
	H_3 = dB_2\,\hspace{10mm}\text{and}\hspace{10mm} |H_k|^2 = \frac{1}{k!} H_{\mu_1\cdots\mu_k} H^{\mu_1\cdots \mu_k}\,.
	\label{eq:defNSNS}
\end{equation}
The action \eqref{action_NS-NS} is presented in the ``Einstein frame'' which has the usual term $\sqrt{-G}\,R$ in the action without dilatonic prefactor. 

There is possible field redefinition in which one is said to be in the ``string frame''. In this second frame, there is a more natural identification between the supergravity fields and the fields entering in the corresponding  string action. This frame is obtained by rescaling the metric as
\begin{equation}
	G_{MN}^E = e^{-\phi/2} G_{MN}^S\,.
\end{equation}
In this frame the action reads
\begin{equation}
\label{action_E-NS-NS}
S^S_{\textrm{NS-NS}}  =  \dfrac{1}{2 \kappa^{2}}  \int d^{10}x \, \sqrt{-G^S} \, e^{-2\phi} \left(   R^S +4 \, \partial^{M} \Phi \partial_{M} \Phi  - \tfrac{1}{2}  |H_{3}|^{2}  \right) \ .
\end{equation}
All contractions are made using $G^S_{MN}$. In the rest of these notes, we will use the Einstein frame. This arbitrary choice also has an impact on the action for the R-R fields: it modifies the dilatonic prefactor in front of the kinetic term of the $p$-forms according to the following formulas. Under a rescaling of a D-dimensional metric:
\begin{equation}
	\tilde{g}_{\mu\nu} = e^{2 \sigma} g_{\mu\nu}
	\label{eq:confTsfMet1}
\end{equation}
The Ricci scalar transforms as \cite{Carroll2013}
\begin{equation}
	R = e^{-2 \sigma} \left(\tilde{R} - 2 (D-1) \square\sigma - (D-2)(D-1) \partial_\mu \sigma \partial^\mu \sigma\right)
	\label{eq:confTsfMet2}
\end{equation}
where contraction on the r.h.s. are made using $\tilde{g}$.
Under the same rescaling, the Maxwell action for a $p$-form changes as
\begin{equation}
	\int \sqrt{-\tilde{g}} |F_{(p)}|_{\tilde{g}} = \int \sqrt{-g} e^{(D-2p)\sigma} |F_{(p)}|_{g}\,.
	\label{eq:confTsfMet3}
\end{equation}
where $|F_{(p)}|_g = \frac{1}{p!} F_{\mu_1\cdots \mu_p} g^{\mu_1\nu_1} \cdots g^{\mu_p \nu_p} F_{\nu_1\cdots \nu_p}$.

\subsubsection{Type IIA supergravity}
\label{subsubsec:TypeIIA}
The field content of type IIA supergravity \cite{Giani84,Campbell84} contains the NS-NS fields ($G$, $B_2$ and $\Phi$), the R-R fields ($C_1$ and $C_3$) as well as fermions (two gravitini $\psi_\mu$ and two dilatini $\lambda$ of opposite chirality). In this case, the SUSY-invariance does not completely fix the action. One can introduce a parameter called the ``Roman's mass''\cite{Romans86}. We will not consider this deformation here. 

The bosonic action of type IIA supergravity is
\begin{equation}
	S_{bos}^{IIA} = S_{\textrm{NS-NS}} + S_{\textrm{R-R}}^{\textrm{IIA}} + S_{\textrm{CS}}^{\textrm{IIA}}\,.
\end{equation}
The NS-NS sector actions is that of \eqref{action_NS-NS}. The R-R sector action and the Chern-Simon terms are
\begin{equation}
	\label{action_R-R_IIA}
	S^{\textrm{IIA}}_{\textrm{R-R}} = \frac{1}{2\kappa^2_{10}} \int d^{10}x \,\sqrt{-G} \left(-\tfrac{1}{2}e^{3\phi/2} |F_2|^2 -\tfrac{1}{2} e^{\phi/2}|\tilde{F}_4|^2\right)\,,
\end{equation}
\begin{equation}
	\label{action_CS_IIA}
	S_{\textrm{CS}}^{\textrm{IIA}} = - \frac{1}{4\kappa^2_{10}} \int d^{10}x \, B_2 \wedge d C_3 \wedge d C_3\,.
\end{equation}
We have used the definitions of \eqref{eq:defNSNS} and the improved field strengths are
\begin{equation}
F_2 = d C_1 \hspace{5mm}\text{and} \hspace{5mm} \tilde{F}_4 = d C_3 + H_3 \wedge C_1\,.
\end{equation}
The bosonic equations of motion are
\begin{align}
	\square \Phi &= -\frac{1}{2} e^{-\phi} |H_3|^2 + \frac{3}{4} e^{3 \phi/2} |F_2|^2 + \frac{1}{4} e^{\phi/2} |\tilde{F}_4|^2\,,\\
	 d \left(e^{3\phi/2} \star F_{2}\right) &= e^{\phi/2} H_3  \wedge \star\tilde{F}_4 \,,\\
	 d \left( e^{\phi/2} \star_E \tilde{F}_4\right) &= - H_3 \wedge \tilde{F}_4\\
	d(e^{-\phi}\star  H_3) &= \frac{1}{2} \tilde{F}_4 \wedge \tilde{F}_4 - e^{\phi/2} F_2 \wedge \star \tilde{F}_4\,.
\end{align}
The Bianchi identities read
\begin{equation}
	dH_3 = 0 \hspace{5mm},\hspace{5mm} d F_2 = 0 \hspace{5mm}\text{and}\hspace{5mm} d\tilde{F}_4=H_3 \wedge F_2
\end{equation}
The Einstein equation of motion is
\begin{align}
R_{MN} =&\phantom{+} \frac{1}{2} \partial_M \phi \partial_N \phi\\
\nonumber& +\frac{1}{2} e^{-\phi} \left(\frac{1}{2} H_{3\,M\,PQ}{H_{3\,N}}^{PQ} - \frac{1}{4\cdot 3!} G_{MN} H_{3\,PQR} {H_3}^{PQR} \right)\\
\nonumber&+ \frac{1}{2} e^{3\phi/2} \left(F_{2\,M\,P}{F_{2\,N}}^P-\frac{1}{8\cdot 2} G_{MN} F_{2\,PQ} {F_{2}}^{PQ}\right)\\
\nonumber&+ \frac{1}{2} e^{\phi/2} \left(\frac{1}{3!}{\tilde{F}_{4\,M\,PQR}}{\tilde{F}_{4\,N}}^{\phantom{4N}PQR} - \frac{3}{8\cdot4!} G_{MN} {\tilde{F}_{4\,PQRS}}{\tilde{F}_{4}}^{PQRS}\right)\,.
\end{align}

\begin{exercise}
Show that the reduction of 11D SUGRA on $S^1$ is type IIA supergravity.
\end{exercise}
We must reduce the metric and the three-form from 11d to 10d. We have that $\beta = \frac{9}{8}$. From equation \eqref{eq:ReductionMetricS1}, we obtain that the E-H action reduces to
\begin{align}
	\mathcal{L}_{\text{metric}} e^{-1}=&\, R_{10} - \frac{9}{8} |d\tilde{\phi}|^2- \frac{1}{2} e^{2 \frac{9}{8} \tilde{\phi}} |F_2|^2 + \text{total derivatives}\,,
\end{align}
and, after rescaling $\tilde{\phi}= \frac{2}{3}\phi$, we get
\begin{align}
	\mathcal{L}_{\text{metric}} e^{-1}=&\, R_{10} - \frac{1}{2} |F_1|^2- \frac{1}{2} e^{\frac{3}{2} \phi} |F_2|^2 + \text{total derivatives}\,,
\end{align}
From equations \eqref{eq:ReductionpFormS1}, the 3-form kinetic term gives the contribution
\begin{align}
	\mathcal{L}_{(3)} =&  \,\sqrt{-g}\, \frac{1}{2}\,e^{\frac{3}{4}\tilde{\phi}}(|\tilde{F}_4|  + e^{-9/4 \tilde{\phi}}|\tilde{F}_3|^2)
\end{align}
After rescaling, and checking that the improved two-form field strength is simply $F_3$ (because we are reducing on a 1-dimensional manifold), we get
 \begin{align}
	\mathcal{L}_{(3)} =&   \,\sqrt{-g}\, \frac{1}{2}\,(e^{\frac{1}{2}\phi}|\tilde{F}_4|  + e^{-\phi}|\tilde{F}_3|^2)
\end{align}
One should still take care of the CS term. There are no factors of $\phi$ to consider this time since this term is topological. We can use the ``going to flat indices trick'' and using the fact that $F_3\wedge F_3 = 0$, by the symmetry properties of the wedge product, we obtain the appropriate CS term \eqref{action_CS_IIA}.

This shows that massless type IIA supergravity is the dimensional reduction of 11D SUGRA on a circle. This fact has an interpretation in the context of string theory, see section \ref{sec:IIA11D}.

\subsubsection{Type IIB Supergravity}
\label{subsec:Type_IIB}

The field content of type IIB supergravity \cite{SCHWARZ1983,HOWE1984181} contains the NS-NS fields ($G$, $B_2$ and $\Phi$), the R-R fields ($C_0$, $C_2$ and $C_4^+$) as well as fermions (two gravitini $\psi_\mu$ and two dilatini $\lambda$ of the same chirality). The e.o.m. of type IIB can be derived as the e.o.m. of the \emph{pseudo}-action supplemented with a \emph{self-duality} condition \cite{Bergshoeff1995,Bergshoeff1995a}\footnote{This extra self-duality condition, that must be imposed after using the variational principle, is why we call the type IIB action a \emph{pseudo}-action.}.
\begin{equation}
\label{S_bosB}
S_{bos}^{\textrm{IIB}} = S_{\textrm{NS-NS}} + S_{\textrm{R-R}}^{\textrm{IIB}} + S_{\textrm{CS}}^{\textrm{IIB}} \ .
\end{equation}
The NS-NS sector action is that of \eqref{action_NS-NS}. The R-R sector action and the Chern-Simon terms are
\begin{equation}
\label{action_R-R_IIB}
S_{\textrm{R-R}}^{IIB}  =  \dfrac{1}{2 \kappa^{2}_{10}}  \int d^{10}x \, \sqrt{-G}  \, \left(  -\tfrac{1}{2}\,  e^{2 \Phi}  |F_{1}|^{2}  -\tfrac{1}{2}\,  e^{\Phi}  |\widetilde{F}_{3}|^{2}  -\tfrac{1}{4}\, |\widetilde{F}_{5}|^{2} \right)  \ , 
\end{equation}
\begin{equation}
\label{action_CS_IIB}
S_{\textrm{CS}}^{IIB}  =  -\frac{1}{4 \,  \kappa^{2}_{10}}  \int   C_{4} \wedge  H_{3} \wedge F_{3}  \ .
\end{equation}
where the tilded field strengths are defined as
\begin{equation}
\label{widetildeFs}
\widetilde{F}_{3}  = dC_2 - C_{0} \wedge H_3 \ , \hspace{5mm}
\widetilde{F}_{5}  =  F_{5} + \tfrac{1}{2} \, \left( B_{2} \wedge dC_2 - C_{2} \wedge H_{3}  \right) \ .
\end{equation}
Additionally, the self-duality condition 
\begin{equation}
\widetilde{F}_{5} = \star \widetilde{F}_{5} \hspace{5mm}
\text{with}\hspace{5mm} \,(\star \widetilde{F})^{MNOPQM} \equiv \dfrac{1}{5!\,\sqrt{-G}} \, \epsilon^{MNOPQM'N'O'P'Q'} \, \widetilde{F}_{M'N'O'P'Q'} \,
\end{equation}
must be supplemented by hand after computing the equations of motion from the variation of \eqref{S_bosB}. We say that $C_4^+$ has to be ``self-dual'', which effectively halves its number of d.o.f. and makes it compatible with supersymmetry.

The bosonic equations of motion read
\begin{equation}
\label{pforms_EOM}
\begin{array}{rll}
d \star \widetilde{F}_{5} & = & \frac{1}{2} \, \epsilon_{\alpha \beta} \, \widetilde{\mathbb{H}}^{\alpha} \wedge  \widetilde{\mathbb{H}}^{\beta} \ , \\[2mm]
d \star (e^{-\Phi} \, H_{3} - e^{\Phi} \, C_{0} \, \widetilde{F}_{3} ) & = & - \widetilde{F}_{5} \wedge (\widetilde{F}_{3} + C_{0} \, H_{3}) \ , \\[2mm] 
d \star (e^{\Phi} \, \widetilde{F}_{3} ) & = &  \widetilde{F}_{5} \wedge H_{3}\ ,  \\[2mm] 
\nabla^{M} (e^{2 \Phi} \, \partial_{M} C_{0} ) & = & - \frac{1}{3!} \, e^{\Phi} \, H_{MNP} \, \widetilde{F}^{MNP} \ , \\[2mm] 
\square\Phi & = & e^{2\Phi} \, |F_{1}|^2 - \frac{1}{2} \, e^{-\Phi} \, 	|H_{3}|^2 +\frac{1}{2} \, e^{\Phi} \, 	|\widetilde{F}_{3}|^2 \ ,
\end{array}
\end{equation}
where $\,\widetilde{\mathbb{H}}^{\alpha} = (H_{3},\widetilde{F}_{3})\,$ and $\,\square\Phi \equiv \nabla^{M} \partial_{M} \Phi \,$. %
The Bianchi identities are
\begin{equation}
\label{pforms_BI}
dH_{3} = 0
\hspace{5mm} , \hspace{5mm}
dF_{1} = 0
\hspace{5mm} , \hspace{5mm}
d\widetilde{F}_{3} = - F_{1} \wedge H_{3}
\hspace{5mm} , \hspace{5mm}
d\widetilde{F}_{5} = H_{3} \wedge F_{3} \ .
\end{equation}
The Einstein equation of motion reads
\begin{equation}
\label{Einstein_EOM}
\begin{array}{rll}
R_{MN} &=& \frac{1}{2} \, \partial_{M} \Phi \,  \partial_{N} \Phi +
\frac{1}{2} \, e^{2\Phi}\, \partial_{M} C_{0} \,  \partial_{N} C_{0} \\[2mm]
&+& \frac{1}{4} \, \frac{1}{4!} \left(\widetilde{F}_{M P_{1}\cdots P_{4}}  \, \widetilde{F}_{N}{}^{P_{1}\cdots P_{4}} - \frac{1}{10} \, \widetilde{F}_{P_{1}\cdots P_{5}}  \, \widetilde{F}^{P_{1}\cdots P_{5}} \, G_{MN} 	\right) \\[2mm]
&+& \frac{1}{4} \, e^{-\Phi} \, \left( H_{M P_{1}P_{2}}  \, H_{N}{}^{P_{1}P_{2}} - \frac{1}{12} \, H_{P_{1}P_{2}P_{3}}  \, H^{P_{1}P_{2}P_{3}} \, G_{MN} 	\right) \\[2mm]
&+& \frac{1}{4} \, e^{\Phi} \, \left( \widetilde{F}_{M P_{1}P_{2}}  \, \widetilde{F}_{N}{}^{P_{1}P_{2}} - \frac{1}{12} \, \widetilde{F}_{P_{1}P_{2}P_{3}}  \, \widetilde{F}^{P_{1}P_{2}P_{3}} \, G_{MN} 	\right) \ .
\end{array}
\end{equation}
Note the equivalence between the first equation of motion in (\ref{pforms_EOM}) and the last Bianchi identity in (\ref{pforms_BI}) due to the self-duality of  $\,\tilde{F}_5\,$.

The type IIB supergravity action admits a SL$(2,\,\mathbb{R})$-symmetry. It is more obvious when rearranging its fields as
\begin{equation}
	\mathbb{H}^\alpha = \begin{pmatrix}
		H_3\\F_3
	\end{pmatrix}
	\hspace{5mm}\text{and}\hspace{5mm} \mathcal{V}(\Phi,\,C_0) =\begin{pmatrix}
		e^{-\Phi/2} & -e^{\Phi/2}C_0\\ 0& e^{\Phi/2}
	\end{pmatrix} \in \textrm{SL}(2,\,\mathbb{R})/\textrm{SO}(2)\,.
\end{equation}
Then the action $S_{bos}^{\textrm{IIB}}$ can be rewritten as
\begin{align}
	S_{bos}^{\textrm{IIB}} = \int d^{10}x \,\sqrt{-G} \big[R &+ \tfrac{1}{4}\text{Tr}\left(\partial_\mu m \partial^\mu m^{-1}\right) - \tfrac{1}{2} H^\alpha_{MNP} m_{\alpha\beta} H^{\beta\,MNP} - \frac{1}{4} |\tilde{F}_5|^2 \big] \\
	&- \epsilon_{\alpha\beta}\, C_4 \wedge H^\alpha \wedge H^\beta\nonumber
\end{align}
\begin{equation}
	m_{\alpha\beta} = \mathcal{V}\mathcal{V}^T = e^\Phi\begin{pmatrix}
		e^{-2 \Phi} + C_0^2  & - C_0  \\ -C_0 & 1
	\end{pmatrix}
\end{equation}
and we rewrite
\begin{equation}
\tilde{F}_5 = F_5 + \frac{1}{2}\,\epsilon_{\alpha\beta}\,\begin{pmatrix} B_2\\C_2\end{pmatrix}^\alpha \wedge \mathbb{H}^\beta\,.
\end{equation}
The SL(2,\,$\mathbb{R}$) acts linearly on $(B_2,\,C_2)$ and by multiplication on the left on $\mathcal{V}$.
The second and third equations of motion in (\ref{pforms_EOM}) are re-expressed in an $\,\textrm{SL(2)}_{\textrm{IIB}}\,$-covariant form
\begin{equation}
d \star (m_{\alpha \beta}\, \mathbb{H}^{\beta}) = -  \epsilon_{\alpha\beta}  \, \widetilde{F}_{5} \wedge \mathbb{H}^{\beta} \ .
\end{equation}
This SL(2, $\mathbb{R}$) global symmetry of the type IIB action is the manifestation of a non-perturbative duality symmetry of the underlying string theory called S-duality.

\begin{figure}[ht!]
\begin{summaryFramed}
\begin{itemize}
	\item Eleven is the maximal number of dimensions for supergravity. The two-derivative action of 11D SUGRA is uniquely fixed by supersymmetry. Its bosonic field content is comprised of the metric and a 3-form.
	\item There are two supergravities in ten dimensions: type IIA and type IIB, each associated with a different supersymmetry algebra, one non-chiral and the other chiral.
	\item The field content of type IIA supergravity consists of the metric, a dilaton and a two-form (the NSNS sector), as well as a 1-form, and a 3-form (the RR sector). It is the dimensional reduction of 11D SUGRA on $S^1$.
	\item The field content of type IIB supergravity consists of the metric, a dilaton and a two-form (the NSNS sector), as well as another scalar, a two-form and a self-dual four-form (the RR sector). It enjoys a $\SL(2,\,\mathbb{R})$ global symmetry.
\end{itemize}
\end{summaryFramed}
\end{figure}
\newpage

\section{Hidden symmetries and dualities}
\label{sec:HiddSym}

In a previous section (\ref{subsubsec:HiddenSym}), we have seen that the reduction of pure gravity down to three external dimensions is somewhat special. Upon dualising the vector fields, it enjoys a ``hidden'' $\SL(d+1)$ symmetry which is not inherited from the symmetries of the original theory. The goal of this section is to show that a similar phenomenon appears for 11D and type II A/B supergravities compactified on tori. We have covered enough background material to perform explicitly the reduction on tori of the bosonic action of the maximal supergravities and you could (but shouldn't) check the main result of this section yourself:
\begin{mdframed}
The hidden symmetry groups of the reduction of type IIA/B on $T^d$ are the $E_{11-d(11-d)}$ groups, the split real form of the exceptional Lie group of rank $11-d$ (see Table \ref{table:ESeries}).
\end{mdframed}

To some, this statement may sound like magic and in this section we will motivate why these specific groups appear. In particular, we will see how they can be derived from the so-called S- and T-dualities of superstrings. While a deep understanding of (super-)string theory is not essential to understand ExFT and the hidden symmetries of supergravities, it is required to fully appreciate this section and to understand the origin of the symmetry groups observed in supergravity. Since doing a full review of string theory is beyond the scope of these notes\footnote{And, frankly, beyond my abilities.}, we will introduce the bare minimum of background material required to describe string dualities and we refer to the standard reference textbooks for more details (e.g. \cite{Becker2007}).

We will start by reviewing a classical duality between the free theory of a $p$-form and the free theory of a $(D-p-2)$-form. We will be particularly interested in the case where $D=2$ and $p=0$ because it will provide a good introduction to T-duality. We will then move on to describing T-duality of bosonic string theory and its consequences in the gravity limit. Then we will present superstring dualities: the T-duality between type IIA and type IIB strings, the S-duality of type IIB, and how the strong coupling limit of type IIA hints at the existence of an 11D M-theory. Finally, we will show how these string dualities translate as hidden symmetries in supergravity compactifications.

\subsection{Electromagnetic dualities}

We say that two classical field theories are dual if the field content and the e.o.m. of one can be mapped to the other. This mapping can be done in a non-local and non-trivial way. The simplest example of this duality is ElectroMagnetic (EM) duality (see \cite{Freedman2012} section 7.8 for more details). Let us start with the Maxwell action for a $p$-form in $D$ dimensions:
\begin{equation}
\label{eq:ActionMaxwellP}
	S_{(p)} = -\frac{1}{2}\int\, \star F_{(p+1)} \wedge F_{(p+1)} = -\frac{1}{2}  \int d^Dx\, \sqrt{-g} \,\,|F_{(p+1)}|^2\,,
\end{equation}
where $F_{(p+1)} = dA_{(p)}$ and $\star$ is the Hodge dual operator. The equations of motion and Bianchi identity for this action are
\begin{equation}
\label{eq:pFormEoMBianchi}
	d\star F_{(p+1)} = 0\hspace{1cm}\text{and}\hspace{1cm}d F_{(p+1)} = 0\,.
\end{equation}
We can make this action first order in the derivatives by adding an auxiliary field $\tilde{A}_{D-p-2}$ and by considering $F_{(p+1)}$ as a fundamental degree of freedom (instead of $A_{(p)}$). We write the action
\begin{equation}
	\hat{S}\left[F_{(p+1)},\,\tilde{A}_{(D-p-2)}\right] = -\int \left(\frac{1}{2}\, \star F_{(p+1)}\wedge F_{(p+1)} + \tilde{A}_{(D-p-2)}\wedge d F_{(p+1)}\right).
\end{equation}
Computing the variation of $\hat{S}$ w.r.t. $\tilde{A}_{(D-p-2)}$, we get the Bianchi equation of \eqref{eq:pFormEoMBianchi}, which we can solve as $F_{(p+1)} = d A_{(p)}$. This reduces $\hat{S}$ to $S_{(p)}$. The other way around, varying $\hat{S}$ w.r.t. $F_{(p+1)}$, we get that $\star F_{(p+1)} = (-1)^{(D-p)}  d\tilde{A}_{(D-p-2)}$. Plugging this back into $\hat{S}$, we get, up to boundary terms, the dual action
\begin{equation}
	\tilde{S}\left[\tilde{A}_{(D-p-2)}\right] = -\frac{1}{2}\int d^Dx \, \star  \tilde{F}_{(D-p-1)}\wedge \tilde{F}_{(D-p-1)},
\end{equation}
where $\tilde{F}_{(D-p-1)} =  d \tilde{A}_{(D-p-2)}$. The e.o.m. and Bianchi identity of the dual theory are
\begin{equation}
	d \star \tilde{F}_{(D-p-1)} = 0 \hspace{1cm},\hspace{1cm} d \tilde{F}_{(D-p-1)} = 0\,.
\end{equation}
We can think of the duality as exchanging the role of the Bianchi identity and the equations of motion.

With this in mind, the field $A$ is no more fundamental than the field $\tilde{A}$, they are both describing precisely the same physics. In even dimensions, when $D= 2p +2$, the free theory of a $p$-form is self-dual, i.e. EM dualisation maps a $p$-form to another $p$-form. This should not obscure the fact that this defines a non-trivial duality, i.e. it is different from the identity map. For $D=4$ and $p=2$, this reduces to the usual EM duality exchanging the role of the electric and magnetic fields. 

\paragraph{An example: $D=2$ and $p=0$} Consider the action of a free scalar field $\phi$ on two-dimensional flat space:
\begin{equation}
	S = -\frac{1}{4\pi\alpha'}\int d^2x\,  \partial_\mu \phi \partial^\mu \phi\,,
	\label{eq:actionPeriodicScalar}
\end{equation}
The equations of motion of this action are
\begin{equation}
	\partial^\mu\partial_\mu \phi = \partial_+ \partial_- \phi =0\,\hspace{1.5cm}\Leftrightarrow\hspace{1cm} \partial_+ \phi = 0 = \partial_{-} \phi\,,
\end{equation}
with $\partial_\pm = \partial_0 \pm \partial_1$. Any solution can be written as 
\begin{equation}
\phi = \phi_+(t+x) + \phi_-(t-x),
\end{equation} for two functions $\phi_+$, called the ``left-moving'' mode, and $\phi_-$, the ``right-moving'' mode which depend only on $\zeta^+ = t+ x$ or $\zeta^- = t-x$. Let us dualise this scalar field. To do so, we introduce a one-form $F_{(1)}= d \phi$ and a Lagrange multiplier $\tilde{\phi}$, and we write the first order action
\begin{equation}
	\hat{S}\left[F_{(1)},\,\tilde{\phi}\right] = -\frac{1}{4\pi\alpha} \int\, \star F_{(1)}\wedge F_{(1)} + 2\,\tilde{\phi}\, d F_{(1)}\,.
	\label{eq:mixed2dEM}
\end{equation} 
Once again, we can solve for either $\tilde{\phi}$ or $F_{(1)}$ yielding the original action or the dual one. Going to the $\pm$ basis, the identification between $\phi$ and $\tilde{\phi}$ reads 
\begin{equation}
\partial_\pm \phi = \pm\, \partial_\pm \tilde{\phi}
\end{equation} 
which only changes the sign of the right-moving modes $\phi_-$ such that
\begin{equation}
\label{eq:TDualityScalar}
	\tilde{\phi} = \phi_+(t+x) - \phi_-(t-x)\,.
\end{equation}
This phenomenon in which a duality acts differently on right- and left-movers will appear again when studying T-duality later on.

\begin{figure}[ht!]
\begin{summaryFramed}
\begin{itemize}
	\item A duality between two classical theories A and B is a map from the field content of A to the field content of B such that the solutions to A are mapped to solutions of B, and conversely.
	\item A free $p$-form in $D$ dimensions is EM dual to a free $(D-p-2)$-form. This duality exchanges the role of the e.o.m. and the Bianchi identities.
	\item For $p = D/2-1$, EM duality is a self-duality (i.e. we recover the usual EM duality for a vector in four dimensions).
	\item For $p=0$, $D=2$ the self-duality is
	\begin{equation}
	\nonumber	\phi = \phi_+ (t+x) + \phi_-(t-x) \hspace{5mm}\Leftrightarrow\hspace{5mm} \tilde{\phi} = \phi_+ (t+x) - \phi_-(t-x)\,.
	\end{equation}
\end{itemize}
\end{summaryFramed}
\end{figure}

\subsection{String dualities}
\label{subsecStringDualities}

The meaning of duality can change a bit in the context of quantum theories. In most cases, we only have access to the perturbative regime of a given theory. In other words, we rely on a perturbative expansion around a classical theory. For a given theory, there might be more than a single classical limit. When a given theory has two such limits, we say that we have a \emph{duality} between the two quantum theories defined perturbatively around each of these limits \cite{polchinski2015dualities}.

\subsubsection{Closed bosonic strings}

Let us start by recalling some facts concerning the closed bosonic string on flat space. The critical bosonic string is a theory of $26$-scalar fields $X^\mu$ on the string worldsheet $\Sigma$, a two-dimensional manifold. The topology of the closed string worldsheet is a cylinder parametrised by $\tau \in \Rbb$ and $\sigma \in [0,\,2\pi[$. The action of the bosonic string is the Polyakov action
\begin{equation}
	S[X] = -\frac{1}{4\pi \alpha'}\int_\Sigma \mathrm{d}\tau\mathrm{d}\sigma \, \sqrt{-\gamma} \gamma^{ab}\partial_a X^\mu \partial_b X^\nu \, \eta_{\mu\nu}\,,
\end{equation}
where $\gamma$ is a metric on $\Sigma$. Using the conformal symmetry of this theory we obtain the equations of motion (varying for $X^\mu$), and an extra set of constraints (varying for $\gamma$), which reads
\begin{equation}
\begin{array}{ll}
	\partial_+ \partial_- X^\mu= 0\, \phantom{aaaaaaaa}&\text{(``Equations of motion''),}\\
	T_{ab} = 0,\,&\text{(``Stress tensor vanishes'')}\,.
	\end{array}
	\end{equation}
As for the free scalar in two dimensions, to solve the e.o.m. we express the scalars $X^\mu$ in terms of left- and right-moving modes:
\begin{equation}
	X^\mu = X_L^\mu(\tau + \sigma) + X_R^\mu(\tau-\sigma)\,.
\end{equation}
Then, we impose the periodic boundary condition on the string:
\begin{equation}
X^\mu(\tau,\,\sigma) = X^\mu(\tau,\,\sigma +2\pi)\,,
\end{equation}
yielding
\begin{align}
	&X_L^\mu(\zeta^+) = \frac{1}{2} \left(x_0^\mu + \alpha' p_0^\mu \,\zeta^+\right) + \text{oscillation modes}\,,\\
	&X_R^\mu(\zeta^-) = \frac{1}{2} \left(x_0^\mu + \alpha' p_0^\mu \,\zeta^-\right) + \text{oscillation modes}\,.
\end{align}
The boundary conditions impose the ``momentum'' $p_0^\mu$ of the left- and right-movers to be identical. Moreover, from the vanishing of the stress-energy tensor, we get a level matching condition and a constraint on the masses of string states. Classically, this corresponds to $(\partial_- X^\mu_L)^2 = (\partial_+ X^\mu_R)^2 =0$. From the quantisation procedure, we can read the spectrum of the strings excitations as
\begin{align}
	&m^2 = \frac{2}{\alpha'} (N + \tilde{N} -2)\,.
\end{align}
Here $N=\tilde{N}$ are the excitation numbers  (integers) of the left- and right-movers. In summary, at the classical level, the e.o.m. of the string separate its excitations into left- and right-movers. The boundary conditions impose the identification of the left- and right-momenta. Through the quantisation procedure, the excitations number of left- and right-movers are identified, and we can compute the mass spectrum.

\begin{figure}
	\centering
		\includegraphics[width=0.5\textwidth]{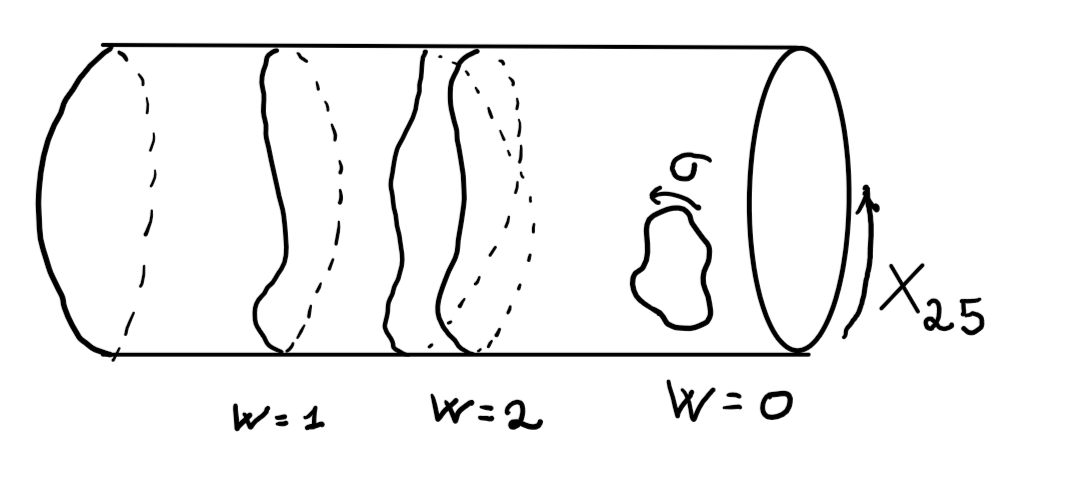}
	\caption{Illustration of strings with winding at $\tau$ fixed.}
	\label{fig:WindingString}
\end{figure}

\paragraph{Bosonic string on $S^1$}
Now that we recalled what was the string quantisation procedure with the scalars taking value in $\mathbb{R}^{1,\,25}$, we can redo it with one of the scalar $X^{25}$ compact (i.e. with target space $\mathbb{R}^{1,\,24}\times S^1_R$ where $R\in\mathbb{R}^+_0$ is the radius of the $S^1$). First, let us discard for a moment the first $25$ modes of the bosonic strings to focus on the $S^1$ parametrised by $X^{25}$. Since $S^1 = \Rbb/(2 \pi R\, \Zbb)$, $X^{25}(\tau,\,\sigma)$ does not have to be periodic and the boundary condition for $X^{25}$ becomes
\begin{equation}
	X^{25}(\tau,\,\sigma + 2 \pi) = X^{25}(\tau,\,\sigma) + 2 \pi w R\,,
\end{equation}
for any fixed $w \in \Zbb$. This integer is called the \emph{winding number} and it represents the number of times the string loops around the $S^1$ before closing (see Figure \ref{fig:WindingString} for an illustration).
The solutions to the e.o.m. with these boundary conditions are now
\begin{align}
\label{eq:solWindingString}
	&X_L^{25}(\zeta^+) = \frac{1}{2}x_0^{25} + (p_0^{25} + \frac{w R}{\alpha'}) \frac{\alpha'}{2} \zeta^+ + \text{oscillation modes}\,,\\
		&X_R^{25}(\zeta^-) = \frac{1}{2} x_0^{25} + (p_0^{25}- \frac{w R}{\alpha'})\frac{\alpha'}{2} \zeta^- + \text{oscillation modes}\,.
\end{align}
Moreover, translation in the $X^{25}$ direction by $2\pi R$ should just be the identity, since $X^{25}$ parametrises a circle. This imposes a quantisation of the momentum
\begin{equation}
	p_0^{25} = \frac{n}{R}\,,\hspace{5mm}\text{for}\hspace{5mm} n\in \mathbb{Z}.
\end{equation}
 This leads us to an expansion of the form
\begin{align}
	&X_L^{25}(\zeta^+) = \frac{1}{2}x_0^{25} + (\frac{n}{R} + \frac{w R}{\alpha'}) \frac{\alpha'}{2} \zeta^+ + \text{oscillation modes}\,,\\
		&X_R^{25}(\zeta^-) = \frac{1}{2}x_0^{25} + (\frac{n}{R} - \frac{w R}{\alpha'})\frac{\alpha'}{2} \zeta^- + \text{oscillation modes}\,.
\end{align}
In this context, the string spectrum and the level-matching condition become
\begin{equation}
	m^2 \alpha' =  \frac{n^2 \alpha'}{R^2} + \frac{w^2 R^2}{\alpha'} + 2 (N + \tilde{N} - 2)\hspace{5mm}\text{and}\hspace{5mm}
	n w + N - \tilde{N} = 0\,,
\end{equation}
given in terms of the number of left- and right-moving number of excitations $N$ and $\tilde{N}$. 
These formulas are completely invariant under the change
\begin{align}
	n \leftrightarrow w\hspace{1cm}\text{and}\hspace{1cm}R \leftrightarrow \frac{\alpha'}{R}\,.
\end{align}
The resulting string spectrum, after applying this duality, is exactly that of a closed bosonic string on a circle of radius $\frac{\alpha'}{R}$! At this level, this only \emph{suggests} that closed strings on a circle of radius $R$ could be related to closed strings on a circle of radius $\alpha'/R$ (in the free limit, i.e. when the dilaton coupling $e^\phi\, R_{(2)}$ is turned off). 
Now, studying the action of T-duality on the coordinate $X^{25}$, one realises that the T-dual theory is the original one in term of a new $\frac{\alpha'}{R}$ periodic coordinate $\tilde{X}^{25}$ defined as
\begin{equation}
	\tilde{X}^{25} = X^{25}_L - X^{25}_R\,.
\end{equation}
This should be reminiscent of \eqref{eq:TDualityScalar}. When we understand this duality as changing the sign of the right-moving modes, it is possible to understand better in which sense this duality can be exact even for perturbative string theory.

\subsubsection{T-duality on $T^d$ with background fields}

Now we make two important generalisations. The first one is to compactify string theory not on $S^1$ but on $T^d$. The second one is to allow for non-vanishing constant background fields $B_{\mu\nu}$, $G_{\mu\nu}$ and $\phi$ in the string action, these correspond to the massless modes of the bosonic string. 

\paragraph{Adding sources and the  gravity approximation}
The string action with sources is
\begin{equation}
	S = \frac{1}{4\pi\alpha'}\int_\Sigma \,-\sqrt{-\gamma}\, \gamma^{ab} G_{\mu\nu} \partial_a X^\mu \partial_b X^\nu + \epsilon^{ab}\,\partial_a X^\mu \partial_b X^\nu\, B_{\mu\nu} + \alpha'\, \sqrt{-\gamma}\,R_{(2)}\,\Phi
\end{equation}
This action is classically invariant under conformal transformation, as was the string in flat space. For consistency, we need for this invariance to remain unbroken at the quantum level. Treating the fields $G$, $B$ and $\Phi$ as couplings, in a perturbative approach, we can compute the associated $\beta$ functions (equivalently, the Weyl anomaly of the theory $\propto T^a_a$). For the theory to be conformal these functions must vanish. We thus obtain the set of equations 
\begin{align}
\nonumber &R_{\mu\nu} + 2 \nabla_\mu \nabla_\nu \Phi - \frac{1}{4} H_{\mu\rho\sigma} H\indices{_\nu^{\rho\sigma}} + O(\alpha') = 0\,,\\
 &\nabla_\mu\left(e^{-2 \Phi} H\indices{^\mu_{\nu\rho}}\right) + O(\alpha') = 0\,,\\
\nonumber & \frac{2}{3\alpha'} (26-D) + R - \frac{1}{2} |H|^2 - 4 e^{\Phi} \square e^{-\Phi}+ O (\alpha') = 0\,.
\end{align}
Which can precisely be obtained by extremising the action
\begin{equation}
\label{eq:ActionClosedBosonicString}
	S_{\text{bos. str.}} = \frac{1}{4\pi\alpha'}\int d^{D}x\sqrt{g} e^{-2\Phi} \left( \Lambda + R + 4 \partial_M\Phi \partial^M \Phi -\frac{1}{2} |H_3|^2 + O(\alpha')\right)\,.
\end{equation}
The presence of the cosmological constant $\Lambda = \frac{2(26-D)}{3\alpha'}$ reveals that the underlying string theory is inconsistent for $D\neq 26$ since the flat target space, on which we quantised the string, is not a solution of the string equations.

It is important to understand in which limit these equations are valid. First, the next order terms are not exactly of the form $O(\alpha')$ but $O(\alpha'\times \mathscr{R})$, where $\mathscr{R}$ denotes quantities built from the Riemann tensor so that $\alpha' \mathscr{R}$ is dimensionless. Thus, the length of the string must be smaller than the characteristic length of the curvature. We also make the approximation of weak string coupling. Since any perturbative string expansion is organised in power of $g_s^{g-2}$ where $g$ is the genus of the surface underlying the string interaction (think of it as generalised Feynman diagrams), we are only taking the lowest order in $g_s$ into account at $g=0$ (which explains the prefactor $e^{-2 \phi}$ in \eqref{eq:ActionClosedBosonicString}). We must thus work at small string coupling $g_s \ll 1$ for these equations to be valid.

\paragraph{Toroidal reduction}
We can now perform a toroidal reduction in the presence of background fields. The target space is $\mathbb{R}^{1,D-1} \times T^d$ (with $d+ D = 26$). We label the external coordinates with the index $\mu = 1,\,\dots,\,D$ and the internal ones with $m=D+1,\,\dots,\,26$. We leave $G_{\mu\nu} = \eta_{\mu\nu}$, $G_{\mu m} = 0$, $B_{\mu \nu } =0$ and $B_{\mu m} = 0$. However, we allow for constant $G_{mn}$ and $B_{mn}$ fields. The periodicities of the target space coordinates $X^m$ are fixed to $2\pi$ and the lengths of the circles are encoded in the metric.

Going properly through the e.o.m. and the definitions of the momenta-densities and momenta, we get that the internal string modes $X^m_{L,\,R}$ receive oscillators contributions, exactly as in the non-compact case, as well as  momenta contributions. These momenta $P^m_{L,\,R}$ are functions of two integers $W^m$ and $K_m$ corresponding to winding modes and the quantised momenta, these generalise the quantities $w$ and $n$ in \eqref{eq:solWindingString}. Explicitly, they are
\begin{align}
p^m_R &= W^m + G^{mk}\left(\alpha' K_k - B_{kl}W^l\right)\,,\\
p^m_L &= -W^m + G^{mk}\left(\alpha' K_k - B_{kl}W^l\right)\,.
\end{align}
Going through the usual exercise of computing the spectrum, we get that the masses of the closed string on tori are
\begin{equation}
M^2  = M_0^2  + \frac{2}{\alpha'} (N + \tilde{N}-2)\,.
\end{equation}
The $M_0^2$ term encodes contributions from the winding and quantised momenta and is defined as
\begin{equation}
M_0^2 = \frac{1}{\alpha'}(W^m\,\,K_m) \begin{pmatrix} \frac{1}{\alpha'}(G_{mn} - B_{mk}G^{kl}B_{ln}) & B_{mk}G^{kn}\\ -G^{mk}B_{kn} & \alpha' G^{mn} \end{pmatrix} \begin{pmatrix}W^m \\K_m\end{pmatrix}\,.
\end{equation}
The level matching condition becomes
\begin{equation}
W^m K_m = N- \tilde{N}\,.
\end{equation}
Introducing and index $_M = (^m,\,_m)$ and the $\SO(d,\,d)$ invariant matrix
\begin{equation}
\eta_{MN} = \begin{pmatrix} 0& 1\\1&0\end{pmatrix}\,,
\end{equation}
we can show that the spectrum is $\SO(d,\,d,\,\mathbb{Z})$ covariant. Indeed, with the definitions
\begin{equation}
	P^M = (W^m,\,K_m) \hspace{5mm} \mathcal{M}_{MN} = \begin{pmatrix} \frac{1}{\alpha'}(G_{mn} - B_{mk}G^{kl}B_{ln}) & B_{mk}G^{kn}\\ -G^{mk}B_{kn} & \alpha' G^{mn} \end{pmatrix}\,,
\end{equation}
the level matching condition becomes
\begin{equation}
P^M P_M = P^M \eta_{MN} P^N = 2(N-\tilde{N})
\end{equation}
and 
\begin{equation}
M_0^2 = \alpha' P^M \mathcal{M}_{MN} P^N\,.
\end{equation} 
Here $P_M$ transforms as a vector under $\SO(d,\,d)$ whereas 
$M_{MN}$ parametrises elements of $\SO(d,\,d)/(\SO(d)\times \SO(d))$. The spectrum is invariant under the action of $\textrm{O}(d,\,d,\,\mathbb{Z})$. One can show that the full string theory is indeed invariant under this $\textrm{O}(d,\,d,\,\mathbb{Z})$ action. This study of the string sigma-model was done in \cite{BUSCHER198759,BUSCHER1988466}. The transformation rules of these fields under T-duality are sometimes called ``Buscher rules''. The elements of $\textrm{O}(d,\,d,\,\mathbb{Z})$ can be understood as either modular transformation of $T^d$ (i.e. $R\rightarrow \alpha'/R$ or large coordinate change) or as a constant shift of the two-form. We will see how this translates in the gravity limit in the next section.

\begin{figure}[h]
\begin{summaryFramed}
\begin{itemize}
	\item The massless fields of bosonic closed strings are a metric $G_{\mu\nu}$, a two-form $B_{\mu\nu}$ and a dilaton $\phi$. Their low-energy effective action is
	\begin{equation}
	S_{\text{bos. str.}} = \frac{1}{4\pi\alpha'}\int d^{D}x\sqrt{g} e^{-2\Phi} \left( \Lambda + R + 4 \partial_M\Phi \partial^M \Phi -\frac{1}{2} |H_3|^2 + O(\alpha')\right)\,.
\end{equation}
where the cosmological constant $\Lambda = \tfrac{2(26-D)}{3\alpha'}$ vanishes for critical strings.
	\item The spectrum of closed bosonic string on $T^d$ is invariant under the group $\SO(d,\,d,\,\mathbb{Z})$ due to ``\emph{T-duality}''. 
	\item For $d= 1$, T-duality sends the radius of $S^1$ to $\alpha'/R$.
\end{itemize}
\end{summaryFramed}
\end{figure}

\subsection{T-duality in the low energy limit}
\label{sec:TDualityInSUGRALimit}

The SO$(d,\,d,\,\mathbb{Z})$ duality group of the bosonic string can also be seen in the gravity limit \cite{Maharana1992}. This is also the simplest example of hidden symmetries in gravity without using the duality between $p$-forms and $(D-p-2)$-forms. The starting point is the toroidal reduction of the gravity limit of the bosonic string action \eqref{eq:ActionClosedBosonicString} :
\begin{equation}
\label{eq:ActionClosedBosonicString2}
	S_{\text{bos. str.}} = \frac{1}{4\pi\alpha'}\int d^{D+d}x\sqrt{g} e^{-2\Phi} \left( \Lambda + R + 4 \partial_M\Phi \partial^M \Phi -\frac{1}{2} |H_3|^2\right)\,.
\end{equation}
Our main focus in these notes is the (super)gravity limit so we will keep the cosmological constant although it vanishes when considering the critical string. (Moreover, I was curious to see what would happen to it).

\subsubsection{The lengthy computation}\label{subsubsec:LengthyComputation} This whole section could be thought of as a long exercise. To perform the dimensional reduction, we must first perform a rescaling of the metric to remove $e^{-2\Phi}$ pre-factor. Otherwise, the ``total derivatives'' in \eqref{eq:ReductionMetricS1} will mess up our nice reduction formulas \eqref{eq:ReductionMetricS1} and \eqref{eq:ReductionpFormS1}. Using the formulas \eqref{eq:confTsfMet1}-\eqref{eq:confTsfMet3}, the precise rescaling is
\begin{equation}
g_{E\,\mu\nu} = e^{4\Phi/(D+d-2)} g_{S\,\mu\nu}
\label{eq:LengthyRescalg}
\end{equation}
and yields the action low-energy closed bosonic string action in the Einstein frame
\begin{align}
	S_{\text{E}} = \frac{1}{2\kappa_{D+d}}\int d^{D+d}x\,\sqrt{g}\,\Big(R -\frac{4}{D+d-2}\partial_M\Phi \partial^M \Phi\,-\,& \frac{1}{2} e^{-8\Phi/(D+d-2)}|H_3|^2\\ 
	\nonumber&\phantom{aaaa}+ e^{4/(D+d-2) \Phi} \Lambda\Big)\,.
\end{align}
We can now compactify this action on $T^d$ and group together the $p$-form terms of the same rank:
\begin{align}
\label{eq:SbosTdSLd}
S =  \frac{1}{2 \kappa^{2}_D} \int d^Dx \sqrt{-g} \Big(& R  -\frac{4}{D+d-2}\partial_\mu\Phi \partial^\mu \Phi - \beta \partial_\mu \phi \partial^\mu \phi\\
\nonumber&+\frac{1}{4} \partial_\mu M_{mn} \partial^\mu M^{mn}  - \frac{1}{4} e^{\frac{4}{D-2} \phi} e^{-4 \beta \phi} e^{-8\tfrac{\Phi}{(D+d-2)}} F_{\mu np}F^{\mu np}\\
\nonumber&- \frac{1}{4} \left( e^{2\beta \phi} M_{mn} {F_{\mu\nu}}^m {F^{\mu\nu\,n}}+ e^{\frac{4}{D-2} \phi}e^{-2 \beta \phi}e^{-8\tfrac{\Phi}{(D+d-2)}}\tilde{F}_{\mu\nu p} \tilde{F}^{\mu\nu p}\right) \\
\nonumber& - \frac{1}{12} e^{\frac{4}{D-2} \phi}e^{-8\tfrac{\Phi}{(D+d-2)}}  \tilde{F}_{\mu\nu\rho} \tilde{F}^{\mu\nu\rho} + e^{4\tfrac{\Phi}{(D+d-2)}} e^{-2\tfrac{\phi}{(D-2)}}\Lambda \Big)\,.
\end{align}
Now we perform the change of variables
\begin{equation}
\begin{cases}
	\Phi &= \frac{1}{4}\left((D-2) \hat{\Phi} + d\,\, \hat{\phi}\right)\,\\[2mm]
	\phi&= \frac{\hat{\Phi} - \hat{\phi}}{2\beta}\,.
	\end{cases}
	\label{eq:redefDilatons}
\end{equation}
With these variables, the prefactor of the scalar kinetic terms only depends on $\hat{\phi}$ whereas the prefactor of the three-form field strength kinetic term only depends on $\hat{\Phi}$. We obtain the action 
\begin{align}
\label{eq:SUGRASOddNotCov}
S =   \frac{1}{2 \kappa^{2}_D} \int d^Dx \sqrt{-g} \Big(& R -\frac{D-2}{4} \partial_\mu \hat{\Phi} \partial^\mu \hat{\Phi} - \frac{d}{4} \partial_\mu \hat{\phi} \partial^\mu \hat{\phi}\\
\nonumber&+\frac{1}{4} \partial_\mu M_{mn} \partial^\mu M^{mn} - \frac{1}{4} e^{- 2\hat{\phi}}F_{\mu np}F^{\mu np}\\
\nonumber&- \frac{1}{4} e^{-\hat{\Phi}}   \left( e^{\hat{\phi}} M_{mn} {F_{\mu\nu}}^m {F^{\mu\nu\,n}}+ e^{-\hat{\phi}}\tilde{F}_{\mu\nu p} \tilde{F}^{\mu\nu p}\right) \\
\nonumber& - \frac{1}{12} e^{-2\hat{\Phi}}  \tilde{F}_{\mu\nu\rho} \tilde{F}^{\mu\nu\rho} + e^{\hat{\Phi}} \Lambda \Big)\,.
\end{align}
The claim is that this action has an $\SO(d,\,d)$ global symmetry group. 

How to observe that with the minimal amount of effort? To reexpress this action as manifestly symmetric, we must understand how the scalars parametrise the $\SO(d,\,d)/(\SO(d)\times \SO(d))$ coset-space in term of a matrix $\hat{M}$. Let us focus on the 1-forms kinetic terms. Embedding the 1-forms $A^m$ and $B_m$ into a vector $A^M = (A^m,\,B_m)$ and defining $F^M = dA^M$ we have that
\begin{equation}
	S_{(1)\,kin} = -\frac{1}{2 \kappa^{2}_D} \int d^Dx \sqrt{-g}\,\tfrac{1}{4}\, {F_{\mu\nu}^M} \hat{M}_{MN} F^{\mu\nu\,N}\,.
\end{equation}
is equivalent to the third line in \eqref{eq:SUGRASOddNotCov} because
\begin{equation}
	\tilde{F}_{\mu\nu\,m} = \partial_{[\mu}B_{\nu]\,m} + \partial_{[\mu}{A_{\nu]}}^n \wedge B_{nm}\,,
\end{equation}
and
\begin{equation} \hat{M}_{MN} =
\begin{pmatrix}
		e^{\hat{\phi}} M_{mn} - 	e^{-\hat{\phi}}B_{mk}M^{kl}B_{ln} &	e^{-\hat{\phi}} B_{mk} M^{kn}\\ -e^{-\hat{\phi}}M^{mk} B_{kn}  & 	e^{-\hat{\phi}}M^{mn}
	\end{pmatrix}\,.
\end{equation}

We compute the scalar kinetic term $\tfrac{1}{4}\partial_\mu \hat{M}_{MN} \partial^\mu \hat{M}^{MN}$\footnote{The inverse scalar coset representative is
\begin{equation} \hat{M}^{MN} =
\begin{pmatrix}
	e^{-\hat{\phi}}M^{mn}	 &	-e^{-\hat{\phi}}M^{mk} B_{kn} \\ e^{-\hat{\phi}} B_{mk} M^{kn} & 	e^{\hat{\phi}} M_{mn} - 	e^{-\hat{\phi}}B_{mk}M^{kl}B_{ln}\,.
	\end{pmatrix}
\end{equation}} to obtain the scalar kinetic term of the action \eqref{eq:SUGRASOddNotCov}. Finally, the two-form $B_{\mu\nu}$ in terms of $d+D$ dimensional quantities has to be slightly modified to make the $\SO(d,\,d)$ symmetry manifest. We choose
\begin{equation}
\label{eq:2FSODD}
	B_{\mu\nu}= C_{\mu\nu} + {A_{[\mu}}^n {B_{{\nu]}\,n}}- {A_\mu}^m B_{mn} {A_{\nu}}^n
\end{equation}
which is a $D$-dimensional two-form whose gauge transformations are
\begin{equation}
	\delta\, B = \frac{1}{2} \left(\Lambda_{KK}^m \partial_{[\mu}B_{\nu]m} + \Lambda_{(1)\,m}\partial_{[\mu}{A_{\nu]}^m}\right).
\end{equation}
This leads to the definition of the invariant \emph{improved} field-strength
\begin{equation}
	\tilde{F}_{\mu\nu\rho}= 3 \partial_{[\mu} B_{\nu\rho]} + \tfrac{1}{2} {A_\mu}^m  \partial_{\nu} B_{\rho]\,m} +\tfrac{1}{2}\,B_{[\mu\,m}  \partial_{\nu} {A_{\rho]}}^m \,.
\end{equation}
With these definitions the action \eqref{eq:SbosTdSLd} reduces to
\begin{align}
\label{eq:SUGRASOddCov}
S =   \frac{1}{2 \kappa^{2}_D} \int d^Dx \sqrt{-g} \Big(& R -\frac{D-2}{4} \partial_\mu \hat{\Phi} \partial^\mu \hat{\Phi} +\frac{1}{4} \partial_\mu \hat{M}_{MN} \partial^\mu \hat{M}^{MN}\\
\nonumber &- \frac{1}{4} e^{-\hat{\Phi}} \hat{M}_{MN} {\tilde{F}_{\mu\nu}}^M {\tilde{F}^{\mu\nu\,N}} - \frac{1}{12} e^{-2\hat{\Phi}}  \tilde{F}_{\mu\nu\rho} \tilde{F}^{\mu\nu\rho} + e^{\hat{\Phi}} \Lambda \Big)\,.
\end{align}
We stress again that only the $\mathfrak{sl}_d \ltimes \Lambda^2 \mathbb{R}^d$ part of the symmetry algebra of the theory has a geometric origin in the $D+d$ dimensional theory. The negative roots of $\mathfrak{so}_{d,\,d}$ do not. Moreover, changing either the coefficient in front of the dilaton kinetic term or the coefficient in the exponential $e^{-2\Phi}$ in \eqref{eq:ActionClosedBosonicString} would prevent us to observe the symmetry enhancement we do.

Let us compare this to results obtained previously, computing the spectrum of closed bosonic string. Following the identification \eqref{eq:LengthyRescalg}-\eqref{eq:redefDilatons}, the scalar matrix $\hat{M}_{MN}$ can be rewritten as
\begin{equation}
	\hat{M}_{MN}= \begin{pmatrix}
		G_S - B.G^{-1}_S.B & B.G^{-1}_S\\
		G^{-1}_S.B & G^{-1}_S
	\end{pmatrix}
\end{equation}
in term of the metric $G_S$ \emph{in the string frame}. From there, it is easy to read the action of $\SO(d,\,d)$ on the internal metric and two-forms. This action matches that obtained from the string up to the $\alpha'$ factors. Those can be re-obtained by the action of the SO($d,\,d$) rescaling
\begin{equation}
\begin{pmatrix}
	\alpha' & 0\\ 0 & \alpha'^{-1}
\end{pmatrix}
\end{equation}
which is the identity in the units where $\alpha' = 1$. 

This computation also gives a prescription for the action of $\SO(d,\,d)$ on $G_{\mu m}$ and $B_{\mu m}$ when they are non-zero. Going thought the chains of definitions, $\SO(d,\,d)$ acts linearly on the vector
\begin{equation}
	({A_\mu}^m,\,C_{\mu m} + {A_\mu}^m B_{mn})
\end{equation}
Finally, the dilaton is not left invariant under the action of T-duality. It transforms in such a way that leaves the normalisation of the action constant and we can compute that $\delta \Phi = -\frac{1}{2} \delta \log(\text{det } G_S)$.

So far we have presented this duality as an $\SO(d,\,d)$ symmetry because it is a connected simple Lie group. However, from the $D$-dimensional action, we actually have a O$(d,\,d,\,\mathbb{Z})$ symmetry. The orientation reversing elements of O$(d,\,d,\,\mathbb{Z})$ are the ones we obtain from acting with an odd number of T-dualities.

\subsubsection{The shorty prediction}

The $\SO(d,\,d)$ symmetry of \eqref{eq:SUGRASOddCov} could have been guessed from our discussion concerning T-duality. The T-duality of closed strings implies that one cannot distinguish the $D$-dimensional theory obtained from the compactification on a circle of radius $R$ or $\alpha'/R$. This implies that there are two possible uplifts of the same $D$-dimensional theory to $D+1$ dimensions. By the same logic, the closed string action compactified on $d$ circles can come from two inequivalent $D+d$ theories.

\begin{figure}[t]
	\centering
		\includegraphics[width=0.7\textwidth]{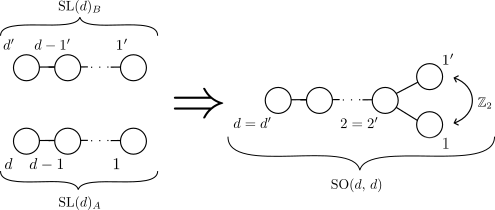}
	\caption{An illustration of the origin of the $\mathfrak{so}_{d,d}$ symmetry group, using T-duality. The nodes are labelled to follow their identification.}
	\label{fig:sodd_hidden}
\end{figure}

What does it imply for the $D$-dimensional symmetry groups? Starting from one $D+d$ realisation, we compactify $d$ times and expect a $\SL(d)_A$ symmetry group. From the dual picture, we also expect a $\SL(d)_B$ symmetry group. These two groups should intersect on a common $\SL(d-1)$ subgroup, identified as the symmetry group of the compactification from $D+(d-1)$ to $D$:
\begin{equation}
\SL(d-1) \hspace{2mm} \subset \hspace{2mm}\begin{array}{l} \SL(d)_A\\[2mm] \SL(d)_B\end{array}  \hspace{2mm} \subset \hspace{2mm} \SO(d,\,d)\,.
\end{equation}
The smallest simple group with that property is $\SO(d,\,d)$ (as can be seen from the Dynkin diagrams Fig. \ref{fig:sodd_hidden}) . The fact that this hidden symmetry group should be simple is not required by our reasoning, and as such it is not a proof. A proper proof requires the explicit computation made in the previous subsection, but it is still a useful prediction that will work in all the examples we will discuss here, this is sometimes called one of the ``silver rules of supergravity'' \cite{Julia1998}.

We can check that the scalar field content of the lower dimensional theory corresponds to the positive roots, and Cartan subalgebra, of $\mathfrak{so}_{d,\,d}$. Under the $\mathfrak{sl}_d$ branching we have
\begin{align}
	\mathfrak{so}_{d,\,d} &\rightarrow \underbrace{(\Lambda^2 \mathbb{R}^d)_{- 2}}_{\text{neg. roots}} \oplus  \underbrace{\mathfrak{gl}_d}_{\text{metric d.o.f.}} \oplus \underbrace{(\Lambda^2 \mathbb{R}^d)_{+ 2}}_{\text{2-form d.o.f.}}\,.
\end{align}
This corresponds to the gravity sector of the higher-dimensional theory coupled to a two-form, which is exactly the massless field content of the bosonic string. The same can be done for the vectors by studying the branching of the vector representation of $\mathfrak{so}_{d,\,d}$ under $\mathfrak{sl}_d$. One gets indeed a $D+d$ vector (from the metric) and a $D+d$ 1-form (from the two-form).

\begin{figure}[h]
\begin{summaryFramed}
\begin{itemize}
	\item The reduction of \eqref{eq:ActionClosedBosonicString2} on $T^d$ is invariant under a global $\SO(d,\,d)$ symmetry group.
	\item The appearance of the $\SO(d,\,d)$ group can be traced back to T-duality using simple group theoretical arguments.
\end{itemize}
\end{summaryFramed}
\end{figure}

\subsection{Superstring dualities}
\label{subsectionSuperStringDualities}

\begin{figure}[h]
	\centering
		\includegraphics[width=0.3\textwidth]{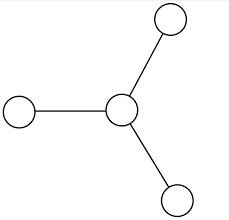}
	\caption{The Dynkin diagram of $\mathfrak{so}_8$}
	\label{fig:D4Dynkin}
\end{figure}

For superstrings, we will not provide as much details as in the case of the closed bosonic string. There are two maximal superstring theories: type IIA and type IIB, each with a different spectrum of massless fields. Both have a geometric interpretation in ten dimensions. The effective action in the supergravity limit (small $\alpha'$ and small string coupling) only depends on the massless modes of these string theories. These modes are the only ones we will describe here. They transform under the little group of $\SO(1,\,9)$ which is SO(8). The group $\SO(8)$ admits three inequivalent 8-dimensional representations because its Dynkin diagram (Fig. \ref{fig:D4Dynkin}) enjoys an $S_3$ symmetry, this fact is sometimes called ``\emph{SO(8)-triality}''. These three representations are called $\mathbf{8}_v$, $\mathbf{8}_c$ and $\mathbf{8}_s$. Because of the triality, what matter is the relative choice of what we call $v,\,c$ or $s$ (much in the same way that we can only distinguish the two spinorial representations of the $\Spin(d,\,d)$ groups with respect to one another).

\paragraph{Type IIA} Its massless states transform as
\begin{equation}
(\mathbf{8}_v \oplus \mathbf{8}_c)_L\otimes (\mathbf{8}_v \oplus \mathbf{8}_s)_R\,.
\end{equation}
The bosonic massless states thus contain the NS-NS sector ($\phi$, $B_{\mu\nu}$ and $G_{\mu\nu}$)
\begin{equation}
\mathbf{8}_v\otimes \mathbf{8}_v = \mathbf{1} \oplus \mathbf{28} \oplus \mathbf{35}
\end{equation}
as well as the ``R-R''sector
\begin{equation}
	\mathbf{8}_c \otimes \mathbf{8}_s = \mathbf{8}_v \oplus \mathbf{56}_v
\end{equation}
corresponding to a massless vector and a 3-form. The fermionic sector transforms in the $\mathbf{8}_v \otimes \mathbf{8}_s \oplus \mathbf{8}_v\otimes \mathbf{8}_c$ and does not have a preferred chirality. The supergravity limit of this theory is type IIA supergravity \ref{subsubsec:TypeIIA}

\paragraph{Type IIB} Its massless states transform as
\begin{equation}
(\mathbf{8}_v \oplus \mathbf{8}_c)_L\otimes (\mathbf{8}_v \oplus \mathbf{8}_c)_R\,.
\end{equation}
This leads to the same NS-NS sector as in the type IIA case:
\begin{equation}
\mathbf{8}_v\otimes \mathbf{8}_v = \mathbf{1} \oplus \mathbf{28} \oplus \mathbf{35}\,,
\end{equation}
but a different R-R sector:
\begin{equation}
	\mathbf{8}_c \otimes \mathbf{8}_c = \mathbf{1} \oplus \mathbf{28} \oplus \mathbf{35}_+\,.
\end{equation}
corresponding the IIB axion, a two-form and a self-dual four-form. The fermionic sector transforms in the $\mathbf{8}_v \otimes \mathbf{8}_c \oplus \mathbf{8}_v\otimes \mathbf{8}_c$ and is chiral. The supergravity limit of this theory is type IIB supergravity \ref{subsec:Type_IIB}.

\subsubsection{T-duality}
As in the case of the bosonic string, T-duality changes the sign of the right-movers of the coordinate on which it is applied. Acting with a single T-duality modifies the chirality of the right-moving fermions, exchanging the representations $\mathbf{8}_c  \leftrightarrow \mathbf{8}_s\,$ for the right-movers. It follows that T-duality maps type IIA to type IIB, and conversely. An even number of T-dualities correspond to self-dualities of type IIA and type IIB. At the level of the supergravity action, this manifests in the fact that type IIA on a circle gives the same D=9 maximal supergravity as type IIB on a circle \cite{Bergshoeff2002}. We will not reproduce this computation here (although you now have all the tools needed to perform it).

\paragraph{The scalar manifold of type IIA/B SUGRA on T$^d$} From the previous computations, we understood that NSNS fields admit an O$(d,\,d)$ duality group. We will focus here on the subgroup of the self-dualities which is $\SO(d,\,d)$. In the supergravity limit (and for $d< 6$), the purely internal NSNS fields parametrise a scalar manifold
\begin{equation}
	\mathcal{M}_{\text{scal},\,\text{NSNS}} = \mathbb{R}_\Phi\times \frac{\SO(d,\,d)}{\SO(d)\times \SO(d)}\,,
\end{equation}
The factor $\mathbb{R}_\Phi$ corresponds to the possible value of the dilaton. Adding the RR fields extends the scalar manifold to
\begin{equation}
\label{eq:scalManifoldRRsodd}
\mathcal{M}_{\text{scal, }d} = \left(\mathbb{R}_\Phi\times \Spin(d,\,d) \ltimes \mathbf{spin}(d,\,d)\right)/(\Spin(d) \times \Spin(d))\,,
\end{equation}
where $\mathbf{spin}(d,\,d)$ denotes one of the two spin irreps of the $\Spin(d,\,d)$ group. These representations are of dimensions $2^{d-1}$. The presence of the spin irreps requires us to extend the duality group from $\SO(d,\,d)$ to $\Spin(d,\,d)$. 

\begin{figure}[h]
	\centering
		\includegraphics[width=0.5\textwidth]{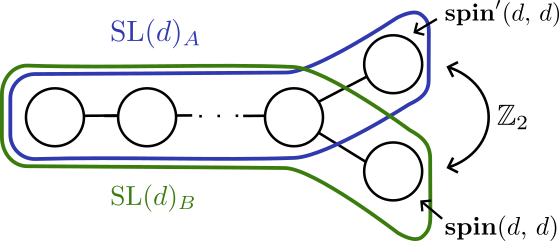}
	\caption{The Dynkin diagram of $\mathfrak{so}_{d,\,d}$. We circled the nodes corresponding to the $\mathfrak{sl}(d)_A$ and $\mathfrak{sl}(d)_B$ subalgebras. The fundamental weight associated to the two-rightmost nodes corresponds to the two $\mathbf{spin}(d,\,d)$ representations. The outer automorphism of $\mathfrak{so}_{d,\,d}$ acts by exchanging those two nodes.}
	\label{fig:SpinRepSOdd}
\end{figure}

To justify heuristically why \eqref{eq:scalManifoldRRsodd} is correct, we must recall some facts about the spin irreps of $\mathfrak{so}_{d,\,d}$. For any $d$, there are two spin irreps corresponding to the Dynkin labels\footnote{See Appendix \ref{app:LieAlgebra} for a reminder.} $\left[0,\,\cdots,\,0,\,1\right]$ and  $\left[0,\,\cdots,\,1,\,0\right]$ (i.e. corresponding to the two rightmost nodes in Fig \ref{fig:SpinRepSOdd}). These two representations are related by the outer automorphism of the Dynkin diagram exchanging the two right-most nodes. 
Let us show how to get the IIA/B RR-fields from the spin representations. To do this we must branch $\mathfrak{so}_{d,\,d}$ representations down to $\mathfrak{sl}_d$. However, after selecting a specific spin representation, there are two non-equivalent embeddings of $\mathfrak{sl}_d$ in $\mathfrak{so}_{d,\,d}$. This depends on whether the spin representation node is one of the  $\mathfrak{sl}_d$ nodes or not (see Fig. \ref{fig:SpinRepSOdd}). We define the subgroup $\SL(d)_B$ as the subgroup of $\Spin(d,\,d)$ corresponding to the $\mathfrak{sl}_d$ algebra containing the spin irrep node. The group $\SL(d)_A$ corresponds to the other $\mathfrak{sl}_d$ algebra. This gives two possible branching rules, corresponding to the appropriate odd/even $p$-form representation of type IIA/B, as summarised in Table \ref{Table:spinddSplit}. When $d\geq 5$ for $\SL(d)_A$ and $d\geq 6$ for $\SL(d)_B$, we must include the EM-duals of the higher-dimensional $p$-forms to fill the spin irreps. This corresponds, in $D$ dimensions, to using the EM duality between scalars and $(D-2)$-forms. Furthermore, for $d\geq 6$, it becomes necessary to also dualise the NSNS fields to obtain the full scalar manifold. For example, when $d=6$, the NSNS two-form $B_{\mu\nu}$ is dual to a scalar field. This yields an extra scalar field, and the expected total of 70 scalars required to build a maximal supergravity in four dimensions. This line of reasoning becomes complicated for $d=7$,  where there are  three external dimensions and scalars are dual to vectors. One must add 14 extra scalar fields coming from the dual KK-vectors and two-forms to obtain the scalar manifold of maximal supergravity in three dimensions.

\begin{table}[h]
	\centering
		\begin{tabular}{|c|c|c|c|c|c|}\hline
			$d$ & $2^{d-1}$ & $\SL_A(d)$  & $\SL_B(d)$ & dim$\left(\mathcal{M}_{\text{scal, NSNS}}\right)$ & $ 1 + d^2 + 2^{d-1}$ \\ \hline
			3& 4 &  $\mathbf{3}\oplus \mathbf{1}$  & $\mathbf{1} \oplus \bar{\mathbf{3}}$ & 10 &  14 \\\hline
			4 & 8 & $\mathbf{4} \oplus \bar{\mathbf{4}}$  & $\mathbf{1}\oplus\mathbf{6} \oplus \mathbf{1}$ & 17 & 25 \\ \hline
			5 & 16 & $\mathbf{5} \oplus \bar{\mathbf{10}} \oplus \mathbf{1}$ & $\mathbf{1}\oplus \mathbf{10} \oplus \bar{\mathbf{5}}$& 26 & 42 \\\hline
			6 & 32 &$\mathbf{6}\oplus \mathbf{20}\oplus \bar{\mathbf{6}}$  &  $\mathbf{1}\oplus \mathbf{15}\oplus \bar{\mathbf{15}} \oplus \mathbf{1}$ & 37 & 69 \\\hline
			7 & 64 & $\mathbf{7}\oplus \mathbf{35}\oplus \bar{\mathbf{21}} \oplus \mathbf{1}$ & $ \mathbf{1} \oplus \mathbf{21}\oplus \bar{\mathbf{35}} \oplus \bar{\mathbf{7}}$ & 50 & 114 \\\hline
		\end{tabular}
		\caption{We write down the branching of the Spin$(d,\,d)$ representation of dimensions $2^{d-1}$ under the two possible inequivalent $\SL(d)$ subgroups. The resulting representation correspond to even/odd $p$-form representations of $\SL(d)$.}
		\label{Table:spinddSplit}
\end{table}

\paragraph{Vectors fields in \emph{D} dimensions} The RR sector also gives contribution to the $D$-dimensional 1-form field content. We can study their representations under Spin$(d,\,d)$ by computing the $\SL(d)_{A/B}$ branchings. Under the $\SL(d)_B$ group, these should transform as even forms, while they transform as odd form under $\SL(d)_A$. In other words, they transform like the dual representation of the scalar fields. This time, for $d\geq 5$ one must be careful to correctly consider the dual NSNS fields. The same ideas hold for higher $p$-forms.

\subsubsection{S-duality} We saw that type IIB supergravity enjoys an $\SL(2,\,\mathbb{R})$ global symmetry. This symmetry originates from a $\SL(2,\,\mathbb{Z})$ duality of the underlying type IIB superstring. The group $\SL(2,\,\mathbb{Z})$ is generated by two elements $S$ and $T$ satisfying $S^2 = -1 = (ST)^3$. They admit the 2-dimensional representation:
\begin{equation}
S = \begin{pmatrix}
	0& -1\\1& 0
\end{pmatrix}\,\,\,\text{and} \,\,\, T = \begin{pmatrix} 1 & 1 \\ 0 & 1 \end{pmatrix}\,.
\end{equation}
We can read their action on the string fluxes from the SUGRA limit. Setting the axion $C_0$ to zero, the inversion element, $S$, of $\SL(2,\,\mathbb{Z})$ sends 
\begin{equation}
g_S = e^\Phi \xrightarrow[]{S} e^{-\Phi} = g_S^{-1}.
\end{equation}
Since amplitudes for perturbative strings are computed as powers of $g_S$, this duality is a ``strong/weak'' duality. It connects strong and weak coupling regimes of type IIB. As such, S-duality can only be proven if one has a fully non-perturbative description of type IIB string theory, which we do not.

In the absence of such a description, we can still argue that it is a good duality by looking at quantities protected (e.g. by supersymmetry) and by making sure that they do transform correctly under S-duality. For example, in the Einstein frame, S-duality exchanges the tension of the fundamental string ($T_{F_1} = e^{\Phi/2}/2\pi \alpha'$) and of the D1-brane ($T_{D1} =  e^{-\Phi/2}/2\pi \alpha'$). Since these are BPS states, we should be able to trust these results at any string coupling. From this computation, it is natural to understand S-duality as exchanging the role of F-string and D-strings at strong coupling. 

For non-vanishing $C_0$ flux, we can extend the action of $S$ to the full $\SL(2,\,\mathbb{Z})$ group. It acts on $\tau = C_0 + i \, e^{\Phi}$ as a linear fractional transformation
\begin{equation}
\begin{pmatrix}
	p & r\\
	q & s\\
\end{pmatrix} \cdot \tau = \frac{p \tau + q}{r \tau + s}\,.
\label{eq:sDualityTau}
\end{equation}
This duality transformation sends the fundamental string to a BPS bound state called the $(p,\,q)$-string with $p$ and $q$ co-prime integers. One can check again that the tension of such objects transforms in the appropriate manner under $\SL(2,\,\mathbb{Z})$.

\subsubsection{An $11^{\text{th}}$ dimension}
\label{sec:IIA11D}
We have seen that type IIA supergravity can be obtained as a compactification of 11D SUGRA on a small circle. This picture is entirely valid in the supergravity limit and could have a naive interpretation as a bookkeeping technique. Can we interpret this as a special limit of string theory? Looking at the supergravity action, we can observe the $S^1$ on which 11D SUGRA is compactified is of radius $R = g_s \,l_s$. Therefore, at weak string coupling, this radius is very small. This suggests that the ``string theory'' whose 11D supergravity is a ``low energy limit'' (whatever this concept might mean) arises in the strong coupling limit of type IIA. 

This interpretation is further supported by the observation that type IIA string theory admits bound states of $n$ particle-like objects called ``D0-branes''. The tension/masses of a $n$-$D0$-brane goes as $n g_s^{-1}$. This begs to be interpreted as a tower of higher KK-modes. This strong coupling limit of type IIA is called ``M-theory'' but its microscopic description is not as well understood as that of superstring theories. This is due to the absence of a weakly interacting perturbative limit.

\begin{summaryFramed}
\begin{itemize}
	\item There are two maximally supersymmetric closed string theories: type IIA and type IIB. These theories enjoy dualities.
	\item T-duality is a duality between type IIA and type IIB superstring on a circle.
	\item S-duality is a self-duality of type IIB string theory. It is a ``strong/weak'' duality.
	\item M-theory is a strong coupling limit of type IIA superstring.
\end{itemize}
\end{summaryFramed}

\subsection{Exceptional hidden symmetries}
\label{subsection:ExceptionalHiddenSymmetries}

We could directly compute the compactification of type IIA/IIB or 11D supergravities on $T^{d(+1)}$ as was done in \cite{Cremmer:1997ct}. Using EM dualities, we would observe that
\begin{mdframed}
\begin{itemize}
\item The symmetry group of type IIA/B SUGRA compactified on $T^d$ is E$_{d+1(d+1)}$.
\item The symmetry group 11D SUGRA compactified on $T^{d+1}$ is E$_{d+1(d+1)}$.
\end{itemize}
\end{mdframed}
However, using superstring dualities, we can predict this result using a bit of group theory. We have already argued that the $\SL(d)_{A/B}$ groups, originating from diffeomorphisms, enhanced to $\SO(d,\,d)$ due to T-duality. This result is not complete. It does not consider the description of type IIA supergravity as 11D SUGRA on $S^1$ or, equivalently, the $\SL(2)$ symmetry of type IIB SUGRA. 

One the one hand, the global symmetry group obtained by compactifying type IIA on $T^d$ is that of 11D SUGRA on $T^{d+1}$ i.e. $\SL(d+1)_{11}$ and not $\SL(d)_A$. On the other hand, the global symmetry group of type IIB on $T^d$ is $\SL(2)_S\times \SL(d)_B$. These two groups intersect on a $\SL(d-1)$ subgroup and we have
\begin{equation}
\SL(d-1) \hspace{2mm} \subset \hspace{2mm}\begin{array}{l} \SL(d)_A \subset \SL(d+1)_M \\[2mm] \SL(d)_B \times \SL(2)\end{array}  \hspace{2mm} \subset \hspace{2mm} \mathrm{E}_{d+1(d+1)}\,.
\end{equation}
The symmetry enhancement can be elegantly captured by the Dynkin diagram of Fig. \ref{fig:Dynkin_diagram_type_E7} showing that the symmetry groups of maximal supergravities on tori are the exceptional Lie groups of Table \ref{table:ESeries}.

\begin{table}[h]
\centering
\begin{tabular}{|l|c|c|}\hline
D & G & H\\\hline
9 & \text{SL}(2) & \text{SO}(2)\\
8 & $\text{SL}(2)\times \text{SL}(3)$ & $\text{SO}(2)\times \text{SO}(3)$\\
7 & \text{SL}(5) & \text{SO}(5)\\
6 & \text{SO}(5,\,5) & $\text{SO}(5)\times \text{SO}(5)$\\
5 & $\text{E}_{6(6)}$ & \text{USp}(8)\\
4 & $\text{E}_{7(7)}$ & \text{SU}(8)\\
3 & $\text{E}_{8(8)}$ & \text{SO}(16)\\\hline
\end{tabular}
\caption{Symmetry groups $G$ of the ungauged maximal supergravities in dimensions $D$. Their scalar manifold are $G/H$, where $H$ is the R-symmetry group.}
\label{table:ESeries}
\end{table}

As a last remark to conclude this section, I want to stress that the usual KK-procedure, summarised by eqs \eqref{eq:ReductionMetricS1} and \eqref{eq:ReductionpFormS1}, does not make any of the $\mathrm{E}_{d+1(d+1)}$ groups apparent. For these groups to appear, we must not only use the appropriate EM dualities, but also perform certain non-linear fields redefinitions. We encountered those already when studying the SO$(d,\,d)$ duality groups of the bosonic string. In that context, the definition of the two-form in equation \eqref{eq:2FSODD} differs from that of the KK reduction \eqref{eq:AnsatzPform}. These redefinitions can be inferred from the gauge transformations of the reduced fields.

\begin{summaryFramed}
A picture is worth a thousand words:

{\centering
      \includegraphics[width=0.6\textwidth]{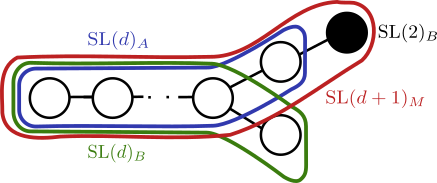}
      \captionof{figure}{The Dynkin diagram of E$_{d+1(d+1)}$. In red we circle the nodes corresponding to the SL($d+1$) diffeomorphism group of M-theory. In blue and green, those of SL$(d)_{A/B}$. We colored in black the node corresponding to the $\SL(2)$ S-symmetry group of type IIB.}
			\label{fig:Dynkin_diagram_type_E7}
  \par}
\end{summaryFramed}

\newpage

\section{\texorpdfstring{$\Es$}{E7}-Exceptional Field Theory}
\label{sec:E7ExFT}

The goal of Exceptional Field Theory (ExFT) is to reformulate the equations of motion of 11D and type IIA/IIB SUGRA as a set of manifestly $\SO(1,\,10-d)\times \mathrm{E}_{d(d)}$ covariant equations. Although the motivation is to reproduce the exceptional symmetry group observed when reducing the maximal supergravities on tori, ExFT encodes the full supergravity, including internal fluctuations. It is not merely a consistent truncation on tori. It is a reorganisation of the SUGRA d.o.f. that makes the action of the exceptional symmetry groups apparent. By making explicit the $\mathrm{E}_{d(d)}$ hidden symmetries, new avenues of research are opened. It becomes possible to build new analytic solutions of 10D/11D supergravities, new consistent truncations, study mass spectra, and to peer into the non-geometric behaviour of strings. These have applications to multiple research directions such as the swampland program, the gauge/gravity correspondence, black-hole physics,$\dots$ In this section we will focus on the case $d=7$ for concreteness.

The rewriting of the e.o.m. is done in two steps. In the first step, we must define what it means to be $\mathrm{E}_{d(d)}$-invariant. By analogy with GR where being ``$\GL(d)$-invariant'' means to be invariant under the action of the Lie derivative, we will introduce notions of generalised geometry such as the ``\emph{generalised Lie derivative}'' and the ``\emph{section constraints}''. In turn, these will allow us to define the ``\emph{generalised internal/external diffeomorphisms}'' which are the proper way to talk about $\text{E}_{d(d)}$-covariance. The second step is to define a field theory, i.e. a field content and equations of motion. We will show that the $\Es$-covariance only admit a unique two-derivatives pseudo-action from which we can write equations of motion. These equations of motion will turn out to be equivalent to both those of type IIB and 11D supergravities (and thus also those of type IIA). 

This section is based on the presentation of the generalised Lie derivative in \cite{Berman2012} (whose geometric interpretation can be found in \cite{Coimbra2011,Coimbra:2012af}). The original $\Es$-ExFT formulation is given in \cite{Hohm:2013uia} where more details are presented. If one wishes to reproduce some results presented here, we have included explicit formulas for the $\es$ generators in App. \ref{app:LieAlgebra}. 

\subsection{Generalised Lie derivative and section constraints}

We start by recalling some properties of the usual Lie derivative, to generalise this notion. For two vector fields $v$ and $w \in \Gamma(TM)$, where $M$ is a manifold, the Lie derivative of $w$ with respect to $v$ is defined as
\begin{equation}
	L_v \,w = \left[v,\,w\right]\,.
\end{equation} 
Choosing a coordinate system, it is simply
\begin{equation}
	(L_v w)^m = v^k\partial_k w^m - \partial_k v^m w^k\,.
	\label{eq:LiederivativeVector}
\end{equation}
The first term, $v^k\partial_k w^m$, encodes translations (i.e. it would also appear for the Lie derivative of a function). The second term $\partial_k v^m w^k$ encodes how the vector $w^k$ rotates under the action of $\mathfrak{gl}(d)$. This can be rewritten in term of a ``projector'' $\mathbb{P}$ defined as
\begin{equation}
	\mathbb{P} : \mathrm{End}(\mathbf{d}) \rightarrow \mathfrak{gl}(d): M_n{}^m \rightarrow (t^\alpha)\indices{^{m}_{n}}(t_\alpha)\indices{^{p}_{q}} M_p{}^q\,,
\end{equation}
where $\mathbf{d}$ is the fundamental representation of $\mathfrak{gl}(d)$, and $t_\alpha$ is a basis of $\mathfrak{gl}(d)$\footnote{The adjoint index, $\alpha$ is raised and lowered using the Cartan-Killing matrix}. This definition allows us to rewrite the Lie derivative as
\begin{equation}
	(L_v w)^m = v^k\partial_k w^m - \mathbb{P}\indices{^m_n^p_q} \partial_pv^q w^n\,.
	\label{eq:LiederivativeP}
\end{equation}
For the algebra $\mathfrak{gl}(d)$, the projector $\mathbb{P}\indices{^m_n^p_q} $ is just the identity $\delta^m_q \delta^p_n$ and one can check that eq. \eqref{eq:LiederivativeP} reduces to the usual formula \eqref{eq:LiederivativeVector}. For a tensor transforming in some representation of $\GL(d)$ indexed by $M$, we can write
\begin{equation}
	(L_v T)^M = v^k \partial_k T^M - \mathbb{P}\indices{^M_N^p_q}\partial_p v^q T^N =v^k \partial_k T^M -(t^\alpha)\indices{^M_N}(t_\alpha)\indices{^{p}_{q}}\partial_p v^q T^N\,,
	\label{eq:LieDerivativeT}
\end{equation} 
where the $(t_\alpha)\indices{^M_N}$ are a basis of $\mathfrak{gl}_d$ in the representation of $T$. This is again equivalent to the usual Lie derivative acting on tensors. Finally, the Lie derivative for a density of weight $\lambda$ is obtained by adding a term $\lambda\, \partial_m v^m T^M$ to \eqref{eq:LieDerivativeT}.

To construct a theory which is diffeomorphisms invariant, we must impose that its equations of motion are covariant under the infinitesimal transformation
\begin{equation}
	\delta_\xi = L_\xi \,\hspace{10mm} \forall \xi \in \Gamma(TM)\,.
\end{equation}
The goal of this section will be to produce a theory invariant under such infinitesimal transformations but for a new operator that generalises the usual Lie derivative. 

\subsubsection{Generalised Lie derivative}
After this unusual presentation of the Lie derivative, let us define the $\Es$-generalised Lie derivative. In this context, a vector field $V^M$ is understood as transforming in the fundamental, $\mathbf{56}$, representation of $\Es$, denoted with indices $M,\,N,\,\dots$. Thus, the partial derivative $\partial_M$ also transforms in the fundamental representation, implying the existence of 56 coordinates $Y^M$.
Let $\Lambda^K$ and $V^M$ be two vector fields, we consider them to be vector fields on $\mathbb{R}^{56}$ (this is not strictly speaking the case, but we will come back to this issue later on). We define:
\begin{equation}
	L_\Lambda V^M = \Lambda^K\partial_K V^M - \alpha_7\, \mathbb{P}\indices{^M_N^P_Q} \partial_P \Lambda^Q V^N + \lambda_V \partial_K\Lambda^K V^M\,,
	\label{eq:genLieDer7}
\end{equation}
where $\alpha_7$ is a constant to be fixed and $\lambda_V$ is the density weight of $V$. This time, $\mathbb{P}$ is not the projector on $\mathfrak{gl}(d)$ but the projector on $\mathfrak{e}_{7(7)}$:
\begin{equation}
\mathbb{P}: \text{End}(\mathbf{56}) \rightarrow \es\hspace{0.5cm}\text{i.e.}\hspace{0.5cm}
\mathbb{P}\indices{^M_N^K_L} = (t_\alpha)\indices{^M_N} (t^\alpha)\indices{^K_L}
\end{equation}
where the generators $t_\alpha$ form a basis of $\es$ in the fundamental representation\footnote{As usual, the adjoint index $\alpha$ is raised and lowered using the Cartan-Killing metric on $\es$.}.
This can be generalised to define the Lie derivative of any tensor, transforming in some representation of $\es$. For example, the generalised Lie derivative on the adjoint representation is
\begin{equation}
	L_{\Lambda} W_\alpha = \Lambda^K \partial_K W_\alpha + 12\, f_{\alpha\beta}{}^{\gamma}\, (t^\beta)_L{}^K\,\partial_K \Lambda^L\, W_\gamma + \lambda\, \partial_K \Lambda^K\, W_\alpha\,,
\end{equation}
where $f_{\alpha\beta}{}^\gamma$ are the $\es$ structure constants and $(t^\beta)_N{}^M$ are the $\es$ generators acting on the fundamental indices.

Following our analogy with the usual diffeomorphisms, this allows us to define the infinitesimal transformation $\delta_\Lambda$ of any vector $V$ as 
\begin{equation}
	\delta_\Lambda V = L_\Lambda V\,.
\end{equation} 
This generalises to any tensor $T$ in a representation of $\Es$ as in \eqref{eq:LieDerivativeT}. When we say that a theory is $\Es$ invariant, we mean that its equations of motion are covariant under the action of generalised infinitesimal diffeomorphisms $\delta_\Lambda = L_\Lambda$, where $L$ is the \emph{generalised} Lie derivative.

\subsubsection{Section constraints} 
For consistency, we must check that the generalised infinitesimal diffeomorphisms, when equipped with the commutator as a Lie-algebra bracket, close and satisfy the Jacobi identity i.e.
\begin{align}
	\exists \,\Lambda_3\,\text{such that}\,\forall\,\Lambda_1,\,\Lambda_2 \hspace{1cm}&[\delta_{\Lambda_1},\,\delta_{\Lambda_2}] = \delta_{\Lambda_3}\,,\label{eq:closureGenLieDer}\\
	\forall\,\Lambda_1,\,\Lambda_2,\,\Lambda_3\hspace{1cm}&[\delta_{\Lambda_1},\,[\delta_{\Lambda_2},\,\delta_{\Lambda_3}]] + \text{(perm.)} = 0 \label{eq:JacobiGenLieDer}\,.
\end{align}
The computations of \eqref{eq:closureGenLieDer} and \eqref{eq:JacobiGenLieDer} are simpler in terms of the tensor
\begin{equation}
	Y\indices{^{MN}_{PQ}} = \delta^M_P \delta^N_Q - \alpha_7\, \Pbb\indices{^M_Q^N_P} + \beta_7 \,\delta^M_Q \delta^N_P\,,
\end{equation}
from which we can rewrite
\begin{equation}
	L_\Lambda V^M = \Lambda^K \partial_K V^M - V^K\partial_K \Lambda^M + Y\indices{^{MN}_{PQ}} \partial_N U^P V^Q\,.
	\label{eq:genLieDerY}
\end{equation}
The first two terms correspond to the usual Lie derivative. The ``$Y$'' term encodes the deviation from the usual Lie derivative. The weight $\beta_7$ is the weight of the parameter of the infinitesimal transformation $\Lambda^M$. For the $\Es$-generalised Lie derivative, one can compute explicitly that \eqref{eq:closureGenLieDer} and \eqref{eq:JacobiGenLieDer} are \emph{not} satisfied! Does this mean that the generalised Lie derivative is terminally ill-defined? Of course not, there is a way out.

The failure to close and to respect the Jacobi identity are (for $\alpha_7=12$ and $\beta_7=\tfrac{1}{2}$) proportional to terms of the form
\begin{equation}
	Y\indices{^{MN}_{PQ}} \partial_M\otimes \partial_N\,.
\end{equation}
As such, (\ref{eq:closureGenLieDer}) and (\ref{eq:JacobiGenLieDer}) can be satisfied if we impose that
\begin{equation}
\label{eq:E7sectionConstraints}
	(t_\alpha)^{MN}\partial_M \partial_N V=0 ,\,\hspace{5mm}(t_\alpha)^{MN}\partial_M V \partial_N W =0 \hspace{5mm} \text{ and }\hspace{5mm} \Omega^{MN}\partial_M V \partial_N W=0\,.
\end{equation}
The fundamental index $M$ is raised and lowered by the symplectic matrix $\Omega$\footnote{Due to the fact that $\Es \subset \text{Sp}(56)$} and $V$ and $W$ represent any field. This signals that fields transforming under the generalised Lie derivative does not depend on all the coordinates $Y^M$ but only on a subset, $y^m$. The section constraints will be satisfied if these $y^m$ satisfy
\begin{equation}
	(t_\alpha)^{mn} \partial_m \otimes \partial_n = 0\hspace{10mm} \text{ and }\hspace{10mm} \Omega^{mn} \partial_m \otimes \partial_n = 0\,.
\end{equation} 
These conditions are known as the ``\emph{section constraints}'' and are the defining feature of ExFTs.

After imposing the section constraints, some transformations $L_\Lambda$ are actually trivial. This is the case for all parameters $\Lambda$ of the form
\begin{equation}
	\Lambda^M = \left(t^\alpha\right)^{MN}\partial_N\chi_\alpha\hspace{4mm} \text{ or }\hspace{4mm} \Lambda^M = \Omega^{MN}\chi_N\,,
	\label{eq:trivialLambda}
\end{equation}
where $\chi_N$ satisfies the same section constraints as $\partial_N$. The fact that those transformations are trivial is what allows the algebra of generalised diffeomorphisms to close. We can check this by computing that, up to the section constraints, we have
\begin{equation}
\label{eq:commDelta}
	\left[\delta_{\Lambda_1},\,\delta_{\Lambda_2}\right] = \delta_{\left[\Lambda_1,\,\Lambda_2\right]_E}\,,
\end{equation}
where the ``E-bracket'' is defined as
\begin{align}
\label{eq:defEBracket}
\left[\Lambda_1,\,\Lambda_2\right]^M_E = 2\,\Lambda^K_{\lb 1\rd} \partial_K \Lambda^M_{\ld 2\rb} &+ 12\, (t_\alpha)^{MN}(t^\alpha)_{KL}\, \Lambda^K_{\lb 2\rd}\partial_N\Lambda^L_{\ld 1\rb}\\
\nonumber & - \frac{1}{4} \,\Omega^{MN}\Omega_{KL}\,\partial_N(\Lambda^K_{2}\Lambda^L_{1})\,.
\end{align}
This E-bracket, although similar to the Lie bracket, does not satisfy the Jacobi identity. It satisfies a weaker identity implying that the failure to satisfy Jacobi identity is a term producing trivial gauge transformations. In other words, we have that
\begin{equation}
	\big[[\delta_{[\Lambda_1},\,\delta_{\Lambda_2}],\delta_{\Lambda_3]}\big] = 0
\end{equation}
only after applying the section constraints, and even though
\begin{equation}
	\big[[\Lambda_{[1},\,\Lambda_{2}]_E,\,\Lambda_{3]}\big]_E \neq 0\,.
\end{equation}
This is still sufficient for the internal diffeomorphism algebra to close.

\subsubsection*{ }
\begin{exercise}
Check that the transformations of parameter \eqref{eq:trivialLambda} are indeed trivial transformations after imposing the section constraints. 
\end{exercise}
\begin{exercise}
Verify the relation \eqref{eq:commDelta} given the definition of the E-bracket in \eqref{eq:defEBracket}.
\end{exercise}

\newpage
\begin{summaryFramed}
\begin{itemize}
	\item The ``$\Es$-\emph{generalised Lie derivative}'' is
	\begin{equation}
	L_\Lambda V^A =  \Lambda^K V_K V^A - 12 \,\mathbb{P}\indices{^A_N^K_L}\partial_K V^L V^N + \frac{1}{2}\, \partial^K\Lambda_K V^A\,,
	\end{equation}
	where $\mathbb{P}\indices{^M_N^K_L} = (t_\alpha)\indices{^M_N}(t^\alpha)\indices{^K_L}$.
	\item The generalised infinitesimal diffeomorphisms are defined as \begin{equation}
	\delta_\Lambda = L_\Lambda\,.
	\end{equation}
	\item For the transformations $\delta_\Lambda$ to close and to satisfy the Jacobi identity, one must impose the ``\emph{section constraints}'':
	\begin{equation}
	{Y^{MN}}_{PQ}\, \partial_M\otimes \partial_N = 0\,,
	\end{equation}
where
\begin{equation}
	Y\indices{^{MN}_{PQ}} = \delta^M_P \delta^N_Q - 12\, \Pbb\indices{^M_Q^N_P} + \frac{1}{2} \,\delta^M_Q \delta^N_P\,.
\end{equation}
This reduces the number of physical coordinates:
\begin{equation}
V(Y^M) \rightarrow V(y^m)\hspace{5mm} \text{ for } \hspace{5mm} y^m \subset Y^M\,.
\end{equation}
\end{itemize}
\end{summaryFramed}

\subsection{Solving the section constraints}

It was shown in \cite{Bossard2015,Bandos2015a} that there are two inequivalent solutions to the section constraints of $\Es$-ExFT. For each of those solutions, the coordinates $y^m$ can be intepreted as either the 6 internal coordinates of IIB or the 7 internal coordinates of 11D SUGRA. The good news is that both these solutions can be built using a bit of group theory and our knowledge of the type IIB and 11d supergravities.

\subsubsection{Type IIB solution} 

Since type IIB supergravity is 10 dimensional, we expect to solve the section constraints by imposing that the fields depend on only six specific internal coordinates: $y^m$, $m=1,\,\dots,\,6$. These coordinates span a subset of the $Y^M$ coordinates. The coordinate dependence of the various fields reduces as
\begin{equation}
	\phi(x^\mu,\,Y^M) \rightarrow \phi(x^\mu,\,y^m)\,.
\end{equation}
The $y^m$ coordinates should transform under the $\GL(6)$ subgroup of internal diffeomorphisms. We know that type IIB supergravity also enjoys an SL$(2)$ global symmetry. We are thus interested in the branching of 
\begin{equation}
\Es\rightarrow \mathrm{GL}(6) \times \mathrm{SL}(2).
\end{equation}
Under this branching we have
\begin{align}
&\textbf{56} \rightarrow (6,\,1)_{+2} + (6',\,2)_{+1}+ (20,\,1)_{0}+ (6,\,2)_{-1}+ (6',\,1)_{-2}\\
&\label{eq:branchingIIB}\textbf{133}\rightarrow (1,2)_{+3} + (15',1)_{+2} + (15,2)_{+1}+ (35+1,1)_{0}+ (1,\,3)_0\\
\nonumber& \hspace{12mm}(1,2)_{-3}  + (15,1)_{-2}+ (15',2)_{-1}  
\end{align}
where the index denotes the $\mathrm{GL}(1)\subset \mathrm{GL}(6)$ weight of each irrep. This branching corresponds to the splitting of coordinates
\begin{align}
\{Y^M\}&\rightarrow \{y^m,\,y_{m\,a},\,y_{[kmn]},\,y^{m\,a},\,y_m\}\label{IIBCoordDec}\,.
\end{align}
where $a= 1,\,2$ is the fundamental SL(2) index.
Selecting the coordinates $y^m$ to be the coordinates of highest $\mathrm{GL}(1)$ weight solves the section constraints \eqref{eq:E7sectionConstraints}. This can be shown easily:
\begin{itemize}
 \item $\Omega^{MN}\partial_MA\partial_NB = 0$ if and only if $\Omega^{mn} = 0$. This is true because $\Omega_{MN}$ is composed of off-diagonal blocks.
\item  $(t_\alpha)^{MN}\partial_M A\, \partial_NB =0$ if and only if $(t_\alpha)^{mn} = 0$. We know that $(t_\alpha)^{MN}\partial_M\partial_N$ transforms in the $\mathbf{133}$ of $\es$. Thus, for $(t_\alpha)^{mn}\partial_m\partial_n$ to be non-zero, $(t_\alpha)^{mn}$ should be of weight $+4$ to compensate for the weight $-2$ of $\partial_m$. We can check that there is no generator of weight $+4$ in the decomposition of the adjoint representation \eqref{eq:branchingIIB}.
\end{itemize}

\subsubsection{11D solution} The 11D solution is obtained through the same method. We expect the solution to the section constraints to single out seven coordinates $y^m$, transforming in the fundamental representation of a GL(7) subgroup of $\Es$. These can be obtained by studying the branching $\Es \rightarrow \GL(7)\,.$
Under this branching we have
\begin{align}
\label{eq:56splitGL7}
\mathbf{56}&\rightarrow 7_{+3} + 21'_{+1} + 21_{-1} + 7'_{-3}\,,\\
\label{eq:branching11D}\mathbf{133}&\rightarrow 7'_{+4} + 35_{+2}+ 1_0 + 48_{0} + 35'_{-2} + 7_{-4}\,,
\end{align}
This corresponds to the splitting of the coordinates
\begin{equation}
\{Y^M\} \rightarrow \{ y^m,\,y_{mn},\,y^{mn},\,y_{m}\}
\end{equation}
By the same argument as in the previous case, selecting $y^m$ to be the coordinates of highest GL(1) weight solves the section constraints. In particular, you can check that $\Omega^{mn}=0$ because it is an off-diagonal block-matrix and ${t_\alpha}^{mn} =0$ because there are no generators of weight $+6$ in \eqref{eq:branching11D}.

\subsubsection{Interpretation} The solutions to the section constraints can be interpreted in term of ``generalised geometry'' \cite{Coimbra2011,Coimbra:2012af}. Here we will focus on the elven-dimensional solution, but similar reasoning holds for the type IIB solution. After imposing the section constraints, we should think of the $y^m$ as local coordinates on a 7-dimensional manifold $M$. This manifold can be identified with the internal manifold of 11D SUGRA compactified to four dimensions. With this identification in mind\footnote{and glossing over important details.}, a vector field in the 56-dimensional representation of $\Es$ should be understood as a section of the space
\begin{equation}
E = TM \oplus \Lambda^2T^*M \oplus \Lambda^5T^*M \oplus (T^*M \otimes \Lambda^7 T^*M)\,.
\end{equation}
You should compare this to the branching in \eqref{eq:56splitGL7}. Sections of this bundle can further be identified with elements in the direct sum of
\begin{itemize}
\item $\Gamma(TM)$: the space of infinitesimal diffeomorphisms; 
\item $\Gamma(\Lambda^2T^*M)$: the space of gauge transformations of $A_3$;
\item $\Gamma(\Lambda^5T^*M)$: the space of gauge transformations of the EM dual of $A_3$: a 6-form;
\item $\Gamma(T^*M \otimes \Lambda^7 T^*M)$: the space dual to $\Gamma(TM)$.
\end{itemize}
In other words, the parameters of the generalised Lie derivative, $\Lambda^M$, encode in a single object all gauge and diffeomorphisms transformations of the internal space of 11D SUGRA. Seen as section of $E$, the generalised Lie derivative can be rewritten using only the usual operators on (co-)tangent bundles, the exterior derivative and the interior product. Although this obscures the role of the $\Es$ group, this construction is completely geometric. 

From this point of view, the $\Lambda^M$ were never vector fields on a 56-dimensional manifold. It is our bad physicist's habit of first computing and then defining what the objects are that mislead us. It was convenient to start with the analogy with the usual Lie derivative and then impose the section constraints. However, if we were doing our work properly, we would first define a specific vector bundle on a six- or seven- dimensional manifold, and only then build a generalised Lie derivative (which closes and satisfies Jacobi identity). Not all is bad with the ``computing first'' philosophy. In the next sections this will allow us to build new pairs of generalised Lie derivative/section constraints that we can use to build different ExFTs.
\newpage
\begin{summaryFramed}
\begin{itemize}
	\item The section constraints admit two inequivalent solutions, corresponding to type IIB/11D SUGRA.
	\item The generalised Lie derivative can be understood as a differential operator on a specific vector bundle on 6 or 7 dimensional manifolds. This removes the need for section constraints but obscures the role of $\Es$.
\end{itemize}
\end{summaryFramed}

\subsection{\texorpdfstring{The equations of motion of $\Es$-ExFT}{The equations of motion of E7(7)-ExFT}}

We will now build an action invariant under the $\Es$-generalised Lie derivative. The field content comprises an  ``external'' metric $g_{\mu\nu}(x^\mu,\,Y^M)$ as well as an ``internal'' metric $\mathcal{M}_{MN}(x^\mu,\, Y^M)$, parametrising elements of $\Es/\text{SU}(8)$. These are the natural generalisations of the external and internal metrics of KK compactification. The generalisation of the KK vectors is the vector field $A_\mu{}^M$, transforming in the $\mathbf{56}$. These will be used as a gauge connection for the generalised internal diffeomorphisms. Finally, the invariance of the action under the generalised Lie derivative will force us to add two-forms $B_{\mu\nu\,\alpha}$ and $B_{\mu\nu\,M}$. They transform in the adjoint and the fundamental representations of $\Es$ respectively. We will impose that $B_M$ also satisfies the section constraints. In summary, the field content of the $\Es$-ExFT is
\begin{equation}
	\left\{g_{\mu\nu},\,\Mcal_{MN},\,{A_\mu}^M,\,B_{\mu\nu\,\alpha},\,B_{\mu\nu\,M}\right\}
\end{equation}

The bosonic equations of motion of the ExFT are completely determined by requiring the invariance under the generalised diffeomorphisms in both the external and internal coordinates. The equations of motion can be computed from the pseudo-action
\begin{align}
\label{E7ExFTAction}
	S = \int \mathrm{d}^4x\, \mathrm{d}^{56}Y\, e\, \biggl(\hat{R} &+ \frac{1}{48} g^{\mu\nu} \mathcal{D}_\mu \mathcal{M}^{MN}\mathcal{D}_\nu \mathcal{M}_{MN}\\
	\nonumber&- \frac{1}{8} \mathcal{M}_{MN} \mathcal{F}^{\mu\nu\,M} \mathcal{F}\indices{_{\mu\nu}^N} +e^{-1} \mathcal{L}_{top} - V \biggl)\,,
\end{align}
supplemented with the so-called ``\emph{twisted self-duality condition}''
\begin{equation}
	\mathcal{F}_{\mu\nu}{}^M = -\frac{1}{2} e \,\epsilon_{\mu\nu\rho\sigma} \,\Omega^{MN}\, \mathcal{M}_{NK} \,\mathcal{F}^{\rho\sigma K}\,,
	\label{eq:SelfDualEqE7ExFT}
\end{equation}
for the ``\emph{improved field-strength}'' ${\cF_{\mu\nu}}^M$. If you are familiar with $\cN=8$ supergravity in four dimensions \cite{deWit:2007mt}, you will notice the similarities between that action and \eqref{E7ExFTAction}. Let us unpack the different terms in the action \eqref{E7ExFTAction}.

\paragraph{Scalar kinetic term} Using the generalised Lie derivative, we can define a covariant derivative w.r.t. the gauge connection ${A_\mu}^M$ defined as
\begin{equation}
	\Dcal_\mu = \partial_\mu - L_{A_\mu}\,.
\end{equation}
By definition, the covariant derivative should transform covariantly, this imposes the gauge transformation of the connection to be
\begin{equation}
	\delta_\Lambda {A_\mu}^M = \Dcal_\mu \Lambda^M\,,
\end{equation}
where the gauge transformation parameter $\Lambda^M$ is a tensor of weight $\lambda=\tfrac{1}{2}$. Defining $\Mcal_{MN}$ to be a tensor of weight zero, gives a definition of $\Dcal_\mu \Mcal_{MN}$ and the scalar kinetic term. 

\paragraph{Improved field strengths}
There is a natural definition for the $A_\mu{}^M$ Yang-Mills field strength:
\begin{equation}
F\indices{_{\mu\nu}^M} = [\mathcal{D}_\mu,\,\mathcal{D}_\nu] = 2 \partial_{[\mu}{A_{\nu]}}^M - [A_\mu,\,A_\nu]_E^M\,.
\end{equation}
However, this field strength is \emph{not} covariant under generalised diffeomorphisms. To compensate for this, we need to add two sets of two-forms and define a fully covariantised field strength as
\begin{equation}
\mathcal{F}\indices{_{\mu\nu}^M} = F\indices{_{\mu\nu}^M} - 12 (t^\alpha)^{MN}\partial_N B_{\mu\nu\alpha} - \frac{1}{2} \Omega^{MK}B_{\mu\nu\,K}\,.
\end{equation}
Importantly, the two-form $B_{\mu\nu\,M}$ is also constrained by the section constraints:
\begin{equation}
	(\Pbb_{1+133})^{MN}B_{\mu\nu\,M}\partial_N = 0 =	(\Pbb_{1+133})^{MN} B_M\, B_N\,.
\end{equation}
If one defines the gauge transformations of the two-forms to be
\begin{align}
&\Delta_\Lambda B_{\mu\nu\,\alpha} = (t_\alpha)_{KL} \Lambda^K {\cF_{\mu\nu}}^{L}\,,\\
&\Delta_\Lambda B_{\mu\nu\,M} = - \Omega_{KL}\left({\cF_{\mu\nu}}^K \partial_M \Lambda^L - \Lambda^L \partial_M {\cF_{\mu\nu}}^K\right)\,,\nonumber 
\end{align}
then, the field strength $\cF$ transforms covariantly as a vector of weight $\tfrac{1}{2}$. Finally, the two two-forms carry their own gauge transformations under which fields transform as
\begin{align}
	\nonumber &\delta_\Xi {A_\mu}^M = 12 (t^\alpha)^{MN} \partial_N \Xi_{\mu\,\alpha} + \frac{1}{2} \Omega^{MN} \, \Xi_{\mu\,N}\,,\\
	&\Delta_\Xi B_{\mu\nu\,\alpha} = 2 \Dcal_{[\mu}\Xi_{\nu]\alpha}\,,\\
	\nonumber &\Delta_\Xi B_{\mu\nu\,M} = 2 \Dcal_{[\mu}\Xi_{\nu]M} + 48 {(t^\alpha)_L}^K\left(\partial_K \partial_M {A_{[\mu}}^L\right)\Xi_{\nu]\alpha}\,,
\end{align}
where $\Xi_{\mu\alpha}$ is an adjoint tensor of weight $1$ and $\Xi_{\mu\,M}$ is a covector of weight $\tfrac{1}{2}$. 

\paragraph{Generalised Ricci scalar}
The generalised Ricci scalar $\hat{R}$ is given by the Einstein-Hilbert term built out of the improved Riemann tensor
	\begin{equation}
\hat{R}\indices{_{\mu\nu}^{ab}} = R\indices{_{\mu\nu}^{ab}}(\omega) + \mathcal{F}\indices{_{\mu\nu}^{M}}e^{a\rho}\partial_Me^b_\rho\,.
\end{equation}
Here, ${R_{\mu\nu}{}^{ab}}$ is the curvature of the spin connection $\omega$. This connection is computed in the usual manner from the vielbein ${e_\mu}^a$, however all partial derivatives are replaced by their covariant counterparts
\begin{equation}
	\Dcal_\mu {e_\nu}^a= \partial_\mu {e_\nu}^a - {A_\mu}^M \partial_M {e_\nu}^a - \frac{1}{2} \partial_M {A_\mu}^M {e_\nu}^a\,.
\end{equation} 

\paragraph{Topological term} The topological term in the ExFT action is more easily understood as the boundary term of a manifestly gauge-invariant exact form in five dimensions:
\begin{align}
\mathcal{S}_{top} =&- \frac{1}{24} \int\limits_{\Sigma_5} d^5x \int d^{56}Y \epsilon^{\mu\nu\rho\sigma\tau} \mathcal{F}\indices{_{\mu\nu}^M}\mathcal{D}_\rho \mathcal{F}\indices{_{\sigma\tau\,M}}\\
\nonumber=&\int\limits_{\partial\Sigma_5}d^4x\int d^{56}Y\mathcal{L}_{top}\,.
\end{align}
This can be viewed as a covariant generalisation of the topological action $\int F\wedge d F$ for a 2-form $F$.

\paragraph{The potential}
We still need to specify the ``potential'' $V$, which is
\begin{align}
V =& -\frac{1}{48} \mathcal{M}^{MN}\partial_M \mathcal{M}^{KL}\partial_N\Mcal_{KL} + \frac{1}{2} \Mcal^{MN}\partial_M\Mcal^{KL}\partial_L\Mcal_{NK} \\
\nonumber&-\frac{1}{2} g^{-1} \partial_M g \partial_N\Mcal^{MN} - \frac{1}{4} \Mcal^{MN}g^{-1}\partial_Mg g^{-1}\partial_Ng - \frac{1}{4} \Mcal^{MN} \partial_Mg^{\mu\nu}\partial_Ng_{\mu\nu}\,.
\end{align}
We used the nomenclature of ``potential'' for this last term because it only contains derivatives w.r.t. internal directions. It does not contain derivatives w.r.t. external directions. For consistent truncations, in which the internal coordinates dependencies factorise out of the equations of motion, this does reduce to a scalar potential, hence the name.

\paragraph{External diffeomorphisms}  We have given definitions for all the terms in the action of (\ref{E7ExFTAction}). Each term of the ExFT action has been built by requiring invariance under generalised \emph{internal} diffeomorphisms. The relative coefficients of these terms are determined by invariance under the generalised \emph{external} diffeomorphisms. The usual external diffeomorphisms are generated by 
\begin{equation}
\Lcal_\xi\hspace{5mm}\text{where}\hspace{5mm} \xi = \xi(x^\mu)^\mu \partial_\mu \in TM_{\text{ext}}.
\end{equation}
We can generalise this by allowing $\xi$ to depend on  the internal coordinates. Using that the weight of the vierbein is $\tfrac{1}{2}$, we define the action of the external diffeomorphisms as
\begin{align}
&\delta_\xi {e_\mu}^a = \xi^\mu \Dcal_\nu \,{e_\mu}^a + \Dcal_\mu\xi^\nu\,{e_\nu}^a\,,\\
&\delta_\xi \Mcal_{MN} = \xi^\mu \Dcal_\mu \mathcal{M}_{MN}\,,\\
&\delta_\xi {A_\mu}^M = \xi^\nu \,\mathcal{F}_{\nu\mu}{}^M + \Mcal^{MN}g_{\mu\nu}\partial_N \xi^\nu\,\,,\\
&\delta_\xi B_{\mu\nu\,\alpha} = \xi^\rho \mathcal{H}_{\mu\nu\rho\,\alpha}\,,\\
&\delta_\xi\,B_{\mu\nu\,M} = \xi^\rho \mathcal{H}_{\mu\nu\rho\,M} + 2 e\,\epsilon_{\mu\nu\rho\sigma} g^{\sigma\tau}\Dcal^\rho \left(g_{\tau\lambda} \partial_M \xi^\lambda\right)\,.
\end{align}
In the case where $\xi$ only depends on the external coordinates this reduces to the usual external diffeomorphisms. When the dependence on the $Y^M$ is non-trivial (still satisfying the section constraints) these transformations mix the different terms in the ExFT action. This is the mechanisms by which we fix the relative coefficients of the different terms in the ExFT action.

\paragraph{Fermionic completion} The E$_{7(7)}$-ExFT action is the unique action compatible with both external and internal generalised diffeomorphisms (at lowest order in the number of derivatives). This bosonic action can be extended to include fermions in a way compatible with supersymmetry \cite{Godazgar2014} (a reformulation using superspace formalisms is performed in \cite{Butter2018}). This can be seen as a surprising feature of the $\Es$-ExFT. Indeed, we have not built this action with supersymmetry in mind. The theory we have written is completely fixed by the external and $\Es$-generalised diffeomorphism and knows, \textit{a priori}, nothing of the supersymmetry of the type IIB or 11D supergravities it is supposed to encode. However, a supersymmetric completion seems to be built-in in the structure of the exceptional groups.

\begin{summaryFramed}
\begin{itemize}
	\item The field content of $\Es$-ExFT is
	\begin{equation}
	\left\{g_{\mu\nu},\,\Mcal_{MN},\,{A_\mu}^M,\,B_{\mu\nu\,\alpha},\,B_{\mu\nu\,M}\right\}
\end{equation}
\item The pseudo-action of $\Es$-ExFT is
\begin{align}
	S = \int \mathrm{d}^4x\, \mathrm{d}^{56}Y\, e\, \biggl(\hat{R} &+ \frac{1}{48} g^{\mu\nu} \mathcal{D}_\mu \mathcal{M}^{MN}\mathcal{D}_\nu \mathcal{M}_{MN}\\
	\nonumber&- \frac{1}{8} \mathcal{M}_{MN} \mathcal{F}^{\mu\nu\,M} \mathcal{F}\indices{_{\mu\nu}^N} +e^{-1} \mathcal{L}_{top} - V \biggl)\,,
\end{align}
supplemented with the twisted self-duality condition
\begin{equation}
	\mathcal{F}_{\mu\nu}{}^M = -\frac{1}{2} e \,\epsilon_{\mu\nu\rho\sigma} \,\Omega^{MN}\, \mathcal{M}_{NK} \,\mathcal{F}^{\rho\sigma K}\,.
\end{equation}
\item The naive field strength $F_{\mu\nu} = \left[\Dcal_\mu,\,\Dcal_\nu\right]$ is not a covariant quantity and needs to be improved by the two-forms.
\item The bosonic $\Es$-ExFT admits a supersymmetric completion.
\end{itemize}
\end{summaryFramed}

\subsection{Type IIB and 11D embeddings}

We have built a theory which is manifestly invariant under both external and internal generalised diffeomorphisms. We must still provide a dictionary between the ExFT fields and the IIB/11D SUGRA fields (after imposing the appropriate section constraints). Then, we must show that the ExFT e.o.m. do reduce to those of the SUGRA.

\paragraph{Type IIB dictionary} Using the IIB solution to the section constraints, we can decompose all the ExFT fields in $\SL(2)\times \GL(6)$ representations. We recall that the ExFT's generalised metric $\Mcal$ can be written as
\begin{equation}
\Mcal = \Vcal \,\Vcal^T\hspace{5mm}\text{for}\hspace{5mm} \mathcal{V}\in \Es/\mathrm{SU}(8)\,.
\end{equation} 
The coset representative $\mathcal{V}$ can be parametrised as a product of terms of the form $\exp(c^\alpha t_\alpha)$ where the $t_\alpha$'s are the positive roots of $\es$. Branching the $t_\alpha$ in $\SL(2)\times \GL(6)$ irreps we can choose the parametrisation
\begin{equation}
	\hspace{-0.1cm}\mathcal{V} = \exp\left(\phi \, t_{(0)}\right) \Vcal_6 \Vcal_2 
	\exp\left(c_{mn\,a} \, t_{(+1)}^{mn\,a}\right) 
	\exp\left(\epsilon^{klmnpq} c_{klmn} \, t_{(+2)\,pq}\right) 
	\exp\left(c_a  \, t^a_{(+3)}\right)\,.
\end{equation}
Here, $\mathcal{V}_6 \in \SL(6)/\SO(6)$ parametrises the internal IIB metric while $\mathcal{V}_2 \in \SL(2)/\SO(2)$ parametrises the IIB axio-dilaton matrix. The $t_{(\lambda)mn\cdots}$ are the generators of $\es$ of weight $\lambda$ under the $\GL(1) \subset \GL(6)$ grading. The indices $m,\,n,\,\dots\in\left\{1,\,\cdots,\,6\right\}$ transform in the fundamental representation of $\SL(6)$. The corresponding $c_{\dots}(x,\,y^m)$ fields match the type IIB bosonic field. There is also a contribution from the dual of the 2-form, i.e. a 6-form in six dimensions ($c_a$). Obtaining the explicit identification between type IIB fields and ExFT fields is not as straightforward as setting $c_{(p)} \rightarrow C_{(p)}$ in the KK ansatz. One needs to perform extra re-definitions and dualities to recover the type IIB equations of motion. This echoes the discussion at the end of section \ref{subsection:ExceptionalHiddenSymmetries}.

The vector fields of ExFT split in the same way as the coordinates do:
\begin{equation}
	{\mathcal{A}_\mu}^M = \left\{{A_\mu}^m,\,A_{\mu\,m\alpha},\,A_{\mu\,kmn},\,{A_\mu}^{m\alpha},\,A_{\mu\,m}\right\}\,.
\end{equation}
We recognise the components of the KK vectors, i.e. the off-diagonal terms of the metric, a vector contribution from the two-forms, and a vector contribution from the four-form. The last two entries are contributions coming from the dual of the 2-form and the KK-vectors. Finally, one must consider the two-forms transforming in the adjoint of $\Es$ which should be understood as dual to the scalar currents.

Setting the vectors and two-forms to zero, the dictionary is \cite{Inverso:2016eet}
\begin{align}
\mathcal{M}^{mn} &= G^{-1/2} G^{mn}\label{IIBExFTcorrespondanceB} \\
\mathcal{M}\indices{^m_{n\alpha}} &=  G^{-1/2}\, G^{mk}\, {B_{kn}}^\beta \epsilon_{\beta \alpha}\\
\mathcal{M}_{m\alpha\,n\beta} &=  G^{-1/2}\, G_{mn}\, m_{\alpha\beta} +  G^{-1/2}\, G^{kl}\, {B_{mk}}^\gamma \,{B_{nl}}^\delta \epsilon_{\alpha \gamma} \epsilon_{\beta \delta}\\
\mathcal{M}\indices{^\rho_{lmn}} &= - 2\, G^{-1/2}\, G^{\rho k} \left( C_{klmn} - \frac{3}{2}\, \epsilon_{\alpha\beta}{B_{k\left[l\right.}}^\alpha\, {B_{\left.mn\right]}}^\beta\right)\label{IIBExFTcorrespondanceE}
\end{align}
These must be supplemented by the self-duality condition for the four-form which determines the external part of the four-form $C_{\mu\nu\rho\sigma}$. These equations can be inverted to obtain the type IIB fields in terms of the ExFT generalised metric.

\paragraph{11D dictionary} As in the type IIB case, we parametrise the generalised internal metric $\mathcal{M}=\mathcal{V}\mathcal{V}^T$ as
\begin{equation}
	\mathcal{V} = \exp\left(\phi\, t_{(0)}\right) \mathcal{V}_7 \exp\left(c_{kmn} t^{kmn}_{(+2)}\right) \exp \left(\epsilon^{klmnpqr}c_{klmnpq} t_{(+4)\,r}\right)\,.
\end{equation}
We have selected only the generators of positive weight under the $\GL(1)\subset \GL(7)$. The element $\mathcal{V}_7 \in \SL(7)/\SO(7)$ encodes the internal contributions to the metric. The field $\phi$ is a dilaton (corresponding to rescaling of the internal metric), $c_{kmn}$ corresponds to the 3-form and $c_{klmnpq}$ encodes contributions to the dual 6-form.
With the same reasoning the vector fields of ExFT split into 
\begin{equation}
	{\mathcal{A}_\mu}^M = \left\{{A_\mu}^m,\,A_{\mu\,mn},\,{A_\mu}^{mn},\,A_{\mu\,m}\right\}
\end{equation}
We recognise the KK vectors coming from the metric, the vector contribution to the three-form as well as a vector contributing to the dual 6-form and the dual KK vector.

\newpage

\section{Other generalised field theories}
\label{sec:genExFT}

In the previous section, we presented the $\Es$-generalised Lie derivative as the fundamental concept underlying the construction of the $\Es$-ExFT. Thus, to build other types of ExFT, we must first build other types of generalised Lie derivatives. It turns out that for any Lie group, $G$, and any representation $R_1$ of $G$, one can build a generalised Lie derivative. The representation $R_1$ specifies the action of $G$ on both $Y^M$, the coordinates, and $\Lambda^M$, the parameter of the infinitesimal generalised transformations. From the pair $(G,\,R_1)$, we can build the projector $\mathbb{P}:\,\text{End}(R_1) \rightarrow \mathfrak{g}$, such that $\mathbb{P}\indices{_M^N_P^Q} \propto (t_\alpha)\indices{_M^N} (t^\alpha)\indices{_P^Q}$, and the associated generalised Lie derivative is
\begin{equation}
\label{eq:GenLieDerSecOther}
	L_\Lambda V^M = \Lambda^K\partial_K V^M - \alpha\, \mathbb{P}\indices{^M_N^P_Q} \partial_P \Lambda^Q V^N + \beta\, \partial_K\Lambda^K V^M\,.
\end{equation}
It depends on the choice of two real constants $\alpha$ and $\beta$. With the operator $L$ being defined, we must still check that the generalised infinitesimal diffeomorphisms $\delta_\Lambda = L_\Lambda$ form an algebra, i.e. we must solve the section constraints. Not all choices of $(G,\,R_1)$ admit interesting solutions. For example, the section constraints might force us to set all the $\partial_M=0$, which is a trivial solution. However, if we can find an interesting solution (which allows $\partial_m \neq 0$ for some subset $\{m\} \in \{M\}$ and fixes $\alpha$ and $\beta$), we can start building generalised field theories, invariant under $(G,\,R_1)$-diffeomorphisms. In this section, we will review the following choices for $(G,\,R_1)$:
\begin{itemize}
	\item $G=\Ed$ and $R_1$ given as in Table \ref{OtherExFTData}. This generalised Lie derivative encodes the gauge transformations of type IIB/11D SUGRA with $11-d$ external dimensions. The associated E$_{d(d)}$-ExFTs encode their equations of motion.
	\item $(G,\,R_1)=(\SO(d,\,d),\,2d)$. This corresponds to the duality group of closed bosonic string theory. The associated field theory is called ``\emph{Double Field Theory}'' (DFT). It can be extended to study half-maximal supergravity whose duality group is ${\SO(d,\,d+N)}$.
\end{itemize}
Furthermore, we will see which modifications have to be considered for ExFT with fewer than three external dimensions (in particular, for the E$_{8(8)}$-, E$_{9(9)}$- and E$_{11(11)}$-ExFTs). We will conclude with some remarks on the general case of the $(G,\,R_1)$-generalised Lie derivative and its associated section constraints.

\subsection{The \texorpdfstring{E$_{d(d)}$}{Ed(d)}-ExFTs}
All the details of these construction are available in either the review \cite{Berman2020} or in the original papers presenting these ExFTs, see Table \ref{OtherExFTData}. Here, we only present a lightning review of the key features common to all the E$_{d(d)}$-ExFTs. 

\paragraph{Generalised Lie derivative and section constraints} For a given $d$ between 3 and 7, we define the generalised Lie derivative as in \eqref{eq:GenLieDerSecOther}. The projector
\begin{equation}
\mathbb{P}\indices{_M^N_P^Q} = (t_\alpha)\indices{_M^N} (t^\alpha)\indices{_P^Q}\,,
\end{equation}
is built out of a basis $(t_\alpha)\indices{_M^N}$ of $\mathfrak{e}_{d(d)}$. The representation $R_1$ is labelled by the indices $M,\,N,\,...$. The coefficients $\alpha_d$ and $\beta_d$ are real constants to be fixed. 

The closure and the Jacobi identity of the generalised Lie derivative will give us the section constraints. In term of the $Y$ tensor
\begin{equation}
	Y\indices{^{MN}_{PQ}} = \delta^M_P \delta^N_Q + \alpha_d \Pbb\indices{_{(d)}^M_Q^N_P} + \beta_d \,\delta^M_Q \delta^N_P\,,
\end{equation}
the constraints reduce to \cite{Berman2012}
\begin{align}
&\phantom{\Big(}Y\indices{^{MN}_{PQ}} \partial_M \otimes \partial_N = 0\\
\nonumber &\left(Y\indices{^{MN}_{TQ}} Y\indices{^{TP}_{RS}} - Y\indices{^{MN}_{RS}} \delta^P_Q \right) \partial_{(N}\otimes \partial_{P)}= 0\\
\nonumber&\left(Y\indices{^{MN}_{TQ}}Y\indices{^{TP}_{[SR]}}+2 Y\indices{^{MN}_{[R|T|}}Y\indices{^{TP}_{S]Q}} - Y\indices{^{MN}_{[RS]}} \delta^P_Q -2 Y\indices{^{MN}_{[S|Q|}} \delta^P_{R]} \right) \partial_{(N}\otimes \partial_{P)}= 0\\
&\nonumber \left(Y\indices{^{MN}_{TQ}}Y\indices{^{TP}_{(SR)}}+2 Y\indices{^{MN}_{(R|T|}}Y\indices{^{TP}_{S)Q}}-Y\indices{^{MN}_{(RS)}} \delta^P_Q - 2 Y\indices{^{MN}_{(S|Q|}}\delta^P_{R)}\right) \partial_{[N}\otimes \partial_{P]}=0
\end{align}
For specific values of $\alpha_d$ and $\beta_d$, it turns out that the first equation implies the others. This is due to properties specific to the exceptional Lie algebras. One can find ``interesting solutions'' when choosing $\alpha_d$, $\beta_d$ and $R_1$ as in Table \ref{OtherExFTData} (notice that, since $\text{E}_{3(3)}$ is not simple, we must specify one representation $R_1$ and one coefficient $\alpha_3$ for each simple factor).
\begin{table}[!h]
\centering
\begin{tabular}{|l|c|c|c|c|c|c|}\hline
$d$ & $\text{E}_{d(d)}$ & $R_1$ & adjoint & $\alpha_d$ & $\beta_d$ & ref\\\hline
3 & $\text{SL}(2)\times \text{SL}(3)$ & ($\mathbf{2},\,\mathbf{3}$) & $(\mathbf{3},\,\mathbf{8})$ & $(2,\,3)$ & $\tfrac{1}{6}$ & \cite{Hohm2015}\\[1mm]
4 & $\text{SL}(5)$ & $\mathbf{10}$ & $\mathbf{16}$ & 3 & $\tfrac{1}{5}$ & \cite{Musaev2015}\\[1mm]
5 & $\text{SO}(5,\,5)$ & $\mathbf{16}$& $\mathbf{45}$ & 4 & $\tfrac{1}{4}$&\cite{Abzalov2015}\\[1mm]
6 & $\text{E}_{6(6)}$ & $\mathbf{27}$ & $\mathbf{78}$ & 6 & $\tfrac{1}{3}$ & \cite{Hohm:2013vpa,Musaev2014}\\[1mm]
7 & $\text{E}_{7(7)}$ & $\mathbf{56}$ & $\mathbf{133}$ & 12 & $\tfrac{1}{2}$ & \cite{Hohm:2013uia,Godazgar2014}\\[1mm]\hline
\end{tabular}
\caption{Relevant data for the $\text{E}_{d(d)}$ generalised Lie derivative and references to their original constructions.}
\label{OtherExFTData}
\end{table}

For each $d$, there are two inequivalent solutions to the section constraints. These correspond to a IIB or a 11D origin. As for the case $d=7$, these solutions are obtained by branching the $\text{E}_{d(d)}$ group down to either $\GL(d)$ or $\GL(d-1)\times \SL(2)$. By selecting the $d$ or $d-1$ highest weight coordinates $y^m$, we solve the section constraints.

\paragraph{Field content and tensor hierarchy}
The next step towards the $\Ed$-ExFT is to build field theories compatible with the generalised diffeomorphisms. They all contain an external metric $g_{\mu\nu}$, an internal metric $\mathcal{M}_{MN}$, and a vector field $A_\mu{}^M$. This field content must be completed by a \emph{tensor hierarchy}: a series of external $p$-forms, each transforming in a specific representation, $R_p$, of $\text{E}_{d(d)}$. We need to include these $p$-forms to build invariant field strengths. As seen for the case $d=7$, the naive field strength for the vector fields ($F_{\mu\nu} = \left[\mathcal{D}_\mu,\,\mathcal{D}_\nu\right]$) does not transform as a tensor for the generalised Lie derivative. This failure can be compensated by the introduction of a $R_2$-valued two-form. Then, we must specify a kinetic term for the 2-form, and thus a field strength $F_{(3)}$ for the 2-form. This field strength will not transform as a tensor and will require an $R_3$-valued 3-form to compensate... You get the idea. If this series of tensors did not stop, this would be an obstruction to the construction of an ExFT. Happily for us, it has been shown that we only need to introduce in the ExFT action these $p$-forms for $p\leq D-2$. In other words, the tensor hierarchy stops, and we can build an invariant action. 

For concreteness, we gathered the representations $R_p$ of the tensor hierarchy in the Table \ref{Table:TensorHierarchy}. For the interested reader, there is a nice systematisation of the process based on a generalised exterior derivative and interior product, see \cite{Cederwall2013,Aldazabal2013,Wang2015}. As a remark, notice that when $d=7$, the constrained two-form $B_{\mu\nu\,M}$ is not included in this formalism. Actually, all ExFT can include constrained $(D-2)$-form although they might not be needed to build an invariant action as in the $D=4$ case.

\begin{table}[h]
\centering
\begin{tabular}{|l|c|c|c|c|c|c|}\hline
$d$ & $\text{E}_{d(d)}$ & $R_1$ & $R_2$ & $R_3$ & $R_4$ & $R_5$\\\hline
3 & $\text{SL}(2)\times \text{SL}(3)$ & ($\mathbf{2},\,\mathbf{3}$) & $(\mathbf{1},\,\mathbf{3}')$ & $(\mathbf{2},\,\mathbf{1})$ & $(\mathbf{1},\,\mathbf{3})$ & $(\mathbf{2}',\,\mathbf{3}')$\\[1mm]
4 & $\text{SL}(5)$ & $\mathbf{10}$ & $\mathbf{5}'$ & $\mathbf{5}$ & $\mathbf{10}'$ & $\mathbf{24}$\\[1mm]
5 & $\text{SO}(5,\,5)$ & $\mathbf{16}$& $\mathbf{10}$ & $\mathbf{16}'$ & $\mathbf{45}$&\ \\[1mm]
6 & $\text{E}_{6(6)}$ & $\mathbf{27}$ & $\mathbf{27}'$ & $\mathbf{78}$ & &  \\[1mm]
7 & $\text{E}_{7(7)}$ & $\mathbf{56}$ & $\mathbf{133}$ &  &  & \\[1mm]\hline
\end{tabular}
\caption{The tensor hierarchy for $\text{E}_{d(d)}$-ExFTs. The $\mathbf{n}'$ denote the dual to the representation $\mathbf{n}$.}
\label{Table:TensorHierarchy}
\end{table}

\paragraph{Equations of motion} We can generalise the method used for $d=7$ to any $d$ and build an action for E$_{d(d)}$-ExFT. These actions are constructed out of a generalised Ricci scalar, a scalar kinetic term, a topological term, and a potential, in a way reminiscent of the $d=7$ case. Since the tensor hierarchy is dimension dependent, the field strength definitions also depend on $d$. We will not present these definitions here and refer to the literature for more details. Using external generalised diffeomorphisms, the relative coefficients of the terms in the action are fixed. Finally, when the number of external dimensions is even, the e.o.m. derived from the action need to be supplemented by a self-duality equation of the type \eqref{eq:SelfDualEqE7ExFT} (of course these self-duality equations, which concerns vector for $d=7$, concerns $\tfrac{(9-d)}{2}$-forms for generic $d$.). 

\subsection{Double Field Theory}
Double Field Theory is the $\SO(d,\,d)$ version of ExFT and historically precedes it. The goal of DFT is to rephrase the gravity limit of the closed bosonic string compactified on $d$ dimensions in a duality invariant way (and beyond the consistent truncation of section \ref{sec:TDualityInSUGRALimit}). The mathematical premises of DFT arose in the works \cite{Hitchin2002, Gualtieri2004} devoted to the study of generalised complex geometry. Then, the DFT was built in \cite{Hohm2010,Hohm2010a,Hohm2010b} (note that some of the ideas underlying DFT can be traced as far back as \cite{Siegel:1993th,Siegel:1993xq}). For a history of the field and further references see the reviews \cite{Aldazabal2013a,Berman2013,Hohm2013a}, as well as \cite{Hohm2013} for more details on the DFT action.

The DFT generalised Lie derivative is based on $(G,\,R_1) = (\SO(d,\,d),\,2d)$. The representation $R_1$ is the $2d$-dimensional vector representation\footnote{Note that the $\SO(5,\,5)$-ExFT is different from the $\SO(5,\,5)$-DFT because they are not built from the same $R_1$.}. Because of this simple choice, and the specifics of the $\mathfrak{so}_{d,\,d}$ algebra, we can write the generalised Lie derivative as
\begin{equation}
	(L_\Lambda V)^M = \Lambda^N\partial_N V^M - \left(\delta^M_Q \delta^P_N - \eta ^{MP}\eta_{NQ} \right)\partial_P \Lambda^Q V^N\,,
\end{equation}
where $\eta$ is the invariant tensor of $\SO(d,\,d)$:
\begin{equation}
	\eta_{MN} = \begin{pmatrix}
		0_{n\times n} & 1_{n\times n}\\ 1_{n\times n} & 0_{n\times n}
	\end{pmatrix}\,.
\end{equation}
We identify that ${Y^{MP}}_{NQ} = \eta^{MP}\eta_{NQ}$ and the section constraints simplify to
\begin{equation}
{Y^{MP}}_{NQ} \partial_M\otimes \partial_P = 0 \,\Leftrightarrow \,\eta^{MP}\partial_M\otimes \partial_P = 0\,.
\end{equation}
These equations can be solved by the usual method. We perform the branching 
\begin{equation}
\begin{array}{rll}
\mathfrak{so}_{d,\,d}\rightarrow \mathfrak{gl}_d:\,&2d &\rightarrow d_1 \oplus d'_{-1}\,\\
&Y^M &\rightarrow (y^m,\,\tilde{y}_m)
\end{array}
\end{equation}
and we keep only the highest weight coordinates $y^m$. This is a solution because $\eta$ is made of off-diagonal blocks, i.e. $\eta^{mn} =0$. Equivalently, we could have chosen the lowest weight coordinates $\tilde{y}_m$ to solve the section constraints (using the fact that $\eta_{mn} =0$). The stringy interpretation of the section constraints of DFT can be understood as choosing what we call ``excitation modes'' and ``winding modes''.

The DFT contains, as before, the external ($g_{\mu\nu}$) and internal ($\mathcal{M}_{MN}$) generalised metrics as well as vector fields ${A_\mu}^M$. The tensor hierarchy is remarkably simple and identical for any number of internal dimensions: it only contains a single singlet two-form $B_{\mu\nu}$. Finally, the DFT also contains a scalar field $\phi$ of weight one. The DFT action reads
\begin{align}
\label{eq:DFTAction}
S = \int d^n x dY^M \sqrt{-g} e^{-2 \phi}\Big(&\hat{R} + 4 \mathcal{D}_\mu \phi \mathcal{D}^\mu \phi + \frac{1}{8} \mathcal{D}_\mu\mathcal{M}_{MN}\mathcal{D}^\mu\mathcal{M}^{MN}\\
\nonumber&-\tfrac{1}{4} \mathcal{M}_{MN} {\mathcal{F}_{\mu\nu}}^M \mathcal{F}^{\mu\nu\,N} - \tfrac{1}{12} \mathcal{H}_{\mu\nu\rho}\mathcal{H}^{\mu\nu\rho}\\
\nonumber&+ \tfrac{1}{4} \mathcal{M}^{MN}\left(\partial_M g_{\mu\nu} \partial_N g^{\mu\nu} + \partial_M \ln|g|\partial_N\ln|g|\right)+ \mathcal{R}_{DFT}(\mathcal{M},\,\phi)\,\Big)\,.
\end{align}
Note that the dilaton of DFT does not identify directly with the NS-NS dilaton because of the series of rescaling of the metric and redefinitions of the scalars necessary to obtain a manifestly SO$(d,\,d)$-invariant action. Finally, DFT can be formulated with only ``internal'' dimensions, where the whole dynamic is encoded by the internal generalised metric. We have not presented it here to make the connection with ExFT and the computations of section \ref{subsubsec:LengthyComputation} clearer.

\paragraph{$\mathbf{SO(d,\,d+N)}$ field theory}

Generalised field theory also covers the half-maximal supergravities in ten-dimensions originating in string theory. There are three half-maximal ($\mathcal{N}=1$) superstring theories. They are called type I, HO\footnote{Heterotic $\SO(32)$} and HE\footnote{Heterotic $\mathrm{E}_8\times \mathrm{E}_8$} superstrings and they all admit a $D=10$ $\Ncal=1$ supergravity limit. There are two possible $\mathcal{N}=1$ supermultiplets: a gravity multiplet and a vector multiplet. The bosonic sector of the gravity multiplet contains a metric, a two-form and a dilaton. The vector multiplet only contains a vector, in the adjoint representation of a Lie group $G$, as well as its spin-1/2 superpartner. The $\mathcal{N}=1$ SUGRA bosonic action in the string frame is \cite{Bergshoeff1982,Chapline1983}
\begin{equation}
\label{eq:N=1SUGRA}
S = \frac{1}{2\kappa^2} \int d^{10}x\,\,e\,e^{-2\phi} \left(R + 4 \partial_\mu\phi \partial^\mu\phi - \frac{1}{12} \tilde{H}_{\mu\nu\rho}\tilde{H}^{\mu\nu\rho} - \frac{1}{4} \beta\,F^\alpha_{\mu\nu} F^{\alpha\,\mu\nu}\right) \,,
\end{equation}
where
\begin{equation}
H = dB - \beta\,\text{Tr}\left(A\wedge dA + \frac{2}{3} A^3\right)\,.
\end{equation}
The trace Tr is taken in the adjoint representation and ${F_{\mu\nu}}^\alpha$ is the field strength of the vector field. At the classical level, this action is invariant under diffeomorphisms and invariant under gauge symmetries for any gauge group $G$. After quantisation, it suffers gravitational, gauged, and mixed anomalies. These can be cured by adding higher order counterterms and fixing $G$ to be either $\SO(32)$ or E$_8\times$E$_8$ \cite{Green1984,Adams2010,Kim2019}. With these groups, the beta-function equations of HO (for $G=SO(32)$) and HE ($G=E_8 \times E_8$) superstring reduce to the e.o.m. of \eqref{eq:N=1SUGRA}. For $G=SO(32)$, performing the following field redefinitions
\begin{equation}
\phi \rightarrow - \phi' \hspace{2cm} g_{\mu\nu}\rightarrow e^{-\phi} g'_{\mu\nu}	
\end{equation}
we obtain the low-energy action of type I superstrings. This suggests a strong/weak duality between HO and type I superstrings \cite{Tseytlin1995}.

The bosonic sector of these SUGRA has been given an $\SO(d,\,d+N)$ ``DFT'' interpretation in \cite{Hohm2011a}. This DFT was later extended to also include a fermionic sector in a supersymmetric manner \cite{Hohm2011}. For an abelian gauge group $G$, the action is straightforwardly that of DFT \eqref{eq:DFTAction} where the index $M = 1,\,\dots,\,2d+N$. The section constraints selects the first $d$ coordinates as non-trivial. This is relevant to the study of $U(1)^N$ invariant sector of any $\mathcal{N}=1$ supergravity. In the non-abelian case, things become more complicated. It is possible to give a formally invariant $\SO(d,\,d+N)$ action although the action will only be $\SO(d,\,d) \times G$ invariant. This is done by introducing a tensor $f^{M}{}_{NK}$ encoding the structure group of $G$. The generalised Lie derivative gets corrections in $f$ and becomes
\begin{equation}
\label{eq:LieDerLprime}
	(L'_\Lambda V)^M = (L_\Lambda V)^M - \Lambda^K f^{M}{}_{KL}V^L\,.
\end{equation}
This change requires to modify the section constraints and to supplement them with
\begin{equation}
	f^{M}{}_{NK}\partial_M = 0\,.
\end{equation}
Since the tensor $f^{M}{}_{NK}$ is constant and does not transform under $\SO(d,\,d+N)$ diffeomorphisms, the $\SO(d,\,d) \times G$ group is in general broken, although the construction is formally invariant (when also acting on the constant $f$).

\subsection{\texorpdfstring{E$_{8(8)}$}{E8(8)}-ExFT}

This ExFT was first presented in \cite{Hohm2014} and its supersymmetric completion was built in \cite{Baguet2016}. To build an E$_{8(8)}$-ExFT, we would assume that we can use the strategy presented in the previous sections. This is not so. The generalised Lie derivative based on the $\mathfrak{e}_{8(8)}$ algebra and its fundamental representation $\mathbf{248}$ does not admit non-trivial solutions to the section constraints. To solve this issue, we must modify the definition of the generalised Lie derivative and add ``ancillary'' parameters. The E$_{8(8)}$-generalised Lie derivative reads
\begin{equation}
	(L_{\Lambda,\,\Sigma} V)^M = \Lambda^K\partial_KV^M - 60 \,\mathbb{P}^M{}_N{}^K{}_L \partial_K \Lambda^L V^N  + \lambda \partial_N \Lambda^N V^M - \Sigma_L {f^{LM}}_N V^N\,.
\end{equation}
The tensor $f^{LM}{}_N$ is the structure constant of $\mathfrak{e}_{8(8)}$ (the adjoint and the fundamental representations of $\mathfrak{e}_{8(8)}$ are isomorphic). There is an extra parameter $\Sigma_L$, an ``\emph{ancillary}'' parameter. For consistency, this parameter must also obey the same section constraints as the coordinates which are
\begin{equation}
	(\mathbb{P}_{1+248+3875})_{MN}{}^{KL}\partial_K \otimes \partial_L = 0\,.
\end{equation}
With these definitions, the E$_{8(8)}$ generalised Lie derivative admits two non-trivial solutions to its section constraints, again connected to a type IIB/11D interpretation.

The bosonic field content of the E$_{8(8)}$-ExFT is given by the vielbein $e_\mu{}^a$, an internal vielbein $\mathcal{V}\in \mathrm{E}_{8(8)}/\SO(16)$, and two vector fields $A_\mu{}^M$ and $B_{\mu\,M}$ used to build covariant derivatives:
\begin{equation}
	\mathcal{D}_\mu = D_\mu + L_{A_\mu,\,B_\mu}\,.
\end{equation}
The action is
\begin{align}
\label{E8ExFTAction}
	S = \int \mathrm{d}^3x\, \mathrm{d}^{248}Y\, e\, \biggl(\hat{R} &+ \frac{1}{240} g^{\mu\nu} \mathcal{D}_\mu \mathcal{M}^{MN}\mathcal{D}_\nu \mathcal{M}_{MN} +e^{-1} \mathcal{L}_{CS} - V \biggl)\,.
\end{align}
This action is simpler than that of the other $\Ed$-ExFT because the three-dimensional EM duality relates vectors to scalars. There is thus no need for a vector kinetic term.

\subsection{Others}

We could extend the argument of section \ref{subsection:ExceptionalHiddenSymmetries} for $d> 8$. It predicts that the algebra of the hidden symmetry groups corresponds to the \emph{extended} Dynkin diagram of $\mathfrak{e}_{d(d)}$. These algebras are infinite dimensional. It turns out that it is possible to build such highly exotic theories. The $E_{9(9)}$-ExFT was built in \cite{Bossard:2017aae,Bossard2018,Bossard2021}. A master ExFT can be defined for zero external dimensions, with the duality group $E_{11(11)}$ \cite{Bossard2019} following construction put forward in \cite{West2001,Tumanov2015}. A systematic study of generalised Lie derivative for Kac-Moody algebras was done in \cite{Cederwall2017}. There, the authors derive several results amongst other the nice classification of non-trivial generalised Lie derivative in the absence of ancillary parameter:
\begin{itemize}
	\item $A_r$ with $R_1$ is one of the $p$-form representation for $p = 1,\,\dots,\,r$;
	\item $B_r$ with $R_1$ the vector representation;
	\item $C_r$ with $R_1$ the symplectic-traceless or $r$-form representations;
	\item $D_r$ with $R_1$ one of the vector or spinor representations;
	\item E$_{6}$ or E$_7$ with $R_1$ the fundamental representation.
	\end{itemize}
In particular this requires $\mathfrak{g}$ to be finite dimensional. Other types of generalised Lie derivative can be built by adding nodes to the Dynkin diagram of the $\mathfrak{sl}_d$ algebra originating from diffeomorphisms. Such a study was performed in \cite{StricklandConstable2013}. 

Finally, we also mention other generalised field theory construction based on the $D=4$ half-maximal sugra duality groups $\SL(2) \times \textrm{O}(d,\,d+n)$ \cite{Ciceri:2016hup}, and the equivalent version in $D=3$ based on $\mathrm{O}(d+1,\,d+1)$ duality group where $d = 7$ or $23$ depending on the embedding in type II supergravity or bosonic string theory \cite{Hohm:2017wtr}. We also mention that massive type IIA supergravity can be encoded in a ExFT by deforming slightly the generalised Lie derivative (along the lines of \eqref{eq:LieDerLprime}) see \cite{Ciceri2016}.

In these lectures, we only covered the role of generalised field theories in the context of closed string theory. However, the open string sector also plays a significant role in the low-energy action of string theory. This requires understanding string sigma-models \cite{Duff:1989tf,Tseytlin1990,Tseytlin:1990va,Hull:2004in,Hull:2006va}, boundary conditions for strings, branes \cite{Hull:2004in,Lawrence:2006ma,Albertsson:2008gq,Blair2019}, ... in a $\Ed$-covariant manner.

\newpage

\section{Applications}
\label{sec:Applications}

Although it is nice to see the e.o.m. of maximal supergravities encoded in $\text{E}_{d(d)}$-covariant theories, ExFT is admittedly a very technical subject. To justify this complexity, it must have useful applications. Happily, ExFT allows us to obtain results that were previously out of computational reach. In this last section, we will briefly present one such application: the construction of consistent truncations of type II/11D supergravities to lower dimensional gauged maximal supergravities \cite{Berman2012a,Aldazabal2013b,Musaev:2013rq,Hohm:2014qga,Ciceri2016}.

Why consistent truncations? Because ExFT is not directly helping us to build new solutions of type II/11D SUGRA. It is just a reshuffling of degrees of freedom, and the e.o.m. of ExFT are no less complicated than that of the original theory. Having said that, the e.o.m.  of ExFT are invariant under the exceptional groups. This invariance under such large symmetry groups can be used to build new consistent truncations. In particular, we will show how to generalise the simple KK-reduction on tori and reduce type IIA/B or 11D SUGRA to lower-dimensional gauged maximal supergravity. Such reductions are useful to find solutions of the form $M_D \times M_{\text{int}}$ where $M_D$ is a $D$-dimensional maximally symmetric space. The reason is that, from a $D$-dimensional perspective, such solutions correspond to extrema of a scalar potential in gauged supergravity. 

These consistent truncation of ExFT we are going to study are called ``\emph{generalised Scherk-Schwarz reductions}'' (gSS). We will start by presenting the original Scherk-Schwarz reduction \cite{Scherk:1979zr}. Then we will describe $D$-dimensional gauged supergravities using the embedding tensor formalism \cite{deWit:2002vt,deWit:2005ub,Nicolai2000,deWit:2004nw,Bergshoeff2007,Samtleben2005}. Finally, we will display the generalised Scherk-Schwarz ansatz and how it connects with gauged supergravities \cite{Hohm:2014qga}. We will conclude with another application, which uses the gSS reduction, and computes the KK-masses for AdS/Mink/dS vacua \cite{Malek:2019eaz,Malek2020}.

\subsection{The Scherk-Schwarz reduction}  
The Scherk-Schwarz reduction ansatz \cite{Scherk:1979zr} provides consistent truncations on group manifolds, i.e. on backgrounds of the type
\begin{equation}
\mathcal{M}_{\text{ext}} \times G\,,
\end{equation}
where $G$ is a Lie group. These manifolds admit both a left- and a right-action of $G$. The fundamental idea is to only keep left-invariant fields. The left-invariant tensors on $G$ can be built out left-invariant 1-forms $U^a \in \Gamma(T^*G)$ for $a=1,\,\dots,\,\text{dim}(\mathfrak{g})$. In coordinates, this corresponds to defining the matrices $U_m{}^a(y^n)$:
\begin{equation}
	U^a(y^n)= U_m{}^a(y^n)\, dy^m\,.
\end{equation}
The index $a$ labels the adjoint of $\mathfrak{g}$ and the $y^m$ are coordinates on $G$ ($m$ obviously has the same range as $a$ and, less obviously, $U$ is an invertible matrix). 

To obtain a consistent truncation, we only allow for $y$ dependencies through the matrices $U^a$. In other words, we expand all the fields on a $U^a$ basis. For example, the metric ansatz is
\begin{equation}
	ds^2 = g_{\mu\nu}(x) dx^\mu \,dx^\nu + M_{ab}(x) (U^a(y) + A^a(x))(U^b(y) + A^b(x))\,.
\end{equation}
The fields $g_{\mu\nu}$, $A^a$ and $M_{ab}$ depend only on the external coordinates $x$ and transform as $D$-dimensional fields. In the same way, the $p$-forms are constrained to be of the form
\begin{equation}
F(x^\mu,\,y^m)= F(x^\mu)_{\mu_1\cdots \mu_r a_{1}\cdots a_{p-r}} dx^{\mu_1} \wedge \cdots\wedge (U^{a_1}+A^{a_1})\wedge \cdots\,.
\end{equation} 
These formulas reduce to the usual KK-ansatz for $G = U(1)^d$ (up to dilatonic factors).
Since all the fields in the ansatz are invariant under the left-action of $G$, any $y$-dependence factors out of the equations of motion. One should still worry about terms of the form $\nabla_a U^b$ which will appear in the e.o.m. and are not, à priori $G$-invariant. We will ignore for now this issue and return to it in section \ref{sec:GStructure}. You can check that the Einstein-Hilbert action reduces to a metric in $D$ dimensions, a Yang-Mills term with gauge group $G$ and scalars charged under $G$. Therefore, we would expect supergravities on group manifold to reduce to \emph{gauged} supergravities. This is the subject of the next subsection. 

We conclude with a last remark on the Scherk-Schwarz ansatz and its limitations: it only works if the internal manifold is a Lie group. This is very restrictive. In particular, it does not provide consistent reductions on spheres, which are coset-spaces and not Lie groups (except $S^3 \cong SU(2)$ and $S^1$). The \emph{generalised} SS ansatz will provide consistent truncation on spheres. It does so by not only taking modes invariant under subgroup of the diffeomorphisms group (in this case $G \subset \text{Diff}(G)$) but by taking modes invariant under subgroups of the generalised internal diffeomorphisms. Since this second group is bigger, it admits more subgroups and thus more consistent truncations.

\subsection{Gauged supergravities and the embedding tensor formalism}
\label{subsec:GaugeSugra}

This part is intended as a crash-course on gauged supergravities, and we refer to \cite{Trigiante:2016mnt} for a delightful review of the subject. We will focus on the embedding tensor formalism for gauged supergravity \cite{deWit:2002vt,deWit:2005ub,Nicolai2000,deWit:2004nw,Bergshoeff2007,Samtleben2005}, with an eye on the $D=4$ case \cite{deWit:2007mt}. We recall that ungauged supergravities are labelled by their field content, the number of SUSY transformations $\mathcal{N}$, and dimension $D$. Their actions and their global symmetry groups $G$ are fixed by $\mathcal{N}$, $D$, and the matter content (at least when $\mathcal{N}$ is sufficiently large). Having an ungauged action and a symmetry group, one would think that it suffices to use the usual gauging procedure: choosing a subgroup $G_g \subset G$ and replacing the derivative with covariant derivatives. This does not work for two reasons. The first reason is that the covariantisation of the derivatives breaks supersymmetry. They introduce new couplings which are not SUSY-invariant. Thus, the gauged action and the transformation rules need to be further modified to ensure gauge and supersymmetry invariance. In theory this could be done order by order in the gauge coupling constant, but it would be extremely complicated without some organising principle. The second issue with the simple covariantisation of the derivatives is that vectors transform in specific $G$-representations $R_v$ which might differ from the adjoint representation of $G$ (once again, $R_v$ is fixed by $\mathcal{N}$, $D$ and the field content). This implies that the vectors cannot be canonically identified with generators of gauge transformations to build a covariant derivative. 

To solve these two problems, we introduce an object called the \emph{embedding tensor} which will play the role of both gauge group structure constants and gauge coupling constants. This tensor, ${\Theta_M}^\alpha$, transforms in the ${R_v}^* \times \mathbf{adj}$ representation\footnote{${R_v}^*$ denotes the dual $R_v$ representation} of the symmetry group $G$. It is equivalent to a map 
\begin{equation}
	\Theta : R_v \rightarrow \mathfrak{g}:\,A^M \rightarrow A^M \Theta_M{}^\alpha t_\alpha\,,
\end{equation} 
whose image is $\text{Im}(\Theta) =\mathfrak{g}_g$, the gauge algebra. It allows us to build covariant derivatives of the form
\begin{equation}
	D_\mu = \partial_\mu + A_\mu^M {\Theta_M}^\alpha t_\alpha\,,
\end{equation}
where $M$ labels the $R_v$ indices. We can now perform the standard covariantisation procedure $\partial_\mu \rightarrow D_\mu$. Unfortunately, the new gauge-covariant action is not SUSY-invariant anymore! To solve this issue, we must correct the action and the transformation rules order by order in $\Theta$ (think about $g^2$ terms in the YM action). Amongst other things, this will introduce a \emph{scalar potential}. The presence of a scalar potential will imply the existence of a rich landscape of AdS/dS/Mink vacua, corresponding to extrema of this potential. 

A priori, there is no reasons for the series of corrections to stop at any order in $\Theta$. However, by imposing extra constraints on the embedding tensor, we can show that it does stop and that we get a well-defined supergravity. The constraints on the embedding tensors are of three types:
\begin{itemize}
\item The \emph{gauge-algebra constraint} requires $\text{Im}(\Theta) = \mathfrak{g}_g$ i.e. the embedding tensor must span a Lie subalgebra of $\mathfrak{g}$. This implies that
\begin{equation}
\label{eq:algeclosuretheta}
	[\Theta_M,\,\Theta_N] = {f_{MN}}^P \Theta_P
\end{equation}
where ${f_{MN}}^P$ are the structure constant of the gauge algebra $\mathfrak{g}_g$. In particular, these structure constants must also satisfy the Jacobi identity. This also implies that the embedding tensor is invariant under the action of $G_g$.
\item The \emph{linear constraint} imposes that the embedding tensor lies in a specific irreducible representations of $\mathfrak{g}$. This is equivalent to requiring that the projection of $\Theta$ on specific irreps is zero. The associated equations are linear in $\Theta$. The specifics depend on $\mathcal{N}$, $D$ and the field content.
\item The \emph{quadratic constraints} are a set of equations, quadratic in $\Theta$, which must be satisfied for the gauge symmetry to be compatible with the supersymmetry. The specifics depend on $\mathcal{N}$, $D$ and the field content.
\end{itemize}
The gauging procedure breaks the global symmetry group of the ungauged SUGRA to the commutant of $G_g$ in $G$, denoted 
\begin{equation}
G_{\text{gl}}= \text{Comm}_{G} (G_g)\,.
\end{equation}
If we acted with any other element of $G$, not in $G_g\times G_{\text{gl}}$, it would act non-trivially on the embedding tensor. This is the reason we say that the embedding tensor is a \emph{spurionic} object: it admits a $G$-action but, in each theory, it is fixed to a certain constant value. In gauged supergravities, $G$ is not a global symmetry group but a \emph{duality} group. It sends a $\Theta$-gauged supergravity to a different $g\cdot \Theta$-gauged supergravity. And, since the embedding tensor constraints are all $G$-covariant, the $g\cdot \Theta$-gauged supergravity is still a well-defined SUGRA. As a last remark, we mention that the gauging procedure sketched here has to be completed by adding a series of $p$-forms forming a \emph{tensor hierarchy} \cite{deWit:2008gc}. It is the SUGRA equivalent of the ExFT tensor hierarchy.

\paragraph{Example: the $D=4$ $\mathcal{N}=8$ SUGRA} We illustrate the usefulness of the embedding tensor formalism in the case $D=4$ with maximal, $\mathcal{N}=8$, supersymmetry. The ungauged maximal supergravity was first built by reducing $D=11$ SUGRA on $T^7$ \cite{Cremmer1978} and its hidden symmetry group, $\Es$, was identified in \cite{Cremmer:1997ct}. The first gaugings of this theory were obtained on a case-by-case basis, e.g. in \cite{deWit:1981eq} for the gauge group $\SO(8)$. Its embedding as $D=11$ SUGRA on $S^7$ was shown in \cite{deWit:1986iy}, well before the first constructions of ExFT. The embedding tensor formalism allows to systematise the gauging procedure, and, as we will see when discussing the gSS reduction, it is the natural language to embed SUGRAs in higher dimensional theories. 

The bosonic sector of the ungauged maximal supergravity in four dimensions admits an E$_{7(7)}$ global symmetry group. It contains 70 scalar fields parametrising the coset space $\Es/\textrm{SU}(8)$ as well as 28 \emph{electric} vectors. In order to make the $\Es$ duality group manifest, we also include 28 \emph{magnetic} vectors. Together, they transform in the fundamental representation of $\Es$. The tensor hierarchy contains two-forms $B^\alpha$ in the adjoint representation of $\Es$.
To gauge this theory, we introduce an embedding tensor $\Theta_{M}{}^\alpha \in \mathbf{56}\times \mathfrak{e}_{7(7)}$. The constraints on this embedding tensor are the followings:
\begin{itemize}
\item The \emph{gauge-algebra constraint} requires $\Theta_M$ to span a Lie subalgebra of $\es$ exactly as in \eqref{eq:algeclosuretheta}.
\item The \emph{linear constraint} imposes that the embedding tensor lies in a specific representation of $\Es$: the $\mathbf{912}$ dimensional representation. The group theoretical computation:
\begin{equation}
	\mathbf{56}\otimes \mathbf{133}= \mathbf{56}\oplus \mathbf{912}\oplus \mathbf{6480}
\end{equation}
shows that we can write explicitly the linear constraint as:
\begin{equation}
	{t_{\alpha\,M}}^N\Theta_N{}^\alpha = 0 \hspace{8mm}\text{and}\hspace{8mm} (t_\beta t^\alpha)_M{}^N \Theta_{N}{}^\beta = -\frac{1}{2} \Theta_M{}^\alpha\,.
\end{equation}
\item The \emph{quadratic constraint} is a \emph{locality constraint}. It imposes that there always exists a frame in which the gauge connection is expressed using solely 28 electric fields, and not their 28 magnetic duals. This is encoded in the equation
\begin{equation}
	\Theta_M  \Omega^{MN}\Theta_N = 0\,.
	\label{eq:quadraticContraints}
\end{equation} 
It implies that the gauge connection $A_\mu{}^M \Theta_M$, which a priori can depend on all 56 vector fields, actually only depends on at most 28 linearly independent vectors. This also imposes constraints on the possible gauge group of theory whose dimension must be less than, or equal to, $28$.
\end{itemize}
Using the linear constraint, the quadratic constraint is equivalent to the closure of the gauge algebra in \eqref{eq:algeclosuretheta}.

\begin{summaryFramed}
\begin{itemize}
	\item The gaugings of supergravities (for $\mathcal{N}$ sufficiently large) is encoded in the \emph{embedding tensor} $\Theta_M{}^\alpha \subset R_v \otimes \mathfrak{g}$. This tensor defines the covariant derivative
	\begin{equation}
	D_\mu = \partial_\mu + A_\mu{}^M \Theta_M{}^\alpha t_\alpha\,.
	\end{equation}
	\item To preserve supersymmetry, the embedding tensor must satisfy \emph{linear} and \emph{quadratic constraints}. One of these constraints is
	\begin{equation}
	\text{Im}(\Theta) = \mathfrak{g}_g\hspace{0.5cm}\text{where } \mathfrak{g}_g \text{ is the algebra of the gauge group } G_g\,.
	\end{equation}
	\item The global symmetry group of the gauged supergravity is $\text{Comm}_G(G_g)$.
\end{itemize}
\end{summaryFramed}

\subsection{The generalised Scherk-Schwarz ansatz}

The gSS ansatz uses the ExFT formulation of maximal supergravity to perform a Scherk-Schwarz-like consistent truncation to maximal gauged supergravity \cite{Hohm:2014qga}. It factors out the internal dependencies from the e.o.m. of the ExFT. We will focus on the $\Es$ case here. Guided by the $\Es$-covariance of the fields, the ansatz is:
\begin{align}
g_{\mu\nu}(x,Y) &= \rho^{-2}(Y)\, g_{\mu\nu}(x)\label{genSSAnsatz}\\
\mathcal{M}_{MN} (x,Y) &= U\indices{_M^K}(Y)\, U\indices{_N^L}(Y)\, M_{KL}(x)\label{SSAnsatz}\\
\mathcal{A}\indices{_\mu^M}(x,Y) &= \rho^{-1}\,A\indices{_\mu^N}(x)\,(U^{-1})\indices{_N^M}(Y)\\
\mathcal{B}_{\mu\nu\,\alpha}(x,Y) &= \rho^{-2}(Y)\, U\indices{_\alpha^\beta}(Y)\, B_{\mu\nu\,\beta}(x)\\
\mathcal{B}_{\mu\nu\,M}(x,Y) &= -2\,\rho^{-2}(Y)\, (U^{-1})\indices{_S^P}(Y)\,\partial_{M}U\indices{_P^R}(Y)\, B_{\mu\nu\,\alpha}(x)\, (t^\alpha)\indices{_R^S}
\end{align}
On the l.h.s. of these equations we recognise the ExFT fields, depending on both internal and external coordinates. On the r.h.s. we recognise the 4D SUGRA fields.
The entire dependency in the internal coordinates is encoded in a E$_{7(7)}$-valued matrix ${U_M}^N(Y)$, called the \emph{twist matrix}, and a function $\rho(Y)$ called the \emph{scale factor}. This $U$ matrix generalises the twist matrix of the Scherk-Schwarz ansatz.

This ansatz will produce consistent compactifications only if, after solving the section constraints, the twist matrix and the scale factors satisfy the following set of differential equations:
\begin{align}
\label{eq:consistencyConstraints}
\left[(U^{-1})\indices{_M^P}(U^{-1})\indices{_N^Q}\partial_P U\indices{_Q^K}\right]_{\textbf{912}} &= \frac{1}{7} \,\rho\, \Theta\indices{_M^\alpha} \,(t_\alpha)\indices{_N^K} \,,\\
\nonumber\partial_N(U^{-1})\indices{_M^N} - 3 \rho^{-1} \partial_N \rho (U^{-1})\indices{_M^N} &= 2\, \rho\, \vartheta_M\,,
\end{align}
where $\Theta$ and $\vartheta$ are constant vectors and $\left[\cdot\right]_{\mathbf{912}}$ is the projector on the $\mathbf{912}$ representation of $\Es$. These equations are called the \emph{consistency constraints}. They should be thought of as a generalisation of the requirement of left-invariance for $U$ in the original Scherk-Schwarz reduction. If the gSS ansatz satisfies such constraints, the 4D gauge transformations can be shown to originate from the generalised diffeomorphism $\delta_\Lambda$ with
\begin{equation}
\Lambda^M = \rho^{-1} \Lambda^P(x) (U^{-1})_P{}^M\,.
\end{equation}
This allows us to identify the embedding tensor of the gauged supergravity with the $\Theta$ tensor in the r.h.s. of \eqref{eq:consistencyConstraints}. The remaining $x^\mu$-dependent equations reduce to the ones of gauged maximal supergravity in four dimensions. The constant vector $\vartheta_M$ is related to ``trombone'' gaugings and has to do with a possible extra $\mathrm{GL}(1)$-gauging\footnote{We have consistently ignored this $\GL(1)$ factor in these notes, and we will continue doing so.}.

These consistent truncations include the simple Scherk-Schwarz reductions but not only. For example, reductions on spheres can be understood as gSS truncations\cite{Lee:2014mla}\footnote{In the language of generalised geometry, the existence of a twist matrix satisfying \eqref{eq:consistencyConstraints} correspond to a notion of ``generalised parallelisability'' which is the one used in that reference.}. On a given manifold, finding pairs $(\rho,\,U)$ satisfying (\ref{eq:consistencyConstraints}) for an unspecified $\Theta$ is not so easy. Amongst other things, it is because they must satisfy the twist conditions \emph{and} respect the section constraints of ExFT. Some examples of gSS truncations include the electric $\SO(p,\,q,\,r)$ \cite{Hohm:2014qga} and dyonic $\SO(p,\,q)\times \SO(r,\,s)$ \cite{Inverso:2016eet} gaugings, which can be uplifted to either IIA/B or 11D SUGRA. For other examples in $D=3$, see e.g. \cite{Galli2022,Eloy2023}. 

There is also the following inverse problem: given an embedding tensor ${\Theta_M}^\alpha$ in maximal supergravity, can we find a twist matrix and a scale factor satisfying \eqref{eq:consistencyConstraints}? It turns out that it is not always possible to find a twist matrix satisfying the consistency constraints (\ref{eq:consistencyConstraints}) \emph{and} the section constraints of $\Es$-ExFT. For example, there exists a class of $\mathrm{SO}(8)_\omega$-gaugings parametrised by an angle $\omega \in [0,\,\pi/4[\,$ \cite{Dall'Agata:2012bb}. At $\omega = 0$, there exists a twist matrix $U$ solving both consistency and section constraints to 11D. It corresponds to the compactification of 11D SUGRA on $S^7$ \cite{deWit:1981eq}. When $\omega\neq 0$, the solutions of (\ref{eq:consistencyConstraints}) always require more than seven extra-coordinates to be solved \cite{Inverso2017,Lee2017} and thus do not have a geometric interpretation as the compactification of a supergravity. Whether or not these theories have an interpretation in string theory remains an open question. The inverse problem, in the context of maximal supergravities, is solved in \cite{Inverso2017,Inverso:2024xok} for $D\geq 3$.

\subsubsection{More on consistent truncations: (generalised) G-structures} 
\label{sec:GStructure}
The Scherk-Schwarz ansatz is only a specific example of a more generic type of consistent truncations based on the notion of ``$G$-structure''. This will allow us to build consistent truncation to theories which are not \emph{maximal} supergravities in lower dimensions. 

\paragraph{G-structure} For our purpose, a \emph{$G$-structure} on a manifold $M$ is a set of tensors 
\begin{equation}
I_{G} = \left\{\,\Xi_i\,|\,i\in I\right\}\,,
\label{eq:defIGS}
\end{equation}
such that, $\forall p \in M$, the group leaving invariant all $\Xi \in I_{G}$ is exactly $G \subset \GL(T_pM)$ (i.e. the \emph{stabilizer} of $I_G$ is $G$).\footnote{In most cases, this definition is equivalent to a more formal one which requires a formalism I do not wish to develop in these notes: ``A $G$-structure is a reduction of the $\GL$-structure group of the frame bundle of $M$ to $G$''. See \cite{kobayashi1963foundations} chapter 5 for more details on the relevant mathematical framework.} Let us give some examples:
\begin{itemize}
\item An $\mathrm{O}(d)$-structure is given by a single rank-two symmetric positive definite tensor: a Riemannian metric $g_{\mu\nu}$. The properties of the metric ensure that for any $p\in M$, there is exactly an $\mathrm{O}(d)$ group leaving $g_{\mu\nu}(p)$ invariant. In the same way, a pseudo-metric of signature $(p,\,q)$ defines an $\mathrm{O}(p,\,q)$-structure.
\item A $\GL(d/2,\,\mathbb{C})$-structure is given by a tensor $I \in TM\otimes T^*M$ such that $I^2 = -\text{Id}$. You can check that, for any $p\in M$, there is a basis such that this tensor is 
\begin{equation}
I_p = \begin{pmatrix}0 & \text{Id}_{d/2}\\ -\text{Id}_{d/2} & 0 \end{pmatrix}\,.
\end{equation} 
The stabilizer of this matrix is $\GL(d/2,\,\mathbb{C})$. This is called an ``almost-complex structure''. We say ``almost'' because, even if the tangent space admits, point by point, a notion of complex structure, it does not mean that $M$ itself is a complex manifold (this would require the ``\emph{Nijenhuis tensor}'' to vanish and we would say that the almost-complex structure is \emph{integrable}).
\item An identity-structure is given by a globally defined basis for $TM$. In the case where $M$ is a Lie group $G$, the set of left-invariant forms, $I_e= \left\{U^a\,|\,a=1,\,\dots,\,\text{dim}(\mathfrak{g})\right\}$, provides such a basis for $TG$.
\end{itemize}
Note that even if $I_G$ and $I'_G$ define structures for the same group $G$, their corresponding $G$-structure might be different. Note also that, given a manifold $M$ and a group $G$, there may exist a topological obstruction to the existence of a $G$-structure on $M$. For example, a $\GL(d/2,\,\mathbb{C})$ structure can only exist if the third Stiefel-Whitney class $w_3(TM)$ vanishes.

As in the Scherk-Schwarz case, after choosing a structure group, we can expand all the fields as $G$-invariant tensors. The equations of motion for this ansatz are invariant under $G$, except for the terms containing covariant derivatives w.r.t. the Levi-Civita connection $\nabla_{LC}$. Indeed, the derivatives of the invariant tensors do not have to be invariant tensors themselves. This problem could be ignore in the case of the Scherk-Schwarz reduction on a group manifold because the left-invariant forms $U^a$ satisfy
\begin{equation}
dU^a = \frac{1}{2}\,f_{bc}{}^a U^b\wedge U^c\,,
\end{equation}
where the $\mathfrak{g}$-structure constants $f_{ab}{}^c$ are a constant $G$-singlet. In the more generic case, the criterion for consistency requires us to do a technical detour and to define the notion of ``\emph{intrinsic torsion}''. 

\paragraph{Intrinsic torsion} Given a $G$-structure, we can study connections 
\begin{equation}
\nabla: \Gamma(TM) \rightarrow \Gamma(T^*M \otimes TM)
\end{equation}
which are \emph{compatible} with the structure group $G$. These are the connections $\nabla$ such that 
\begin{equation}
\nabla \, \Xi = 0 \hspace{5mm}\forall\, \Xi \, \in \,I_{G}\,.
\label{eq:defCompConn}
\end{equation}
We stress that a compatible connection might not be the Levi-Civita connection. It might not even be torsionless. For a given $I_G$, there may exist more than a single compatible connection. In fact, the space of compatible connections is modelled on the affine vector space $K_{G} = T^*M \otimes \mathfrak{g}$. We will now build, out of the torsion, quantities which only depend on a choice of $G$-structure. We recall that the torsion is defined as
\begin{equation}
    T^\nabla(X,\,Y) = L^\nabla_X Y - L_X Y\,,
\end{equation}
where $L^\nabla$ is the Lie derivative with all partial derivatives replaced by $\nabla$. You can check that the torsion is a tensor. The space of all torsions for compatible connections is denoted $W$ and can be identified with a subspace of $T^*M \otimes \mathfrak{gl}_d$. The torsion of a compatible connection does not only depend on the choice of $G$-structure but also on the choice of connection. We must project out this arbitrary choice, this will lead to the notion of \emph{intrinsic} torsion.

Given two compatible connections $\nabla$ and $\nabla'$, the difference between the two $\nabla-\nabla' = \Omega \in K_{G}$ is a tensor. We define the map
\begin{equation}
\tau_{G}: K_G \rightarrow W : \Omega \rightarrow T^\nabla - T^{\nabla + \Omega} \hspace{5mm} \text{for any compatible connection }\nabla\,.
\end{equation}
This definition is independent of the choice of compatible connection $\nabla$. 
Given a compatible connection $\nabla$, its \emph{intrinsic} torsion is its torsion modulo the image of $\tau_G$. It is an element of
\begin{equation}
W^{int}_{G} = W/\text{Im}(\tau_{G})\,.
\end{equation}
It turns out that with these definitions, intrinsic torsion is independent of the choice of connection $\nabla$. It only depends on the choice of $G$-structure $I_G$. As an example, you can check that for $\mathrm{O}(d)$-structure, $W = \text{Im}(\tau_{O(d)})$. In other words, $W^{int} = \{0\}$. You can also check that for an identity structure, $\text{Im}(\tau_{Id}) =0$. This implies that the intrinsic torsion for the left-invariant one-forms on a Lie group $G$ is an element of $\mathfrak{g} \otimes \text{End}(\mathfrak{g})/\{0\}$. The intrinsic torsion can be identified with the structure constants of $G$. 

\begin{exercise}
Show that the space of compatible connection for a $G$-structure is modelled, as an affine vector space, on $K_G \cong T^*M \otimes \mathfrak{g}$.
\end{exercise}
\begin{exercise}
Show that the torsion is a tensor. 
\end{exercise}
\begin{exercise}
Show that the map $\tau_G$ is well-defined and independent on the choice of connection $\nabla$.
\end{exercise}
\begin{exercise}
Show that the space of intrinsic torsion for an $\mathrm{O}(d)$-structure is $W^{int} = \{0\}$
\end{exercise}
\begin{exercise}
Show that the space of intrinsic torsion for the identity-structure defined by the left-invariant forms on $G$ can be identified with the structure constants on $\mathfrak{g}$. 
\end{exercise}

\paragraph{Consistent truncations} These definitions allow us to give the criterion for the consistency of a truncation: 
\begin{mdframed}
Given a $G$-structure $I_G = \{\Xi_i\,|\,i\in I\}$, the expansion in terms of the invariant tensors $\Xi_i$ yields a consistent truncation if the intrinsic torsion is a constant $G$-singlet. 
\end{mdframed}
Indeed, given a compatible connection $\tilde{\nabla}$, the Levi-Civita connection can be written as
\begin{equation}
\nabla_{LC} = \tilde{\nabla} + \Omega\,\hspace{5mm}\text{where}\hspace{5mm}\Omega \in T^*M \otimes \mathfrak{gl}_d
\end{equation}
Thus, the action of the covariant derivative on the invariant tensor will only produce terms of the form $\Omega \cdot \Xi_i$. Only the terms in the intrinsic torsion defined by $\Omega$ produce non-trivial terms. This proves the consistency.

\paragraph{Examples} When $G = \mathrm{O}(d)$, the intrinsic torsion is always null because $W^{int}_{O(d)} = 0$. Thus any $\mathrm{O}(d)$-structure gives a consistent truncation. The caveat is that these consistent truncations only contain a single $\mathrm{O}(d)$-invariant scalar from the metric. This scalar corresponds to an overall scaling factor of the internal space. This makes the resulting theory quite uninteresting. In the case of the identity-structure, there is a unique compatible connection, the Weitzenb{\"o}ck connection. Its intrinsic torsion is always a singlet under the trivial group. If the torsion is constant then this reduces to the usual Scherk-Schwarz reductions. 

\paragraph{Generalised G-structures}
This whole procedure can be extended to generalised geometries with very few changes. This time, the structure group $G \subset \Ed$, the generalised connection acts on generalised vectors and the torsion $T$ must be redefined in terms of generalised Lie derivatives as
\begin{equation}
(L^\nabla_\Lambda - L_\Lambda) V =: T(\Lambda) V\hspace{5mm}\text{for}\hspace{5mm} \Lambda\text{ and }V\text{ any generalised vectors.}
\end{equation}
With these definitions, we can define the notion of intrinsic torsion and study more systematically truncations to non-maximal gauged supergravity, possibly with extra matter multiplet. The gSS consistent truncation we presented here is an example of generalised identity-structure. The first consistency constraints \eqref{eq:consistencyConstraints} is exactly the constraint that the intrinsic torsion must be a constant tensor\footnote{The second constraint is due to an $\GL(1)$ factor we have systematically omitted to discuss in these notes.}. As you can see, the intrinsic torsion is identified with the embedding tensor of the lower-dimensional theory. The $G-$singlets can be identified with the fields of the lower-dimensional theory. For a systematic study of consistent truncations from generalised $G$-structures, see \cite{Cassani2019}. We refer to \cite{Malek2017a} for a focus on consistent truncation to half-maximal supergravities and other examples.

\begin{summaryFramed}
\begin{itemize}
	\item A $G$-structure with constant, singlet, intrinsic torsion defines a consistent truncation. This can be extended to generalised geometry with ``generalised $G$-structures''.
	\item The generalised intrinsic torsion is identified with the embedding tensor.
\end{itemize}
\end{summaryFramed}

\subsection{KK spectrometry}

Given a gSS reduction on $M_{int}$, finding new solutions of the form 
\begin{equation}
M_D\times M_{\text{int}}\,
\end{equation}
where $M_D$ is either AdS$_D$, dS$_D$ or $\mathbb{R}^{1,\,D-1}$, is equivalent to the extremisation of the scalar potential of a $D$-dimensional gauged supergravity. It is then possible to compute the masses of the modes kept in that truncation, just by looking at the quadratic fluctuations about a given vacuum. However, for many purposes, it is also important to compute the masses of the full tower of KK-modes. These masses can teach us about 
\begin{enumerate}
\item The conformal dimensions of the dual operators through the AdS/CFT dictionary, and when $M_D = \text{AdS}_D$ .
\item Perturbative stability of the solution, allowing us to check that the masses of the KK-modes are positive or, when $M_D =\textrm{AdS}_D$, stay above the Breitenlohner-Freedman (BF) bound \cite{Breitenlohner1982}
\begin{equation}
m^2_{BF} L_{\text{AdS}_D}^2 = - \frac{(D-1)^2}{4}\,.
\end{equation}
\end{enumerate}
In the original example of section \ref{sec:DimRed}, where we studied a free scalar on $S^1$, these masses where easy to compute
\begin{equation}
m_n = \frac{|n|}{2\pi\,R}\,.
\end{equation}
They are labelled by an integer $n$, labelling the corresponding Fourier modes, i.e. the corresponding scalar harmonic on the circle. In that specific example, the KK-mode of lowest mass was the one kept in by the truncation. This is far from being an universal behaviour. Since consistent truncations are not EFT, there are scenarios where the higher KK-modes have masses below those of the low-dimensional theory. These scenarios are called ``\emph{space invaders scenario}'' \cite{Duff1986}. Some examples include \cite{Awada:1982pk,Cesaro2020}.

Before the use of ExFT, the procedure to obtain the masses of the KK-modes was very labour-intensive, if not completely out of computational reach. First, it required to find the scalar and $p$-form harmonics on $M_{int}$. This can already be a very complicated problem if $M_{int}$ has few or no symmetries. These harmonics were then used to provide a basis of the possible deformations around the solution. Finally, one had to plug this ansatz into the e.o.m., study the second order terms and reorganise them to obtain mass-like terms. At this point, a last matrix diagonalisation problem would yield the KK masses. 

Using ExFT, the masses of the higher KK-modes can be obtained using a technique called ``\emph{KK-spectrometry}'' \cite{Malek:2019eaz,Malek2020}. This method was used for example in \cite{Bobev:2020lsk,Giambrone:2021zvp,Giambrone2022,Eloy2023,Eloy2024} all examples with $D=5,\,4$ and $3$. By using the ExFT formulation, its fields and its e.o.m., rather than the SUGRA e.o.m., KK-spectrometry simplifies the computation of masses in two ways. The first improvement is the possibility to use only scalar harmonics around very symmetric field configurations. For example, all the KK-masses on a deformed $n$-sphere can be obtained using only the scalar harmonics on the \emph{round} $n$-sphere (provided the round $n$-sphere can be reached in the truncation). The second improvement is that ExFT reorganises nicely the e.o.m. of the supergravity. This allows us to obtain easily the mass matrices. As an illustration, we will provide the formulas for the spin-2 mass matrices of $\mathrm{E}_{6(6)}$-ExFT.

We start with an extremum of the scalar potential $M_{\underline{MN}}$ which parametrises an element $\mathrm{E}_{6(6)}/\mathrm{USp}(8)$. Recall that, using the $\mathrm{E}_{6(6)}$ duality group, it is always possible to place the solution at the origin of moduli space. We will thus assume that the point $M_{\underline{M}\underline{N}} = \delta_{\underline{M}\underline{N}}$ is an extremum of the scalar potential for some gauged supergravity admitting a gSS uplift. Therefore, we can write a twist matrix $U_{M}{}^{\underline{N}} \in \mathrm{E}_{6(6)}$ and the gSS ansatz gives us the ExFT generalised internal metric:
\begin{equation}
	\mathcal{M}_{MN} = U_{M}{}^{\underline{M}}\,\delta_{\underline{MN}}\, U_{N}{}^{\underline{N}}\,.
\end{equation}
We can expand the solution around the vacuum as
\begin{align}
\mathcal{M}_{MN} &= U_M{}^{\underline{A}}\,U_N{}^{\underline{B}}\,\left(\delta_{\underline{AB}}+ j_{\underline{AB}\,\Sigma}(x)\,\mathcal{Y}^\Sigma\right)\,,\\
\mathcal{A}_\mu{}^M &= \rho^{-1}\left(U^{-1}\right)_{\underline{A}}{}^{M} \, \mathcal{Y}^\Sigma \, A_{\mu}{}^{\underline{A}\,\Sigma}\,,\\
\mathcal{B}_{\mu\nu\,\alpha} &= \rho^{-2} U_M{}^{\underline{A}} \,B_{\mu\nu\,\underline{A}\,\Sigma} \,\mathcal{Y}^{\Sigma}\,,\\
g_{\mu\nu} &= \rho^{-2}\left(\mathring{g} + h_{\mu\nu\,\Sigma} \mathcal{Y}^\Sigma\right)\,.
\end{align}
The parameters $j_{\underline{AB}\,\Sigma}$, ${A_\mu}^{\underline{A}\Sigma}$, $B_{\mu\nu\,\underline{A}\,\Sigma}$ and $h_{\mu\nu\,\Sigma}$ are small perturbations around the solution. The functions $\mathcal{Y}^\Sigma$ are scalar harmonics w.r.t. to any metric in the gSS consistent truncation. If the consistent truncation includes a particularly simple metric with a compact isometry group $G_{max}\subset \mathrm{E}_{6(6)}$, it can be used to build a basis of scalar harmonics $\mathcal{Y}^\Sigma$. The index $\Sigma$ labels $G_{max}$-representations. From these harmonics we can define $\mathscr{T}_{\underline{N}}{}^\Sigma{}_\Omega$ as
\begin{equation}
	L_{U_{\underline{M}}} \mathcal{Y}^\Sigma = - \mathscr{T}_{\underline{N}}{}^\Sigma{}_\Omega \mathcal{Y}^\Omega\,.
\end{equation}
The matrices $\mathscr{T}_{\underline{N}}{}^\Sigma{}_\Omega$ correspond to the generators of $G_{max}$ in the representation of the $\mathcal{Y}^\Sigma$.
 
With these definitions, the masses of the spin-2 particles are the eigenvalues of the matrix
\begin{equation}
M_{\Sigma\Omega} = - \delta^{\underline{MN}} (\mathscr{T}_{\underline{M}}\mathscr{T}_{\underline{N}})_{\Sigma\Omega} \,.
\end{equation}
The mass matrices for the other fields are more involved and we refer to the original paper \cite{Malek2020} for their derivation and expression. Notice that this expression is quite simple and that, once a basis $\mathcal{Y}^\Sigma$ and $\mathscr{T}_{\underline{M}}$ has been computed, it is easy to compute the masses of any other solution in the same consistent truncation, which can be simply done by acting with the duality group to obtain the masses at another point $M_{MN}$. 

\newpage

\acknowledgments
I would like to thank first and foremost the XIX Modave Summer School's organisers: Maria Knysh, Ludovico Machet, Romain Vandepopeliere, and, in particular, No\'emie Parrini who helped me review these notes. I would also like to thank Adolfo Guarino and Gianluca Inverso for their comments on these notes and for many interesting discussions on the subject of exceptional field theory and its applications. These notes were started when the author was supported by IISN-Belgium (convention 4.4503.15), and finished supported by an INFN postdoctoral fellowship, Bando 24736.

\newpage

\appendix

\section{Lie Algebra}
\label{app:LieAlgebra}

We collect here a few properties concerning finite dimensional Lie algebras that might be useful for the reader. The goal is to refresh the memory of the reader, not to learn the subject here. We do not present any proof and refer to the textbook \cite{Georgi2018} and the shorter notes \cite{Lorier2012} for more details. 

\subsubsection*{Definitions and main results}
\begin{definition}
A \emph{real Lie algebra}, $\mathfrak{g}$, is a vector space endowed with a $\mathbb{R}$-linear alternating Lie bracket 
\begin{equation}
	\left[\bullet,\,\bullet\right]: \mathfrak{g}\times \mathfrak{g}\rightarrow \mathfrak{g}
\end{equation} satisfying the Jacobi identity
\begin{equation}
	\left[x,\,\left[y,\,z\right]\right]+\left[y,\,\left[z,\,x\right]\right]+\left[z,\,\left[x,\,y\right]\right] = 0\,.
\end{equation}
\end{definition} 

\noindent We are interested in understanding the structure and classification of the finite dimensional Lie algebras. We introduce a few definitions.
\begin{definition}
An \emph{ideal} $\mathfrak{i} \subset \mathfrak{g}$ of a Lie algebra $\mathfrak{g}$ is a sub-algebra such that $\left[\mathfrak{i},\,\mathfrak{g}\right] \subset \mathfrak{i}\,.$
\end{definition}
\begin{definition}
A \emph{solvable} Lie algebra $\mathfrak{s}$ is a Lie algebra such that $\left[\mathfrak{s},\,\mathfrak{s}\right]$ is nilpotent.
\end{definition}
\begin{definition}
A \emph{simple} Lie algebra is a non-abelian Lie algebra which contains no ideal different form zero and itself (i.e. no ``proper'' ideals). A semi-simple Lie algebra is a Lie algebra containing no proper solvable ideals. A semi-simple Lie algebra is always isomorphic to the direct sum of simple Lie algebras. 
\end{definition}
\noindent (Semi)-simple real Lie algebras are the well-behaved Lie algebra in the sense that they can be completely classified in terms of Dynkin diagrams. To check if a Lie algebra is semi-simple, we introduce a pseudo-metric on Lie algebras called the Cartan-Killing form:
\begin{definition}
The Cartan-Killing form is a symmetric bilinear form 
\begin{equation}
	B: \mathfrak{g}\otimes \mathfrak{g}\rightarrow \mathbb{R}: x\otimes y \rightarrow \mathrm{Tr}\left[\mathrm{Ad}(x)\circ \mathrm{Ad}(y)\right]
\end{equation}
where $\mathrm{Ad}(x)$ is the adjoint action of $\mathfrak{g}$ on itself.
\end{definition}
\noindent The signature and degeneracy of this form encodes information on the (semi-)simplicity of the Lie algebra. For (semi-)simple Lie algebra, the Cartan-Killing form is non-degenerate. The Cartan-Killing form also characterises solvable algebra since for any solvable algebra $\mathfrak{s}$, $x \in \mathfrak{s}$ and $y\in \left[\mathfrak{s},\,\mathfrak{s}\right]$ we have $B(x,\,y) = 0$. Solvable Lie algebras also play an important role in characterising Lie algebras because of the following theorem:
\begin{theorem}[Levi decomposition]
Any finite-dimensional Lie algebra $\mathfrak{g}$ is isomorphic to the semi-direct product of a semi-simple Lie algebra $\mathfrak{h}$ and a solvable ideal $\mathfrak{r}$.
\end{theorem}
\noindent The solvable ideal $\mathfrak{r}$ in this proposition is a maximal solvable ideal of $\mathfrak{g}$ and is called its ``radical''. These types of algebras often arise when studying the symmetries of dimensionally reduced gravity coupled to $p$-forms. The algebras of these groups are of the form 
\begin{equation}
	\mathfrak{g} = \underbrace{\mathfrak{sl}_n}_{\text{simple}} \ltimes \underbrace{(\mathbb{R} \oplus \Lambda^p \mathbb{R}^n)}_{\text{radical}}\,.
\end{equation}

\subsubsection*{Simple Lie algebra}

\noindent All simple $\mathbb{C}$ Lie algebra can be classified by the means of Dynkin diagrams. Let $\mathfrak{g}$ be a simple Lie algebra.
\begin{definition}
A maximal commutative Lie subalgebra of $\mathfrak{g}$ is called a \emph{Cartan subalgebra} of $\mathfrak{g}$. The dimension of a Cartan subalgebra is called the \emph{rank} of $\mathfrak{g}$. 
\end{definition}
\noindent Let $\mathfrak{h}$ be a Cartan subalgebra of $\mathfrak{g}$ and $\{H_i\,|\,i=1,\,\dots,\,\textrm{rank}(\mathfrak{g})\}$ be a basis of $\mathfrak{h}$. Given an irreducible representation $\rho: \mathfrak{g} \rightarrow \text{End}(V)$ of $\mathfrak{g}$, the matrices $\rho(H_i)$ can be simultaneously diagonalised. This implies that they share eigenvectors which can be uniquely identified by their eigenvalues $\mu_i$. These eigenvalues are called \emph{weights}. Given a representation, the set of all weights is called the \emph{weights of the representation}. These weights completely characterise the representation. 
\begin{definition}
The weights of the adjoint representation are called the \emph{roots} of $\mathfrak{g}$. By extension, the root also refers to the element of $\mathfrak{g}$ which is the eigenvector associated to a given weight.
\end{definition}
\noindent  Weights are $r$-tuples of real numbers it is thus possible to order them. Therefore, the roots of a Lie algebra can be split into positive and negative roots. The full algebra is generated, as a vector space, by the sum of the Cartan generators, the negative $\mathfrak{r}_-$ and the positive $\mathfrak{r}_+$ roots:
\begin{equation}
\mathfrak{g} = \mathfrak{r}_- \oplus \mathfrak{h} \oplus \mathfrak{r}_+\,.
\end{equation}
Amongst the positive roots, certains are called \emph{simple}. They are defined as the positive roots which cannot be written as the sum of two other positive roots. There are as many simple roots as the rank of $\mathfrak{g}$. From the simple roots, it is possible to reconstruct the whole algebra $\mathfrak{g}$ as well as its Lie bracket. Sets of simple roots can be characterised by \emph{Dynkin diagrams}. The different Dynkin diagrams each corresponds to a different Lie algebra. The use of weights also allows to give a unique label to any possible irreducible representation of a Lie algebra. To do so, we define the \emph{fundamental weight} as the dual of the roots. They are defined as the vector $\lambda_j$ satisfying $\tfrac{\langle\mu^k \alpha_i\rangle}{\alpha_j^2} = \delta^k_{j}$. Any weight can be written using integer coefficients as $\sum\limits_k \lambda_k \mu^k$. The $\lambda_k$s of the highest weight of a representation are the \emph{Dynkin label} of the representation and identify it uniquely.
\begin{figure}[h]
	\centering
		\includegraphics[width=0.50\textwidth]{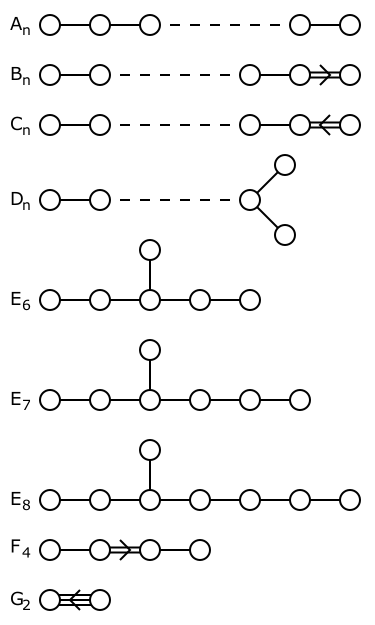}
	\caption{List of Dynkin diagrams and their corresponding Lie algebra}
	\label{fig:Connected_Dynkin_Diagrams}
\end{figure}

Weights can also be used to classify induced representations of subalgebra through ``\emph{branching rules}''. Let $\mathfrak{h}\subset \mathfrak{g}$ be two simple Lie algebras and $(\rho,\,V)$ be an irreducible representation of $\mathfrak{g}$. This implies that $(\rho_{|\mathfrak{h}},\,V)$ induces a representation of $\mathfrak{h}$. However, $\rho_{|\mathfrak{h}}$ might not be a simple representation. Writing $V_i$ the $\mathfrak{h}$-invariant subspaces of $V$, $\rho$ induces a series of irreducible representations $(\rho_i,\,V_i)$ $i\in I$. \emph{Branching rules} describe how irreducible reprensentation of $\mathfrak{g}$ induces irreducible representations $\rho_i$ of $\mathfrak{h}$. We refer to \cite{Yamatsu2015} for a few examples.

\subsubsection*{Real forms}

\noindent Given a complex Lie algebra $\mathfrak{g}$, there is more than one corresponding real Lie algebra $\mathfrak{g}_0$, i.e. more than a single real Lie algebra such that 
\begin{equation}
\mathfrak{g} = \mathfrak{g}_0 \otimes_\mathbb{R} \mathbb{C}.
\end{equation}
There always exists a compact form of $\mathfrak{g}$, which is obtained by taking the anti-Hermitian elements of the algebra. The associated Lie group is compact. There always exists a ``split-real'' form which is a ``maximally non-compact'' real form of the algebra. This is the case of the $\es$ Lie algebra presented in the next paragraph. The different real forms can be identified by studying the signature of the associated Cartan-Killing form. Roughly speaking, the generators with positive norm will correspond to ``non-compact'' generators whereas those of negative norm will correspond to ``compact generators'' (i.e. in a given representation, non-compact generators generate the group $\mathbb{R}$ whereas compact ones generate the group $\Urm(1)$).

You already know several examples of real forms. Let us take the algebra $\mathfrak{so}_n(\mathbb{C})$, generated by anti-symmetric matrices of dimension $n\times n$. A possible real form is $\mathfrak{so}_n(\mathbb{R})$ the $\mathbb{R}$-algebra of antisymmetric matrices. Another possibility would be to choose $\mathfrak{so}_{p,\,q}$, the real Lie algebra of matrices preserving a metric of signature $(-1_p,\,1_q)$ with $p+q=n$. You can check that 
\begin{equation}
\mathfrak{so}_{p,\,q} \otimes_\mathbb{R} \mathbb{C} \cong \mathfrak{so}_n \cong \mathfrak{so}_n(\mathbb{R})\otimes_\mathbb{R} \mathbb{C}\,.
\end{equation}

\subsubsection*{The exceptional Lie group \texorpdfstring{$\Es$}{E7(7)}}
\label{app:E7-conventions}

\noindent The group $\Es$ is the split-real form associated with the $\es$ Lie algebra. It is a group of dimension 133 whose maximal compact subgroup is $\SU(8)/\Zbb_2$\footnote{We often disregard the $\mathbb{Z}_2$ factor and write, wrongly, $\SU(8) \subset \Es$}. The fundamental representation of $\Es$ is the $\mathbf{56}$, its adjoint representation is the $\mathbf{133}$. Another representation of interest for supergravity is the $\mathbf{912} \subset \mathbf{133}\otimes \mathbf{56}$. The Cartan-Killing metric on $\es$ induces a pseudo-metric on $\Es$. The pair $(\Es,\,\SU(8))$ is reducible which means that we can split the algebra $\es$ as 
\begin{equation}
	\es = \mathfrak{su}(8) \oplus \mathfrak{K}\,,
\end{equation}
with 
\begin{equation}
[\mathfrak{su}(8),\,\mathfrak{su}(8)] \subset \mathfrak{su}(8)\hspace{5mm},\hspace{5mm}[\mathfrak{su}(8),\,\mathfrak{K}] \subset \mathfrak{K}\hspace{5mm} \text{and}\hspace{5mm}[\mathfrak{K},\,\mathfrak{K}]\subset \mathfrak{su}(8)\,.
\end{equation}
This means that $\mathfrak{K}$ admits an $SU(8)$ representation. Under the branching $\Es \rightarrow \SU(8)$ the adjoint branches as
\begin{equation}
	\mathbf{133} \rightarrow \mathbf{63}\oplus \mathbf{70}\,.
\end{equation}
 where $\mathbf{63}$ spans the $\mathfrak{su}(8)$ algebra and the $\mathfrak{K}=\mathbf{70}$ is a representation of $\mathfrak{su}(8)$. We can choose $\mathfrak{K}$ such that it is orthogonal to $\mathfrak{su}_8$ w.r.t. the Cartan-Killing metric. This allows us to endow the quotient space $\Es/\SU(8)$ with a Riemannian structure induced by the Cartan-Killing metric.

Another interesting way to study $\Es$, is through the branching to its subgroup $\SL(8)$. The adjoint representation of $\,\textrm{E}_{7(7)}\,$ splits into $\,\textbf{133} \rightarrow \textbf{63} \oplus \textbf{70}\,$ under $\,\textrm{SL}(8)\subset \textrm{E}_{7(7)}\,$. This implies a splitting of generators of the form 
\begin{equation}
\,t_{\alpha} \rightarrow \underbrace{t_{A}{}^{B}}_{\mathfrak{sl}_8} \oplus \underbrace{t_{ABCD}}_{\mathbf{70}}\,\text{ with }\,t_{A}{}^{A}=0\,\text{ and }\,t_{ABCD} = t_{[ABCD]}\,.
\end{equation}
The index $A,\,B,\,\dots = 1,\,\cdots,\,8$ and label the fundamental representation of $\mathfrak{sl}_8$.
 The fundamental representation of $\,\textrm{E}_{7(7)}\,$ branches as $\,\textbf{56} \rightarrow \textbf{28} \oplus \textbf{28}'\,$ so the fundamental $\,\textrm{E}_{7(7)}\,$ index splits as $\,_{\mathbb{M}} \rightarrow _{[AB]} \oplus ^{[AB]}\,$. Then, the $\,\textbf{63}\,$ generators of $\mathfrak{sl}_8$ correspond to the $\es$ generators of the form
\begin{equation}
\label{E7_gen_63}
[t_A{}^B]_\mathbb{M}{}^\mathbb{N} = \frac{1}{\sqrt{12}}
\begin{pmatrix}2 \, \delta^{[E}_{[C} \, [t_A{}^B]_{D]}{}^{F]} & 0 \\ 
0 & - 2 \,\delta^{[C}_{[E} \,\, [t_A{}^B]_{F]}{}^{D]}  \end{pmatrix} 
\end{equation}
with
\begin{equation}%
[t_A{}^B]_C{}^D = 4 \, \delta_{A}^C \, \delta_D^B - \frac{1}{2} \,  \delta_A^B \, \delta_C^D \ ,
\end{equation}
whereas the remaining $\,\textbf{70}\,$ generators, extending $\mathfrak{sl}_8$ to $\es$, take the form
\begin{equation}
\label{E7_gen_70}
[ t_{ABCD} ]_\mathbb{M}{}^\mathbb{N} = \sqrt{12} \begin{pmatrix}0 & \epsilon_{ABCDEFGH} \\ 
4! \,\delta^{EFGH}_{ABCD} & 0\end{pmatrix} \ .
\end{equation}
They are normalized such that $\,\text{Tr}(t_\alpha t_\beta{}^t) = \delta_{\alpha\beta}\,$. The Cartan-Killing matrix is then given by 
\begin{equation}
\label{E7_KC}
\mathcal{K}_{\alpha \beta} =\text{Tr}(t_\alpha t_\beta) = 
\left\lbrace 
\begin{array}{l}
1 \hspace{5mm} \textrm{ if } \hspace{5mm} \beta = \alpha^t \\[2mm]
0 \hspace{5mm} \textrm{ otherwise} 
\end{array} 
\right.  \ ,
\end{equation}
whereby $\,\alpha^t\,$ we refer to the generator $\,t_{\alpha^t} \equiv (t_\alpha)^t\,$. With the generators in (\ref{E7_gen_63}) and (\ref{E7_gen_70}) one has that
\begin{equation}
(t_A{}^B)^t = t_B{}^A
\hspace{8mm} \textrm{ and } \hspace{8mm} 
(t_{ABCD})^t = \frac{1}{4!} \, \epsilon^{ABCDEFGH} \, t_{EFGH} \ .
\end{equation}
Note that if $\,t_{\alpha}\,$ is a positive root of the $\,\mathfrak{e}_{7(7)}\,$ algebra then $\,t_{\alpha^{t}}\,$ is the corresponding negative root.

\newpage

\bibliography{references}
\end{document}